\documentclass[preprint2]{aastex}

\usepackage{epsfig}

\begin{document}

\title{GASPS - a Herschel survey of gas and dust in Protoplanetary Disks:  Summary and Initial Statistics}


\author{
W.R.F. Dent\altaffilmark{1}, W.F. Thi\altaffilmark{2,4}, I. Kamp\altaffilmark{2}, J.P. Williams\altaffilmark{3}, F. Menard\altaffilmark{4}, 
S. Andrews\altaffilmark{5}, D. Ardila\altaffilmark{6}, G. Aresu\altaffilmark{2}, J-C. Augereau\altaffilmark{4},
D. Barrado y Navascues\altaffilmark{7,8}, S. Brittain\altaffilmark{9},
A. Carmona\altaffilmark{4}, D. Ciardi\altaffilmark{10}, W. Danchi\altaffilmark{11}, J. Donaldson\altaffilmark{12},
G. Duchene\altaffilmark{4,13}, C. Eiroa\altaffilmark{14}, D. Fedele\altaffilmark{15}, C. Grady\altaffilmark{16},
I. de Gregorio-Molsalvo\altaffilmark{1}, C. Howard\altaffilmark{17}, N. Hu\'elamo\altaffilmark{7}, A. Krivov\altaffilmark{18}, J. Lebreton\altaffilmark{4}, R. Liseau\altaffilmark{19}, 
C. Martin-Zaidi\altaffilmark{4}, G. Mathews\altaffilmark{3}, G. Meeus\altaffilmark{14},
I. Mendigut\'ia\altaffilmark{7}, B. Montesinos\altaffilmark{7}, M. Morales-Calderon\altaffilmark{7}, A. Mora\altaffilmark{20}
H. Nomura\altaffilmark{21}, E. Pantin\altaffilmark{31}, I. Pascucci\altaffilmark{22}, N. Phillips\altaffilmark{1}, C. Pinte\altaffilmark{4}, L. Podio\altaffilmark{1}, S.K. Ramsay\altaffilmark{23}, B. Riaz\altaffilmark{24},
P. Riviere-Marichalar\altaffilmark{7}, A. Roberge\altaffilmark{11},  G. Sandell\altaffilmark{17}, E. Solano\altaffilmark{7}, I. Tilling\altaffilmark{25}, J.M. Torrelles\altaffilmark{26},
B. Vandenbusche\altaffilmark{27}, S. Vicente\altaffilmark{2}, G.J. White\altaffilmark{28,29}, P. Woitke\altaffilmark{30}
}


\altaffiltext{1}{ALMA SCO, Alonso de Cordova 3107, Vitacura, Santiago, Chile}
\altaffiltext{2}{Kapteyn Astronomical Institute, Postbus 800, 9700 AV Groningen, The Netherlands}
\altaffiltext{3}{Institute for Astronomy, University of Hawaii at Manoa, Honolulu, HI, USA}
\altaffiltext{4}{UJF-Grenoble 1 / CNRS-INSU, Institut de Planétologie et d'Astrophysique de Grenoble (IPAG) UMR 5274, Grenoble, F-38041, France}
\altaffiltext{5}{Harvard-Smithsonian Center for Astrophysics, 60 Garden Street, Cambridge, MA, USA}
\altaffiltext{6}{NASA Herschel Science Center, Caltech, 1200 E. California Blvd., Pasadena, CA, USA}
\altaffiltext{7}{Centro de Astrobiolog\'ia -– Depto. Astrof\'isica (CSIC–-INTA), ESAC Campus, PO Box 78, E-28691 Villanueva de la Ca\~nada, Spain}
\altaffiltext{8}{Calar Alto Observatory, Centro Astron\'omico Hispano Alem\'an, C/Jes\'us Durb\'an Rem\'on, E-04004 Almer\'{\i}a, Spain}
\altaffiltext{9}{Dept. of Physics \& Astronomy, 118 Kinard Laboratory, Clemson University, Clemson, SC, USA}
\altaffiltext{10}{NASA Exoplanet Science Institute/Caltech, Pasadena, CA, USA}
\altaffiltext{11}{NASA Goddard Space Flight Center, Exoplanets and Stellar Astrophysics Laboratory, Code 667, Greenbelt, MD 20771, USA}
\altaffiltext{12}{Department of Astronomy, University of Maryland, College Park, MD, USA}
\altaffiltext{13}{Astronomy Department, University of California, Berkeley, CA, USA}
\altaffiltext{14}{Dep. de Fisica Teorica, Fac. de Ciencias, UAM Campus Cantoblanco, Madrid, Spain}
\altaffiltext{15}{Max Planck Institut f\"ur Extraterrestrische Physik, Giessenbachstrasse 1, Garching, Germany}
\altaffiltext{16}{Eureka Scientific, 2452 Delmer, Suite 100, Oakland, CA, USA}
\altaffiltext{17}{SOFIA-USRA, NASA Ames Research Center, MS 232-12, Building N232, Rm. 146, P. O. Box 1,
Moffett Field, CA 94035-0001, USA}
\altaffiltext{18}{Astrophysikalishes Institut, Friedrich-Schiller-Universit\"at Jena, Schillerg\"asschen 2-3, Jena, Germany}
\altaffiltext{19}{Dept. of Earth and Space Sciences, Chalmers University of Technology, Onsala Space Observatory, Onsala, Sweden}
\altaffiltext{20}{ESA-ESAC Gaia SOC. PO Box 78. 28691 Villanueva de la Canada, Madrid, Spain}
\altaffiltext{21}{Department of Astronomy, Graduate School of Science, Kyoto University, Kyoto 606-8502, Japan}
\altaffiltext{22}{Lunar and Planetary Laboratory, The University of Arizona, Tucson, AZ 85721, USA}
\altaffiltext{23}{European Southern Observatory, Karl-Schwarzschild-Strasse 2, 85748 Garching bei M\"unchen, Germany}
\altaffiltext{24}{Centre for Astrophysics Research, University of Hertfordshire, Hatfield, UK}
\altaffiltext{25}{University of Edinburgh, Royal Observatory, Edinburgh, Blackford Hill, Edinburgh, UK}
\altaffiltext{26}{CSIC-UB/IEEC, Universitat de Barcelona, Mart\'i Franqu\`es 1, Barcelona, Spain}
\altaffiltext{27}{Instituut voor Sterrenkunde, Katholieke Universiteit Leuven, Celestijnenlaan 200D, Heverlee, Belgium}
\altaffiltext{28}{Department of Physical Sciences, The Open University, Milton Keynes, UK}
\altaffiltext{29}{RALSpace, Rutherford Appleton Laboratory, Chilton, Didcot OX11 0NL, UK}
\altaffiltext{30}{SUPA, School of Physics \& Astronomy, University of St. Andrews, North Haugh, St. Andrews, KY16 9SS, UK}
\altaffiltext{31}{Laboratoire AIM, CEA/DSM - CNRS - Université Paris Diderot, IRFU/SAp, CE Saclay, France 
}



\begin{abstract}
We describe a large-scale far-infrared line and continuum survey of protoplanetary disk through to young debris disk systems carried out using the PACS instrument on the Herschel Space Observatory. This Open Time Key Program, known as GASPS (Gas Survey of Protoplanetary Systems), targeted $\sim$250 young stars in narrow wavelength regions covering the [OI] fine structure line at 63$\micron$, the brightest far-infrared line in such objects.  A subset of the brightest targets were also surveyed in [OI] 145$\micron$, [CII] at 157$\micron$, as well as several transitions of H$_2$O and high-excitation CO lines at selected wavelengths between 78 and 180$\micron$.  Additionally, GASPS included continuum photometry at 70, 100 and 160$\micron$, around the peak of the dust emission. The targets were SED Class II-III T Tauri stars and debris disks from 7 nearby young associations, along with a comparable sample of isolated Herbig AeBe stars. The aim was to study the global gas and dust content in a wide sample of circumstellar disks, combining the results with models in a systematic way.
In this overview paper we review the scientific aims, target selection and observing strategy. We summarise  some of the initial results, showing line identifications, listing the detections, and giving a first statistical study of line detectability.

The [OI] line at 63$\micron$ was the brightest line seen in almost all objects, by a factor of $\sim10$. Overall [OI]63$\micron$ detection rates were 49\%, with 100\% of HAeBe stars and 43\% of T Tauri stars detected. A comparison with published disk dust masses (derived mainly from sub-mm continuum, assuming standard values of the mm mass opacity) shows a {\em dust} mass threshold for [OI]63$\micron$ detection of $\sim10^{-5}M_{\odot}$. Normalising to a distance of 140pc, 84\% of objects with dust masses $\geq 10^{-5}M_{\odot}$ can be detected in this line in the present survey, along with 32\% of those of mass $10^{-6} - 10^{-5}M_{\odot}$, and only a very small number of unusual objects with lower masses. This is consistent with models with a moderate UV excess and disk flaring. For a given disk mass, [OI] detectability is lower for M stars compared with earlier spectral types.
Both the continuum and line emission was, in most systems, spatially and spectrally unresolved and centred on the star, suggesting emission in most cases was from the disk. Approximately 10 objects showed resolved emission, most likely from outflows.

In the GASPS sample, [OI] detection rates in T Tauri associations in the 0.3-4Myr age range were $\sim$50\%. For each association in the 5-20Myr age range, $\sim$2 stars remain detectable in [OI]63$\micron$, and no systems were detected in associations with age $>$20Myr. Comparing with the total number of young stars in each association, and assuming a ISM-like gas/dust ratio, this indicates that $\sim$18\% of stars retain a gas-rich disk of total mass $1M_{Jupiter}$ for 1-4Myr, 1-7\% keep such disks for 5-10Myr, but none are detected beyond 10-20Myr. 

The brightest [OI] objects from GASPS were also observed in [OI]145$\micron$, [CII]157$\micron$ and CO J=18-17, with detection rates of 20-40\%. Detection of the [CII] line was not correlated with disk mass, suggesting it arises more commonly from a compact remnant envelope. 

\end{abstract}



\section{Introduction}

One of the most significant astronomical discoveries of the past decade has been the realisation that roughly 20\% of
main-sequence FGK stars harbour planets \citep{Fis05, Bor11}.
Moreover, at least 16\% of FGK main-sequence stars are found to have a debris disk more massive than the dust in our own Solar System, indicating an unseen population of colliding planetesimals \citep{Tril08, Wya08}. 
These two independent results imply that the planet and planetesimal formation process is common and robust, and can lead to a wide diversity of systems. However, it is not clear how young gas-rich disks - where planet formation is either still occuring or has recently completed - evolve into mature planetary and/or debris disk systems.

Both debris disks and main sequence planetary systems are gas-poor. Debris systems, composed of grains in a collisional cascade, have dust masses of $\leq10^{-7} M_{\odot}$, although the mass of planetesimals - thought to be the starting point of the cascade - may be $10^{-4} M_{\odot}$ or more \citep{Wya02}.  In most cases, no molecular gas is detected \citep{Den95}; however, in a very few nearby young debris systems such as $\beta$ Pic, a small mass of mostly atomic gas is seen \citep{Lag98, Olo01, Ro06}. Possible gas formation mechanisms in such systems are secondary release during grain-grain collisions \citep{The05}, photodesorption from dust grains \citep{Che07}, or sublimation from comets \citep{Zu12}.

By contrast, the material around young, pre-main-sequence (but optically-visible) stars is gas-rich. Such so-called protoplanetary disks are found towards $\sim$10\% of stars aged 5~Myr, and at least 80\% of stars aged $<1$~Myr \citep{HLL01}. Similar in size to debris disks, their {\em dust} masses are typically $10^{-5} - 10^{-3}M_{\odot}$. The assumption normally made is that 99\% of the disk mass is gas, the same as that of the natant interstellar cloud, leading to these disks being described as `primordial'. Their total masses would then be similar to that of the minimum mass Solar Nebula \citep{WC11}. In the even younger, so-called `protostellar' stage (typically $\leq 0.1$Myr), disk masses may be still larger, approaching that of the protostar itself. Systems at this phase are usually optically obscured, as the remnant cloud and infalling envelope have not yet dispersed.

Dissipation of the primordial disk gas limits the timescale for giant planet formation, affects the dynamics of planetary
bodies of all sizes during their formation, and determines the final architecture and constitution of planetary systems. No planet formation will take place without gas to damp the particle velocities.
The methods for removal of the gas and dust components are generally different.  Photoevaporation is thought to be important in gas dissipation \citep{Gor04, GH09}, and bipolar outflow jets may also play a role. Molecular species will also be depleted in regions exposed to the photodestructive effects of UV \citep{Kam04}. Selective removal of the dust can be caused by the interaction with a planet \citep{Rice06}, ice or refractory grain sublimation \citep{Thi05}, or by grain growth and settling \citep{DD05}. As noted by these authors, these effects can be very rapid, occuring on timescales $\sim 10^4$yrs - significantly shorter than the disk ages.

Statistical studies show that the presence of Jupiter-mass planets in mature systems is strongly influenced by stellar metallicity, mass, and binary companions \citep{Fis05, Jo07}.  But do the stellar characteristics also affect disks? Age clearly affects the fraction of stars with primordial disks \citep{HLL01} and, on a longer timescale, debris disks \citep{Carp09}.  Disks may be affected by binary companions in debris systems \citep{Tril07} but not substantially by stellar metallicity in debris or protoplanetary systems \citep{Gre06, Dor11, Mald12}. And although there appears to be no direct correlation between debris disks and planets \citep{Bry09}, there are clearly some systems which have both, and where the planet creates a gap or affects the disk shape \citep{Ka08, Tha10, Hu11}.
At present, there is no clear observational evidence that primordial disk lifetimes are significantly affected by the stellar mass \citep{Bo11, Er11}.

To study disk evolution and look for general trends, many large and unbiased dust continuum surveys have been carried out, in the near-infrared \citep{KH95, HLL01}, the mid to far-infrared (most complete out to a wavelength of 70$\micron$ in the Spitzer projects FEPS \citep{Hill08} and c2d \citep{Eva09}), and in the sub-mm \citep{AW05}.
They indicate primordial disk lifetimes of a few Myr, but there is a broad distribution of dust mass at any particular age, with notable outliers. For example, although the fraction of disk-bearing stars in the $\sim$1Myr-old Taurus star-forming region is as high as 75\%, a significant minority of its stars have no detectable dust excess \citep{Luh10}.

Dust emission is ostensibly easy to interpret, as the normally optically-thin sub-mm continuum can be used to directly estimate the dust mass, $M_d$, by employing a mass opacity, $\kappa_{\nu}$ and emissivity power law, $\beta$. Typical values adopted in the literature are $\kappa_{\nu}=1.7 g cm^{-2}$ and $\beta=1.0$ \citep{AW05}. However, $\kappa_{\nu}$ depends on the grain size distribution \citep{Dal06}, and most of the solid body mass may be in large grains contributing little to the observed flux \citep{Wya02, Kriv08}. One option is to define a dust mass which only includes solid material smaller than 1mm \citep{Thi10}; this is reasonably consistent with the standard literature value of  $\kappa_{\nu}$. Deriving the {\em total} disk mass requires an assumption of the gas:dust ratio; normally the interstellar medium value of 100 is used, but it is unclear whether this value is maintained in disks (and it is certainly not valid for debris-dominated systems). Throughout most of this current work, we have quoted disk masses in terms of the {\em dust} mass, $M_d$, allowing comparisons to be made independently of the gas/dust ratio.

Although gas dominates the mass (at least for protoplanetary disks), emission lines are generally more difficult than the continuum to both observe and interpret.  The bulk of the gas is in H$_2$ which has no dipolar moment. The observed intensity is affected by abundance variations due to complex chemistry, molecular photodissociation, or freezeout in the cool disk midplane, as well as high optical depths and uncertain excitation processes.  Unlike the continuum, a more limited number of gas surveys have been carried out. Mid-IR studies of H$_2$O show emission from the inner 10AU in many T Tauri but few HAeBe stars \citep{Pont10}.  Spitzer surveys of [NeII] at 12.8$\micron$ show warm gas in many systems, thought to arise from winds from the disk surface at radii up to $\sim$10AU \citep{Lah07, Pasc09}. At sub-mm wavelengths, limited surveys of low-J rotational lines of CO have been carried out, both with single-dish telescopes \citep{Zuc95, Den05}, and with interferometers \citep{Obe10}. Most of this emission arises from 30-300AU radii. In the far-IR, scans of a few bright embedded Class 0 YSOs (e.g. L1448-mm) and massive young stars (e.g. Orion-KL) with the long-wavelength spectrometer (LWS) on ISO showed rich spectra, including fine structure lines of [OI] and [CII], and many transitions of H$_2$O and CO \citep{Bene02, vanD04}. Class I YSOs also show similar lines, albeit fainter than the Class 0s. 
Among less embedded systems, the bright `prototypical'  Class I-II object T~Tauri\footnotemark~also has many FIR lines \citep{Spin00}. However, more typical optically-visible Class II-III objects were not detected in FIR lines due to the relatively low sensitivity and large beams.

\footnotetext{There is some discussion as to whether T~Tauri should actually be classified as a Class I YSO with a massive envelope: typically, the prototypical object in a class actually turns out to be rather unusual!}

Both gas and dust observations suffer from problems in interpretation, and both are needed for the best understanding of disks.
The motivation for GASPS was to conduct a relatively large, systematic study of gas and dust in the far-infrared, utilising the sensitivity improvements available with the Herschel Space Observatory\footnote{{\it Herschel} is an ESA space observatory with science instruments provided by European-led Principal Investigator consortia and with important participation from NASA.}. The survey covers a broad sample of optically-visible young systems, from Class II gas-rich protoplanetary disks, through to Class III objects and gas-poor debris disks. It focusses on the brightest lines and the FIR peak of the continuum emission, and is complemented by data at other wavelengths. In this paper, we describe the survey and observing techniques ($\S$2), discuss the target selection criteria and give the complete target list ($\S$3 and Table~\ref{tab:slist}). $\S$4 summarises the  origins of FIR line emission in these objects, and outlines the modeling used by GASPS. In $\S$5, we give an overview of the results, with the lines detected and the detection statistics.

\section{The GASPS survey}

GASPS (Gas Survey of Protoplanetary Systems) uses the Photodetector Array Camera \& Spectrophotometer (PACS) \citep{Pog10} on the Herschel spacecraft \citep{Pil10} to study a predefined set of the brightest lines and dust continuum in the far-infrared from a relatively large sample of targets. The aim was to allow the detection of gas in systems with a disk mass limit similar to, and possibly lower than, existing sub-mm dust surveys. The wavelength
coverage of the spectroscopic observations was tailored to include the [OI] $^3$P$_1$-$^3$P$_2$ and $^3$P$_0$-$^3$P$_1$ lines at 63 and 145$\micron$, [CII] $^2$P$_{3/2}$-$^2$P$_{1/2}$ at 157$\micron$, several H$_2$O lines, particularly those at 78 and 180$\micron$, along with adjacent transitions of CO and OH observable without incurring a significant penalty on the total required time.
In addition, GASPS provides accurate far-infrared photometry at 70 and 160$\micron$ and, in most cases, 100$\micron$\footnotemark.
The project was awarded 400 hours of time to survey up to 250 young systems (in several cases, multiple systems were covered in the same observation), and observations were taken at various times between Dec 2010 and July 2012. The spectrometer was used with up to 4 settings per target, each of which covered a relatively small wavelength range (typically $\Delta \lambda / \lambda \sim 5\%$) simultaneously in two PACS grating orders. Most objects were observed in the setting covering the [OI]63$\micron$ line, with a subset of the brighter ones observed in the other settings, resulting in a two-phase survey strategy (see \S\ref{sec:phased}).

\footnotetext{FIR fluxes for the brighter objects are available from IRAS, ISO or Spitzer, although in many cases the fluxes at $\lambda > 70\micron$ are unreliable because of the large beams and confusion levels involved - particularly in star forming regions. }

To help maintain the unbiased nature of the survey, targets were chosen with a wide range of spectral type, disk dust mass, age and other stellar parameters (see \S3). They were located in 7 well-studied young clusters and associations, with a distance range of 40 - 200~pc (with the majority around 150~pc). Assuming typical disk sizes of 100-300AU \citep{WC11} and with the angular resolution of Herschel/PACS of 5 arcsec at the shortest wavelength, line and continuum emission from the disk itself is unlikely to be spatially resolved. With the highest spectral resolution of PACS (88 km s$^{-1}$ at 63$\micron$), disk emission will also not be spectrally resolved (the Keplerian rotation velocity of most of the disk mass is $\sim$10-50 km s$^{-1}$). However, non-disk components such as outflow jets or ambient cloud emission may be resolved (see \S\ref{sec:lines}). In most cases, all we have is a single measurement of the line flux on each target, yielding highly degenerate solutions to the underlying disk physics and chemistry. In the absence of resolution, a survey covering a wide range of target parameters is required, along with detailed modeling and data from other wavelengths.

\subsection{Observational technique and survey strategy}

\subsubsection{PACS Spectroscopy}
The PACS instrument \citep{Pog10} offers resolutions of 1500-3400 (200-88 km s$^{-1}$) and the ability to observe most of the 60-200$\micron$ wavelength range. In spectroscopic mode, PACS provides an IFU with a 
5$\times$5 array of spaxels, and a pixel size of 9.2~arcsec. By comparison the instrumental PSF ranges from 4.5~arcsec (FWHM) at 63$\micron$ up to 13~arcsec at 180$\micron$.
For the GASPS project, spectral observations were taken using line-scan or range-scan modes, whereby the grating is scanned over a small wavelength range, taking data from all detector pixels simultaneously. Wavelengths around the central region of the spectra are observed by all of the 16 detector pixels to minimise flat-fielding problems due to inter-pixel variations. Line-scan observations have small wavelength coverage, and are designed to cover a single spectrally-unresolved line and immediately adjacent continuum with the full sensitivity. Range-scan observations have an arbitrary wavelength coverage and for GASPS were set up to include several close lines of interest by scanning up to 2$\micron$. Table 1 shows the settings of the four wavelength scans A through D. Each has a primary line targetted in one of the grating orders; the secondary simultaneous grating order (given in brackets) was used to observe other useful lines (the full list of lines detected during the course of the survey is given in $\S$5.1).  For a few individual targets, integration times longer that given in Table~1 were used for followup of marginal detections.
Note that with the array spectral scanning technique, not all wavelengths are being observed by a detector at all times, so the noise level increases towards the spectrum edges. However, the rms values in the scan centre in Table 1 are in good agreement with predictions.  The observed sensitivity at 63$\micron$ is equivalent to a $3\sigma$ line luminosity sensitivity limit of $6\times10^{-6}  ($D$/140)^2$ L$_{\odot}$, where D is the source distance in pc. 

\begin{table*}[htbp]
\begin{center}
\caption{\sf PACS wavelength settings and sensitivities in the primary grating order. }
\begin{tabular}[t]{clcccccl}
\hline\
Grating & Primary & Primary $\lambda$ & Grating & Time$^1$ & Predicted & Observed & Notes \\
setting & line  &   ($\micron$)  & orders & (s) & rms$^2$ & rms$^2$ & \\
\hline
\hline\\[-3mm]
A & [OI]63.2 & 63.08-63.29 & 3 (1) & 1760 &4.6  & 2.5-3 & LineScan  \\
B & [CII]157.7 & 157.10-158.90 & 1 (2) & 1500 & 1.3 & 0.8-1.4 & RangeScan \\
C & H$_2$O 180 & 178.90-181.00 & 1 (2) & 2000 & 1.7 & 2.0 & RangeScan\\
D & [OI]145.5 & 144.00-146.10 & 1 (2) & 1630 & 1.4 & 1.6 & RangeScan\\
Phot & (Blue) & 60-85 & 3 & 180 & 2.3 & 2.6-3.0 & \\
Phot & (Green) & 85-130 & 3 & 180 & 2.7 & 2.6-3.0 & \\
Phot & (Red) & 130-210 & 1 & 360 & 3.6 & 4.7-9.0 &(background)\\
\hline
\end{tabular}
\end{center}
\label{tab:pacs}
\footnotesize{(1) Approximate times for most targets in the main survey, not including overheads.

(2) Note that noise rms levels are given in units of $10^{-18}Wm^{-2}$ for the spectroscopic observations, and in mJy for the continuum photometry.}
\end{table*}

Observations were performed in chop-nod mode with a small throw (1.5 arcmin), primarily to remove telescope and background variations. This chops out smooth background emission from around the source, but equally it may result in confusion from chopping onto extended emission. In some targets, this could be seen in the [CII] line (\S\ref{sec:cii-confus}).

\subsubsection{PACS Photometry}

Photometric data were obtained using the fast scanning mode of the PACS imaging photometer, operating at central wavelengths of 70 and 160$\micron$ simultaneously (and repeated at 100 and 160$\micron$ in most objects). This technique scans the telescope over the source, using relatively short scan lengths of 3.5~arcmin, and small (4~arcsec) orthogonal steps between each scan. Two scans were performed, at 70 and 110 degrees to the array, to improve the final image fidelity and avoid striping effects in the scan direction. 
The photometer array field-of-view is 3.5$\times$1.75 arcmin and, although the resultant image does not have constant signal:noise over the field, the noise level in the central $180 \times 80$~arcsec region varied by less than 20\%. The technique was found to be more sensitive than the chop-nod method and in some cases, several objects could be covered in the same field. It also enabled searches for faint companions in the radius range  $\sim$1500-7000~AU.
The required sensitivity for the photometric observations was better than 5mJy rms at 100 and 160$\micron$, and a factor of $\sim2$ lower at 70$\micron$. Although the FIR continuum flux from disks is dependent on the stellar luminosity and mean disk temperature as well as the dust mass (and may be optically thick), disks of dust mass $10^{-5} - 10^{-3} M_{\odot}$ have IRAS 60$\micron$ fluxes of typically 1Jy at the fiducial distance of 150~pc. So the survey should detect dust in systems 1-2 orders of magnitude fainter than this. The noise level of the observations was generally close to the original prediction (see Table 1), although in some cases it was limited by galactic background emission at 160$\micron$.

Although the spectroscopic data could in principle be used to give narrowband continuum fluxes from the line-free parts of the spectra (albeit with a factor of $\sim$30 less sensitivity than the full photometry), it was found that the photometric accuracy of these data was lower than the broad-band photometry, and generally they were not used for SED fitting.

\subsubsection{Phased survey strategy and data reduction}
\label{sec:phased}

The spectroscopic observations were carried out in one or more of the wavelength settings in Table 1. Phase I of the project consisted of [OI]63$\micron$ observations of most targets using grating setting A, concatenated observations of the brightest $\sim 10\%$ of targets in settings B-D (in order to reduce spacecraft slew overheads), plus photometry\footnotemark. Note that not all objects from the initial survey list were observed in the lines; based on early survey results, a number of targets were dropped as they were deemed too faint in continuum to have likely emission in any line. In addition a few Taurus objects were dropped from both continuum and line observing based on updated re-classification as field stars \citep{Luh09}. Phase II of the project consisted of flexible followup of the brightest [OI] targets using grating settings B, C and/or D, as well as some deeper observations of a few individual sources.

\footnotetext{Some bright targets were dropped from the photometric list, as suitable data was available from other Herschel surveys, e.g. some of the Taurus and ChaII objects were covered by the Goult Belt Survey.}

During the course of the survey, GASPS photometric and spectroscopic data were reduced using prevailing versions of the standard Herschel data processing pipeline, HIPE \citep{Ott10}. This provides calibrated FITS images and datacubes; further photometry and spectroscopic extraction were performed with packages such as STARLINK Gaia. However, the released version of HIPE evolved during the course of the mission and different versions were used to reduce GASPS datasets in different publications, ranging from v2.3 in early data \citep{Meus10} to v7.0 \citep{Riv12a} and v9.0 (Howard et al., submitted). Later HIPE releases generally have improved calibration as well as better flat-fielding, and the complete GASPS survey is to be re-reduced using a single mature version before being made publicly available as a systematically-calibrated dataset.
The current work makes use of data extracted from the Herschel science archive during 2012, but the detection statistics presented here are unlikely to change significantly in the final data release.

\section{Target selection}

GASPS targets were selected from the 7 well-known nearby young star formation regions and associations listed in Table 2 and described in \S3.1.  The complete target list, with system parameters from the literature, is given in Table \ref{tab:targets}. For completeness, we list all the initial targets in this table, although some were not observed in spectral lines in the final survey (see above). The criteria used to select the targets were:

\begin{itemize}
\item{ Age range 0.3 to 30 Myr. As discussed by several authors (e.g. \citet{Har01}), stellar ages are uncertain - particularly for $\geq 10$Myr - and in these cases it may be better to take the ensemble age for a cluster rather than ages of individual stars.  Systems of age $\leq$0.3 Myr were considered more likely to include non-disk emission components such as remnant ambient material, infalling envelopes, or energetic outflows (see \S\ref{sec:lines}). Those older than $\sim$30Myr were expected to have very little circumstellar gas.}
\item{Optically visible stars. This means mostly SED Class II, III, Transition Objects or debris disks. Targets have optical extinctions less than $\sim 3^m$. We avoided embedded objects (i.e. Class 0 - I), because of potential confusion from extended surrounding gas.}
\item{Disk {\em dust} masses\footnotemark~mostly in the range $10^{-3}\leq M_d\leq10^{-7}M_{\odot}$. Also included were a number of coeval stars with Class III SEDs, or upper limits for $M_d$ of $\sim10^{-6}M_{\odot}$, where the lack of continuum excesses in the IR or sub-mm suggested negligible warm or cool dust. Some of these still had gas accretion signatures and were included as they potentially could be associated with moderate masses of gas. In addition a number of debris disks in the young associations were also observed, with $M_d$ as low as $10^{-11}M_{\odot}$.}
\footnotetext{Values of $M_d$ were mostly based on published mm wavelength continuum observations, with estimates based on shorter wavelength data in some cases.}

\item {Stellar spectral type A0 through M5. A similar range of stellar
spectral types was chosen in each region where possible, although to increase the numbers of early spectral type objects, we also identified a sample of isolated well-studied Herbig AeBe stars with a similar age spread to that of the clusters. The resulting stellar mass range was $\sim0.2 - 3M_{\odot}$, based on published HR diagrams.}
\item{Nearby regions, with distances of $<$ 180~pc for low-mass and $<$200~pc for HAeBe stars.}
\item{Low confusion level (from Herschel Confusion Noise Estimator). Confusion noise was $<$100mJy at 100$\micron$. This meant that several dense star-forming regions such as $\rho$ Ophiuchus were excluded from the survey.}
\item{Extensive photometric and, in many cases, spectroscopic datasets available at other wavelengths.}
\item{A range of accretion rates (based initially on H$\alpha$ equivalent width, EW), X-ray luminosity, and binary separation.}
\end{itemize}

\begin{table*}[htbp]
\label{tab:assoc}
\begin{center}
\caption{\sf Summary of Clusters and associations in GASPS}
\begin{tabular}[t]{lcccccc}
\hline\
Group & Distance & Age & Disk fraction$^1$ & GASPS & Notes/Main population \\
 & (pc) & (Myr) & \% & targets & \\
\hline
\hline\\[-3mm]
Taurus & 140 & 0.3-4 & 90 & 106 & Class I-III T Tauri stars\\
Cha II & 178 & 2-3 & 75 & 19 & Class II T Tauri stars\\
$\eta$~Cha & 97 & 5-9 & 56 & 17 & T Tauri and debris disks\\
TW Hya & 30-70 & 8-10 & $\geq$30 & 13 & T Tauri and debris disks\\
Upper Sco & 145 & 5/11 & 20 & 44 & Class II-III T Tauri stars.\\
$\beta$ Pic & 10-50 & 10-20 & $\geq$37 & 18 & Debris disks\\
Tuc Hor & 20-60 & 30 & $\geq$26 & 16 & Debris disks\\
HAeBe stars & 20-200 & $\sim$0.5-30 & 100 & 24 & Includes debris disks\\
\hline
\end{tabular}
\end{center}
\footnotesize{(1) Note: disk fractions are based on published photometric excesses. They are the fraction of stars with any measured disk, so include mostly debris disks in the older associations and protoplanetary disks in the younger star forming regions.}
\end{table*}

\subsection{Individual associations}

\subsubsection{Taurus}

The Taurus star formation complex lies at 140~pc with a depth along the line of sight of $\sim$20pc \citep{To09}. Taurus contains mostly low-mass stars with an age range of $<0.1$ up to $\sim$10Myr \citep{PS02, Gu10}. Stars in this region have been extensively studied at many wavelengths, and the census of Class 0-II YSOs is essentially complete, with a significant fraction of the Class III YSOs also known \citep{Reb10}. Stellar parameters in Table \ref{tab:targets} are from \citet{KH95} and \citet{Ken98}, updated where appropriate by values in \citet{Fur06}, \citet{Gu07} and \citet{Reb10}, and with disk masses from \citet{AW05} and \citet{Cur11}.

\subsubsection{Cha II}

Chamaeleon II is a nearby (178\,pc; \citet{Whi97}) star forming region included in the Spitzer Legacy cores-to-disks program \citep{Eva09} and the Gould Belt key program \citep{And10}.
It contains a lightly clustered distribution of low mass
YSOs in a range of evolutionary states with spectral types K--M
and ages estimated from protostellar evolutionary tracks of 4$\pm$2 \,Myr \citep{Spez08}.
The GASPS subsample consists of 19 targets from Cha II, generally with infrared colors
of Class II objects and/or H$\alpha$ equivalent widths indicative of gas accretion.
Although there exists no deep sub-mm survey to give disk mass estimates (the survey of \citet{You05} only detected DK~Cha and possibly IRAS12500-7658, with a dust mass limit on the other objects of $2\times 10^{-4}M_{\odot}$), \citet{Alc08} have estimated masses based on SED fits to far-IR data, and so we give these values (using their D01 models) in Table \ref{tab:targets}. However, it should be noted that these entail higher uncertainties compared with estimates from the sub-mm.
Spectral types are from \citet{Spez08} (who cite Hartigan (1993) for SpT), binarity is discussed in \citet{Alc08} and L$_x$ comes from the ROSAT survey of \citet{Alc00}.

\subsubsection{Upper Sco}

The mean cluster age and distance of Upper Sco was estimated as 5~Myr and 145pc by \citet{deZ99} although more recent estimates suggest it may be as old as 11Myr \citep{Pec12}, which is consistent with the rather low disk fraction observed. 
The 8-70$\micron$ SED has been used to identify disks as Class II, Class III or debris \citep{Carp09} and these classifications are given in Table \ref{tab:targets}. The disk masses and system parameters are from the sub-mm observations of \citet{Mat12} and references therein.

\subsubsection{$\eta$~Cha}
This is a compact grouping of $\sim$19 stars, first identified as a young association through X-ray observations \citep{Ma99}. One of the reasons for interest in this cluster is its' age, at an estimated 8~Myr, and relatively close distance (97~pc). The disk fraction, based on Spitzer observations at 24 and 70$\micron$ \citep{Gau08}, is 56\%, which is relatively large for the age of the association (c.f. \citet{HLL01}). A number of the stars are active accretors, and at least two are identified as Class II T Tauri stars \citep{Sic09}. In Table \ref{tab:targets}, spectral types, H$\alpha$ EW, and the presence of an infrared excess are based on \citet{Sic09}, and information on binarity is from \citet{Bou06}. X-ray luminosities are taken from \citet{Ma99} and \citet{Lop10}, and disk mass estimates are mostly based on FIR measurements \citep{Cur11}.

\subsubsection{TW Hya association}
First recognised as a group of nearby young stars by \citet{Kas97}, the number of members in the TW Hya association (TWA) is now at least 25 \citep{Webb99, Ma05}. It is the closest association with accreting T Tauri stars, and includes two classical T Tauri stars (TW~Hya itself, and Hen 3-600), and two bright debris disks (HD~98800, a hierarchical multiple system, and HR4796A). \citet{Low05} used Spitzer to measure 24 and 70$\micron$ excesses around TWA members and found these four systems have a 24$\micron$ excess a factor of $\sim$100 larger than the other members. However, several of the other stars also have evidence of dust disks, from weak excesses at longer wavelengths. The age of this system is confirmed at $\sim 10$Myr \citep{Bar06}. Parameters in Table \ref{tab:targets} are taken from \citet{DLR04} and \citet{S07}, with disk masses mostly from \citet{Mat07}. Note that TWA member HR4796A (TWA11) is listed under the HAeBe stars as A-12.

\subsubsection{$\beta$~Pic}
The moving group associated with $\beta$~Pic was identified by \citet{Bar99}, and membership extended by \citet{Zu01} and others (see \citet{Tor08} for a summary). With a derived mean age of 12Myr, and range in distance of 10-50~pc, many of its' members have been extensively studied over a wide range of wavelengths, including 24 and 70$\micron$ with Spitzer \citep{Reb08}, as well as the submm \citep{Nil09}. The disk fraction is $\geq 37\%$ \citep{Reb08}, and includes a number of debris disks in addition to $\beta$~Pic itself\footnotemark. Data in Table \ref{tab:targets} are mostly from the above references. Detailed results from GASPS, including model fits to the photometry, have been presented for HD~181327, one of the brightest debris disks in this group, and HD~172555 \citep{Leb12, Riv12b}.

\footnotetext{$\beta$~Pic was observed as part of the Herschel GT program `Stellar Disk Evolution' (P.I. G.Olofsson).}

\subsubsection{Tuc Hor}
This stellar association was first recognised by \citet{ZW00} and \citet{Tor00}, who derived an age of 20-40~Myr and distance range of 20-60~pc. No N-band excesses were seen around any stars in Tuc Hor \citep{Mam04}, however, a Spitzer study at 24 and 70$\micron$ \citep{Smi06} showed 5/21 stars with a measurable excess at 70$\micron$. \citet{Zu11} subsequently extended the search and found several more stars with IR excesses. All such systems in Tuc Hor are thought to be debris disks, and this is the oldest association in GASPS. The photometric data have been presented in \citet{Don12}.

\subsubsection{Herbig Ae Be stars}

The survey includes 25 IR-excess stars of spectral types late B to F, improving the statistics at the higher end of the stellar mass range (around 2-4$M_{\odot}$)\footnotemark. \footnotetext{Note that the HAe star AB Aur is listed under the Taurus subsample as T-101 in Table \ref{tab:slist}.} This sample also includes some A-type stars with excesses where the classification is less clear: the peculiar Be star 51 Oph \citep{Thi05}, and 5 systems which may be classified as debris or HAe (including 49~Cet, where the age was recently revised to 40Myr \citep{Zu12}). Like the lower-mass counterparts, the program HAeBe stars are biased toward isolated systems which have published IR excesses and ancilliary data (particularly UV spectra, resolved coronagraphic images and/or millimeter interferometry). HD~97048 - one of the brightest targets in our sample - had prior evidence from ISO of [OI] and possibly [CII] emission \citep{Lor99}. The HAe sample includes several disks with large gaps and/or cavities, as well as 2-3 systems with jets.  Unlike the T Tauri stars, which are represented in sufficient number to permit statistical evaluation of association ages, the HAe stars represent extremes in stellar and disk properties, and have more uncertain ages except where there are common proper motion late-type companions.  Stellar parameters in Table \ref{tab:targets} are mostly taken from \citet{Mo09} and \citet{Meu12}, with disk masses from \citet{Ack04} and \citet{Sa11}. Note that the SED classifications in the Table are different to the T Tauri class, and are based on the mid-IR slope as suggested by \citet{Meus01}: group I has an SED rising to longer wavelengths in the mid-IR, and group II has a falling SED. Results from the GASPS HAeBe subsample have been presented in \citet{Meu12}.

\subsection{Ancilliary data}

Many of the GASPS targets are well-known systems, with photometry in optical through to mid-IR (including Spitzer fluxes at wavelengths as long as 70$\micron$), and sub-mm (mostly 850$\micron$). In addition, H$\alpha$ or Br$\gamma$ line strengths are published for many targets. Derived parameters such as stellar spectral type, T$_{eff}$, disk dust mass and SED Class are also mostly available, and the most recent published estimates are given in Table \ref{tab:targets}. 
As part of the GASPS project, we have endeavoured to obtain such data in cases where it is missing, and to make the target sample uniform both in data and in derived parameters.
One additional issue is that much of the published photometry is not contemporaneous; in some cases, photometric points in the optical and NIR have been taken 20 years apart. For time-variable objects, SED fitting under these circumstances may be significantly affected, and more recent optical and near-IR photometry is being obtained for a number of the targets in order to improve the reliability of SED fits.

\section{Origins of far infrared lines from young stars}
\label{sec:lines}

At far-infrared wavelengths, common species such as C, O and N have several prominent fine-structure transitions. [OI] lines at 63 and 145$\micron$ and the [CII] line at 157$\micron$ are important cloud coolants, on a galactic scale \citep{Sta91}, in photodissociation regions \citep{Holl91}, and in circumstellar disks \citep{Kamp03, Gor04}. In star formation regions, both models and observations indicate that [OI]63$\micron$ is the single brightest emission line in the FIR/sub-mm.
Abundant molecules such as CO and H$_2$O also have numerous rotational lines throughout the FIR with energy levels of a few hundred K, and can trace the `warm' gas components.
Other FIR-emitting species such as OH are photodissociation products of H$_2$O, and are therefore predicted to be abundant \citep{Naj10}. 
Around individual young stars, FIR lines can arise from several different regions. For example, CO and OH emission from young highly luminous HAeBe stars was thought to arise from dense regions of size $\sim 200~AU$ \citep{Gia99}, but it was unclear whether these were disks or remnant envelopes. Lines are also seen from high-velocity jets and low-velocity photoevaporating disk winds, and the relative contributions of disk, outflow, disk wind, and envelope will depend on the SED class, stellar radiation field, disk structure, mass loss rate and the environment. 

The GASPS project involves both in-depth studies of individual targets using multi-wavelength data, as well as a statistical analysis of the full FIR sample. Interpretation of the results generally requires detailed comparison with models and in the following sections, we outline methods of estimating the contributions to the FIR line emission, focussing on Class II-III YSOs, which form the bulk of our targets.

\subsection{Disks}
Fine structure atomic line emission arises from the surface of disks at A$_V$$\sim$1 over a wide range of radii, where the stellar UV or X-ray photons ionise the exposed gas to produce a mainly atomic extended disk atmosphere \citep{Mei08, Gor08, Woit09}. [OI]63$\micron$ is predicted to be the brightest line from disks at any wavelength, with line luminosities as high as $\sim10^{-4}$L$_{\odot}$ from T Tauri systems \citep{Gor08}. It becomes optically thick relatively easily, and traces the mean gas temperature on the disk surface rather than the mass directly.  
FIR molecular lines such as CO and H$_2$O also arise from the warm heated surface of dense disks \citep{Woit09b}. However, molecular photodissociation in more tenuous debris disk systems may mean the atomic fine structure lines will dominate the FIR \citep{Kamp03, Zag10}. Line fluxes depend strongly on the disk structure (for example, a flared disk has a larger exposed surface area, resulting in brighter lines - \citet{Jonk04}), the radiation field from the central star, as well as the details of chemistry and gas/dust ratio in the disk atmosphere.  In the following we summarise the methods used for modeling the emission.

\subsubsection{Disk modeling}

{\bf MCFOST} and {\bf ProDiMo} are the two main codes used in GASPS to model the protoplanetary disk structure and appearance.  {\bf MCFOST} is a three-dimensional Monte Carlo continuum and line radiative transfer code \citep{Pin06, Pin09}. The parametrized input disk density distributions can accomodate structures such as holes, gaps and dust settling. The calculation of the dust temperature and radiation field takes into account non-isotropic scattering, absorption and re-emission based on the local dust properties. The code uses a large variety of grain size distributions and compositions, e.g. porous grains and icy grains. SEDs, thermal and scattered light images, visibilities as well as line emission are derived by a Monte-Carlo method and ray-tracing of the final physical disk structure. {\bf ProDiMo} is a two-dimensional thermo-chemical disk code that calculates the vertical hydrostatic equilibrium, gas phase (e.g. neutral-neutral, ion-molecule, photochemistry, X-ray chemistry) and gas-grain chemistry (ad- and desorption processes), 2D continuum radiative transfer (with isotropic scattering), detailed gas heating/cooling processes (including X-rays) using 2-directional escape probability, and spatial decoupling of gas and dust (e.g. settling) \citep{Woit09, Kamp10, Ares11, Woit11}. The observables derived from the resulting chemo-physical disk structure include SEDs, line fluxes, profiles and images.
For optically thin cases such as debris disks, we also use {\bf GRaTer}, a ray-tracing code incorporating a large variety of grain compositions which fits SEDs, images and interferometric visibilities using parametrized optically thin disk models \citep{Aug99}.

For modeling individual sources, grids of {\bf MCFOST} or {\bf GRaTer} models were run over a broad parameter space to identify the best fitting dust model, based on SEDs, images and interferometric data when available. {\bf MCFOST} results were passed to {\bf ProDiMo} for detailed gas modeling. Examples of this approach are \citet{Meus10} and \citet{Thi10} and for a debris disk, \citet{Leb12}. Another approach employs a genetic algorithm minimisation strategy with the {\bf ProDiMo} models to find local minima in the parameter space constrained by observations. Examples are found in \citet{Till12} and \citet{Woit11}.

\begin{figure}[htbp]
\plotone{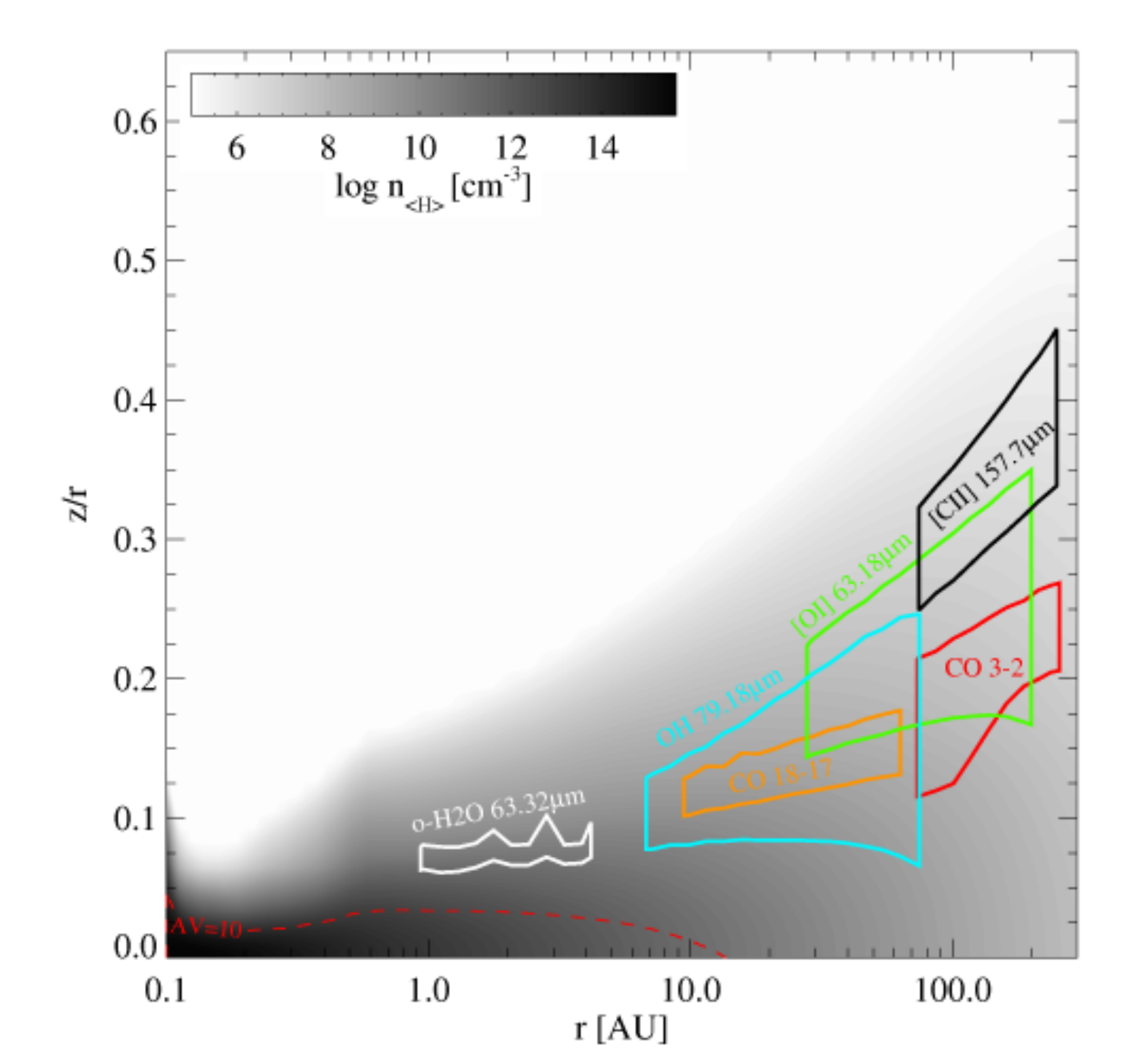}
\plotone{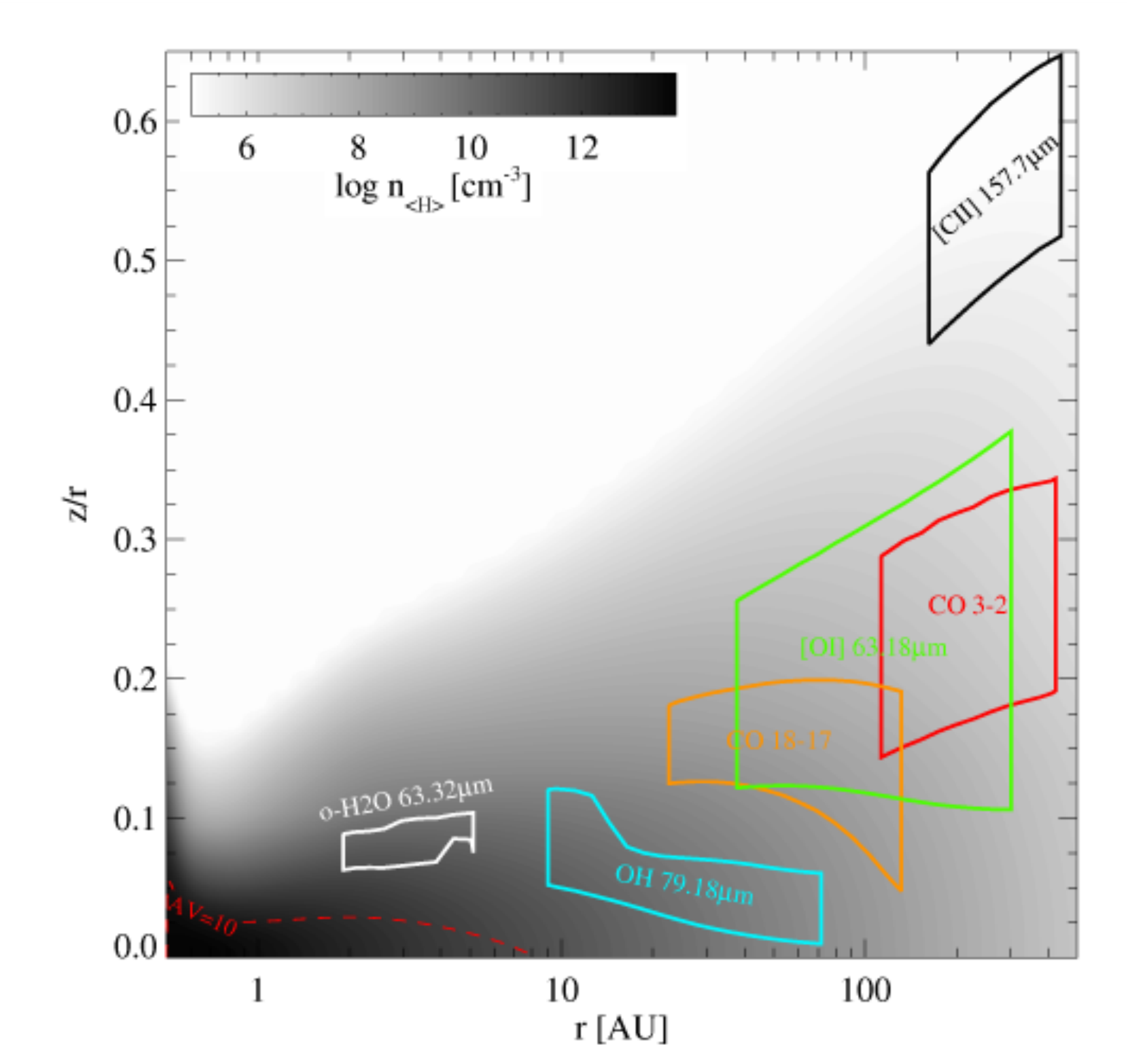}
\caption{Cross-sections through ProDiMo models of a T Tauri disk (upper) and a HAeBe disk (lower), with the density structure as a greyscale, and the A$_V$=10$^m$ surface shown as a dashed red line. The primary emitting regions for GASPS lines are given by the coloured boxes, and indicate where 80\% of the emission arises. The lines are [OI]63$\micron$ (green), CO J=18-17 (orange), [CII]157$\micron$ (black), OH79$\micron$ (blue) and the 63.3$\micron$ H$_2$O line (white). Also shown is the CO J=3-2 emitting region (red). Model parameters are given in Table~3. Note the larger outer radius displayed in the HAeBe model.}
\label{fig:models}
\end{figure}

In Figure~1 we use {\bf ProDiMo} to illustrate the regions where most FIR line emission is expected to arise in T~Tauri and HAeBe disks, using the model parameters given in Table~3. Note that these are relatively massive disks - towards the high end of the range of the GASPS sample. For these models we have proscribed the disk vertical struture by the scale height and flaring index. The results indicate that the [OI] lines are seen mostly from the disk surface at 20--200AU radius around a T Tauri star, and a factor of 1.5 further out in the more luminous HAeBe star. The [CII] line comes from the tenuous outer atmosphere at radii $>$100AU, whereas high-J CO emission (e.g. the J=18-17 line at 144.78$\micron$) is predicted to arise only from within a few tens of AU for T~Tauri disks. The model predicts emission from the 63.3$\micron$ line of H$_2$O mostly from within a few AU.

\begin{table*}
\begin{center}
\caption{\sf Parameters used by ProDiMo to illustrate the region of line emission from disks around T Tauri and HAeBe stars in Figure~1.}
\begin{tabular}[t]{lccl}
\hline\
Parameter & T~Tauri & HAeBe & Notes\\
\hline
\hline\\[-3mm]
SpT & K4 & A3 &\\
M$_*$(M$_{\odot}$) & 1.4 & 2.2 & \\
T$_{eff}$ (K) & 4500 & 8600 & \\
L$_*$ (L$_{\odot}$) & 2.0 & 32.0 & \\
T (Myr) & 2.3 & 4.6 & Age \\
f$_{UV}$  & 0.01 & 0 & Additional UV fraction, L$_{UV}$/L$_*$\\
L$_X$ (erg/s) & 10$^{30}$ & 0 & Additional X-ray luminosity\\
Rin (AU) & 0.1 & 0.5 & Set by the grain sublimation radius \\
Rout (AU)& 300 & 500 & Outer disk radius\\
$M_d$ (M$_{\odot}$) & $10^{-4}$ & $10^{-4}$ & Disk dust mass\\
g/d & 100 & 100 & gas/dust mass ratio \\
$\epsilon_d$ & 1.0 & 1.0 & Surface density power law exponent \\
a$_{min}$/a$_{max}$ ($\micron$) & 0.05/1000 & 0.05/1000 & min/max grain size \\
p & 3.5 & 3.5 & grain size power law \\
f$_{PAH}$  & 0.01 & 0.01 & PAH mass fraction\\
H$_0$ (AU)& 10 & 10 & scale height at 100AU radius \\
$\beta$ & 1.1 & 1.1 & disk flaring index \\
\hline
\end{tabular}
\end{center}
\label{tab:models}
\end{table*}

\subsubsection{Model grids}
\label{sec:grids}
To support a broader statistical analysis of the GASPS data, we have produced a grid of models covering a wide parameter space \citep{Woit10}\footnote{The grid was calculated on the FOSTINO computer cluster financed by ANR and operated by SCCI at OSUG.}. Stellar masses between 0.5 and 2.5~M$_\odot$ and  pre-main sequence evolutionary tracks at 1, 3, 10 and 100~Myr are used to define $T_{\rm eff}$, $R_\ast$ and hence L$_\ast$. The UV excess, $f_{\rm UV}$, is treated as a power-law that is added on top of the photospheric spectrum in the wavelength range 912 to 2500~\AA. Dust masses range from $10^{-7}$ to $10^{-3}$~M$_\odot$, and the gas/dust mass ratio runs from~$10^{3}$ (10$\times$ ISM) to 0.1 (0.001$\times$ISM). Geometries include young flaring disks, flat evolved systems, as well as `transition' disks with inner holes up to 100 times the sublimation radius. The grid also contains models with a settled dust distribution, where larger grains have a smaller vertical scale height than smaller ones. The observables calculated from these models include SEDs and integrated line fluxes.

\citet{Woit10} show that the fine structure line fluxes of [OI] and [CII] depend strongly on the stellar UV excess and disk flaring. Using the entire grid of parameter space (not folding in the likelihood of these disks occuring in nature) about 70\% of the high-mass models (dust mass, $M_d\geq10^{-5}$~M$_\odot$) with a strong UV excess were predicted to be detected in [OI] 63~$\mu$m line by GASPS, and 55\% detected in [CII]157$\micron$. Without a UV excess, the percentage drops to $\sim$30\% (14\% for [CII]).  An initial statistical comparison between the early GASPS line fluxes and the grid \citep{Pin10} shows that some of the disks around low-mass stars ($\lesssim 2 L_\odot$) do require additional heating from a moderate UV excess (with $f_{\rm UV}=0.1$) or X-rays (which were not included in this first grid).  However, results from T Tauri disk modeling with stellar X-rays indicate that the [OI] 63~$\mu$m line flux is only affected by X-ray heating for $L_{\rm X} \gtrsim 10^{30}$~erg/s \citep{Ares11}.

Although most gas emission lines sample a thin, warm surface layer (see Figure \ref{fig:models}), combining the FIR data with results from other wavelengths (e.g. the [OI]63{\micron}/CO(2-1) ratio) and with physically-plausible models does allow us to break model degeneracies, giving approximate estimates for gas parameters independent of the dust \citep{Kamp11}. But it is clear that the reliability of derived values such as the gas mass relies on the accuracy of the models.

\subsection{Outflow jets}
\label{sec:outflows}
Highly embedded Class 0-I YSOs are known to have prominent outflows, and early observations using the KAO as well as more recent observations with Herschel/PACS show bright FIR lines around Herbig-Haro objects and high-velocity CO outflow lobes \citep{Coh88, vanK10}. As well as fine structure lines such as [OI]63$\micron$, many CO and H$_2$O transitions are readily detectable from Class 0-I objects \citep{Lor00, Mol00, vanD04, Goi12}. The stars are young ($\leq$0.1~Myr), optically obscured, embedded in an envelope, and located near dense cloud cores.  Their dense environments and high outflow luminosities suggest that the FIR line emission is dominated by outflow shocks \citep{Mol00, Nis00, Nis02, Frank08}. These shocks also affect molecular abundances, for example, releasing H$_2$O from grains and increasing its gas-phase abundance to as much as $10^{-4}$ - comparable with that of CO \citep{Bene02}. 
In most cases the FIR lines dominant the shock cooling, and line fluxes may be used to estimate the outflow luminosity \citep{Holl85, Nis02, Pod12}.

Evolved, isolated objects such as optically-visible Class II-III T Tauri stars have mass accretion rates at least 1-2 orders of magnitude lower than Class I objects \citep{Hart98, AS06}, which are themselves an order of magnitude lower than the Class 0 objects \citep{Bon96, Pod12}.  Class II are pure disk systems, and are generally isolated with no dense ambient gas. Consequently we assume the fraction of outflow luminosity deposited in shocks near the star, $\eta_s$, is given by the geometric fraction of the initially broad wind intercepted by the disk. So ${\eta_s\sim \mathrm{H_0} /100}$, where H$_0$ is the scale height (in AU) at 100AU radius. In the same way as embedded objects, the jet mass loss rate $\dot{M}_{out}$ can be estimated from the [OI]63$\micron$ luminosity, $L_{OI}$ by: 

$\dot{M}_{out} = 2 . L_{OI} / [v^2_w . f_{OI} . \eta_s . (v_a/v_w) ]$

where $f_{OI}$ is the fraction of FIR line luminosity in the [OI]63$\micron$ line, $v_w$ the outflow jet velocity, and $v_a$ the ambient shock velocity \citep{Nis02}.  For embedded objects the dominant emission is from the integrated CO and H$_2$O lines, thought to be from C shocks, and \citet{Goi12} finds $f_{OI}\sim0.12$ in the Class 0 YSO Serpens SMM1. In fast dissociative J shocks, [OI]63$\micron$ emission may dominate the luminosity, and $f_{OI}$ is found to be 0.5 or greater \citep{Pod12}.
Assuming a jet velocity of 100$km s^{-1}$ \citep{Pod12}, $v_a/v_w=0.2$ \citep{Nis02}, with $H_0=10AU$ (Table~3), then the GASPS sensitivity (\S2.1.1) may allow the detection of outflow mass loss rates of $\sim 3\times 10^{-9}$M$_{\odot}$/yr for stars at a distance of 140pc. However, if C shocks dominate, the
contribution to the [OI] line from the jet may be lower.
In Table \ref{tab:slist}, we have indicated the stars with published mass loss rates greater than this value \citep{Hart95}.

Although most are isolated disks, a few of the GASPS targets are somewhat embedded Class II objects and have extended optical jets; for these we may expect some outflow contribution to the FIR lines. The spatial resolution of PACS is modest, but can help investigate this contribution, resolving jets on scales of $>$1000~AU.  For spatially-unresolved outflows the situation is less clear. However, shocked outflow emission may be broadened to $\sim$200~km s$^{-1}$ or velocity-shifted by more than a few tens of km s$^{-1}$, similar to the high-velocity component seen in optical lines in a few high-accretion objects \citep{Hart95, Ack05}. In these cases, the PACS spectral resolution of 88 km s$^{-1}$ at $63 \micron$ may also be used to help discriminate between outflow and disk.

\subsection{Remnant envelope gas}
Low-density PDRs in the remnant envelope gas centred on the stars may contribute to the [CII] flux from some objects, as the [CII]157$\micron$ critical density is only $\sim3\times10^3cm^{-3}$. To mitigate this, targets were selected to be SED Class II-III with low A$_V$, i.e. optically-visible stars where the envelope mass is at least 1 or 2 orders of magnitude less than the disk mass \citep{Fuen02, AS06}. The [OI]63$\micron$ critical density is $\sim$100 times higher than [CII], and the mean envelope density is small compared to the disk, so the envelope contribution to the total [OI] flux should be small.

\subsection{Disk winds}
A photoevaporative UV-driven wind \citep{Pasc09} will serve to extend the scale height of a disk atmosphere, and may enhance FIR emission lines. The effect of this on the [OI]63$\micron$ flux is under investigation.

\subsection{Extended ambient gas}
\label{sec:cii-confus}
Observations of star-forming clouds in FIR fine structure lines using the LWS on ISO showed bright emission in regions containing luminous Herbig AeBe and FU Ori stars \citep{Lor99, Lor00, Cree02}.  The [CII] emission in many of these objects was spatially extended, and commonly the line fluxes at reference positions many arcminutes from the stars were as bright as the on-source position. In these cases, the dominant [CII] source was thought to be extended low-density PDRs, with optical depths of 1-2$^m$ \citep{Holl97}.
The GASPS targets were chosen to avoid the densest clouds, and we used the Herschel Confusion Noise Estimator (HCNE) to estimate the $100 \micron$ continuum
confusion noise ($F_c$) for all targets. From this we estimate a [CII] confusion noise level by adopting a ratio of I$_{[CII]}/$F$_c = 1.2\times10^{-19}$W m$^{-2}$/mJy, found for large-scale Galactic clouds by \citet{Shib91}. During the course of the survey, 10 objects having a relatively high continuum confusion level ($F_c > 30$mJy) were observed in [CII] (8 of which were in the Taurus cloud). Based on the HNCE, the predicted [CII] confusion level for these was $>3.6\times10^{-18} W m^{-2}$. An examination of the initial data shows no extended [CII] over the PACS footprint in 9 of these objects, with an rms level of $\sim2\times10^{-18} W m^{-2}$. Either the confusion level is lower than predictions from the HCNE, or the [CII] emission is smooth over the PACS IFU field ($\sim$arcmin) and emission is being chopped out. One high background confusion source (HD~163296) had evidence of extended [CII] at a level of $\sim10^{-17} W m^{-2}$ in the PACS field of view and in the chopped reference beam. From the HCNE, this object has the highest value of $F_c$ in the GASPS sample (85 mJy), which would predict, based on the above ratio, a [CII] confusion noise of $I_{[CII]}=10^{-17} W m^{-2}$, consistent with the observations.
 
The 100$\times$ larger critical density of [OI]63$\micron$ compared with [CII] implies that extended [OI] emission from diffuse ambient gas is expected to be negligible \citep{Lis06}.  ISO found that the 63$\micron$ line flux is mostly higher towards highly luminous YSOs than off-source.

\subsubsection{Line-of-sight absorption}
As well as emission, dense clouds may have significant optical depth and be self-absorbed in the [OI] $63 \micron$ line. However, estimates suggest the line optical depth may not become significant until $Av > 10$ \citep{Lis06, Abe07}. Moreover the linewidths of the cool line-of-sight clouds are $<$ 1 km s$^{-1}$, small by comparison with the
5-20 km s$^{-1}$ widths predicted for Keplerian disk emission. Combined with the extinction
limit of $Av < 3$ in the GASPS survey means this effect should be small in most cases.

\section{First results}

Results from some subsets of the GASPS study have been presented in previous papers. A summary of the `science demonstration' observations of a small number of targets was given in \citet{Math10}, and a comparison of these data with a broad grid of disk models was shown by \citet{Pin10}. More detailed comparisons of the line and continuum data with individual tailored models were carried out based on the detections of [OI]63$\micron$ in the T Tauri star TW Hya \citep{Thi10} and the HAeBe stars HD~169142 \citep{Meus10} and HD~163296 \citep{Till12} . The T Tauri star ET~Cha was detected in both [OI]63$\micron$ and FIR continuum, and modeling indicates the disk is unusually compact \citep{Woit11}. CH$^+$ was detected in one of the brightest targets - the HAeBe system HD~100546 \citep{Thi11}. An emission line of H$_2$O at 63.3$\micron$ found in a number of the T~Tauri stars indicates warm ($\sim$500K) H$_2$O, possibly from the inner few AU of the disks \citep{Riv12a}. In most of the older gas-poor systems the lines were not detected, however, the far-IR photometry has been used to improve the SEDs and dust modeling \citep{Don12, Leb12}.

In the following sections we summarise some of the overall results from GASPS, including identification of the lines found in the survey, and an initial comparison of the spectra of different types of objects (\S5.1). In Table~\ref{tab:slist} (column 12) we indicate which of the four primary lines ([OI]63$\micron$, [CII]157$\micron$, CO J=18-17 and H$_2$O) were observed and detected in the targets. For these purposes, a detection is regarded as $>$3-$\sigma$ above the noise. In \S5.2 we give the overall line detection statistics from the survey, and discuss the [OI]63$\micron$ and [CII] emission characteristics in \S5.3 and 5.4.
Finally in \S5.5, we show the effects of other system parameters on the line detection statistics. It should be noted that these data are mostly based on results from early versions of the reduction pipeline HIPE, consequently the flux calibration and flat-fielding is not finalised and some detections are subject to re-analysis. Final values of the fully-calibrated fluxes and detailed flux correlations will be given in subsequent papers.

\subsection{Summary of lines detected}

To illustrate and compare the lines detected in the richer GASPS targets, spectra from the central spaxel in three objects from the survey are shown in Figures~2-5 (note that the spectra are scaled to enable comparison in these figures). T~Tau \citep{Pod12} (shown in red) is a K0V star with a massive disk, compact outflow, some surrounding reflection nebula and possibly a PDR. FIR lines may arise from a mixture of these components, although the molecular transitions seen in the ISO LWS spectra were attributed mainly to shock emission \citep{Spin00}. HD~100546 \citep{Meu12} (shown in blue) is a young Herbig AeBe star with a bright disk but without a prominent outflow, but which also has a rich FIR spectrum. AA~Tau (in green) is perhaps a more typical isolated T Tauri star, with a luminosity of $\sim 1L_{\odot}$, weak outflow and a relatively massive disk. In Table~4 we identify all the lines observed in these three objects.

Both T~Tau and HD~100546 have similar strengths in the fine-structure atomic lines. AA~Tau is $\sim$200 times weaker, but is detected in [OI] with a comparable line/continuum ratio to the others. However, it shows no evidence of [CII]. In HD~100546, molecular transitions have a line/continuum ratio which is considerably lower than both T Tau and AA Tau.  The PACS data cover four transitions of CO: J=18-17, 29-28, 33-32 and 36-35. T Tau shows emission in all four CO lines, and comparison with the CO rotational diagram of \citet{Spin00} shows that the 3 highest transitions are new detections, requiring an additional hot gas component ($> 1000$K) to account for the emission. AA Tau is detected only in the two lower-level CO lines, most likely because of sensitivity limits.

The OH doublet around 79$\micron$ is detected  in all three sources (Fig.~3). Several H$_2$O lines with upper energy levels from 115-1300K are seen towards both T~Tau and AA~Tau, and in AA~Tau, H$_2$O is the only line detected, other than [OI]63$\micron$, CO and OH. By contrast, HD100546 has no clear evidence of H$_2$O emission, although other lines (atomic species, OH and CH$^+$) are relatively bright. The highest energy level H$_2$O transition covered by GASPS is the ortho 8$_{18}$ - 7$_{07}$ line at 63.3$\micron$ (Figure~2); this was detected in T~Tau, AA~Tau and several other T Tauri stars \citep{Riv12a}.  Finally, both
HD~100546 and T~Tauri show clear CH$^+$ emission at 72.14$\micron$, with possible blends of CH$^+$ and H$_2$O around 90.0 and 179.5$\micron$; this species was also identified at several other wavelengths in HD~100546 \citep{Thi11}.

\begin{table*}[htbp]
\begin{center}
\caption{\sf Lines identified in HD100546, T~Tau and/or AA~Tau. An X indicates a detection.}
\begin{tabular}[t]{clccccc}
\hline\
Wavelength & Line ID & Transition & E$_{upper}$ & HD100546 & T Tau & AA Tau\\
($\micron$) &&& (K) &&& \\
\hline
\hline\\[-3mm]
63.18 & [OI] & 3P1-3P2 & 228 & X & X  & X\\
63.33 & o-H$_2$O & 8$_{18}$ - 7$_{07}$ & 1293 &...& X & X\\
71.94 & o-H$_2$O & 7$_{07}$ - 6$_{16}$ & 685 &...& X  & X\\
72.14 & CH$^+$ & J=5-4 & 600 & X &...&...\\
72.84 & CO & J=36-35 & 3700 &...& X & ...\\
78.74 & o-H$_2$O & 4$_{23}$ - 3$_{12}$ & 432 &...& X & X \\
78.92 & p-H$_2$O & 6$_{15}$ - 5$_{24}$ & 396 &...& X &... \\
79.11/79.18 & OH & 1/2 - 3/2 hfs & 182 & X & X & X \\
79.36 & CO & J=33-32 & 3092 & X & X  &... \\
89.99 & p-H$_2$O & 3$_{22}$ - 2$_{11}$ & 297 & (blend with CH$^+$) & X  & ...\\
90.02 & CH$^+$ & J=4-3 &   297 & X &...&... \\
90.16 & CO & J=29-28 & 2400 & X & X  & X\\
144.52 & p-H$_2$O & 4$_{13}$ - 3$_{22}$ & 396 &...& X & ...\\
144.78 & CO & J=18-17 & 945 & X & X  & X\\
145.52 & [OI] & 3P0 - 3P1 & 326 & X &X &... \\
157.74 & [CII] & 2P3/2 - 2P1/2 & 91 & X & X & ... \\
158.31 & p-H$_2$O & 3$_{31}$ - 4$_{04}$ & 410 &...&...&... \\
179.53 & o-H$_2$O & 2$_{12}$ - 1$_{01}$ & 115 & (blend with CH$^+$) & X   &...\\
179.6 & CH$^+$ & J=2-1 & 114 & X &...& ...\\
180.49 & o-H$_2$O & 2$_{21}$ - 2$_{12}$ & 194 &...& X  &...\\
\hline
\end{tabular}
\end{center}
\label{tab:lines}
\end{table*}

\begin{figure}[htbp]
\plotone {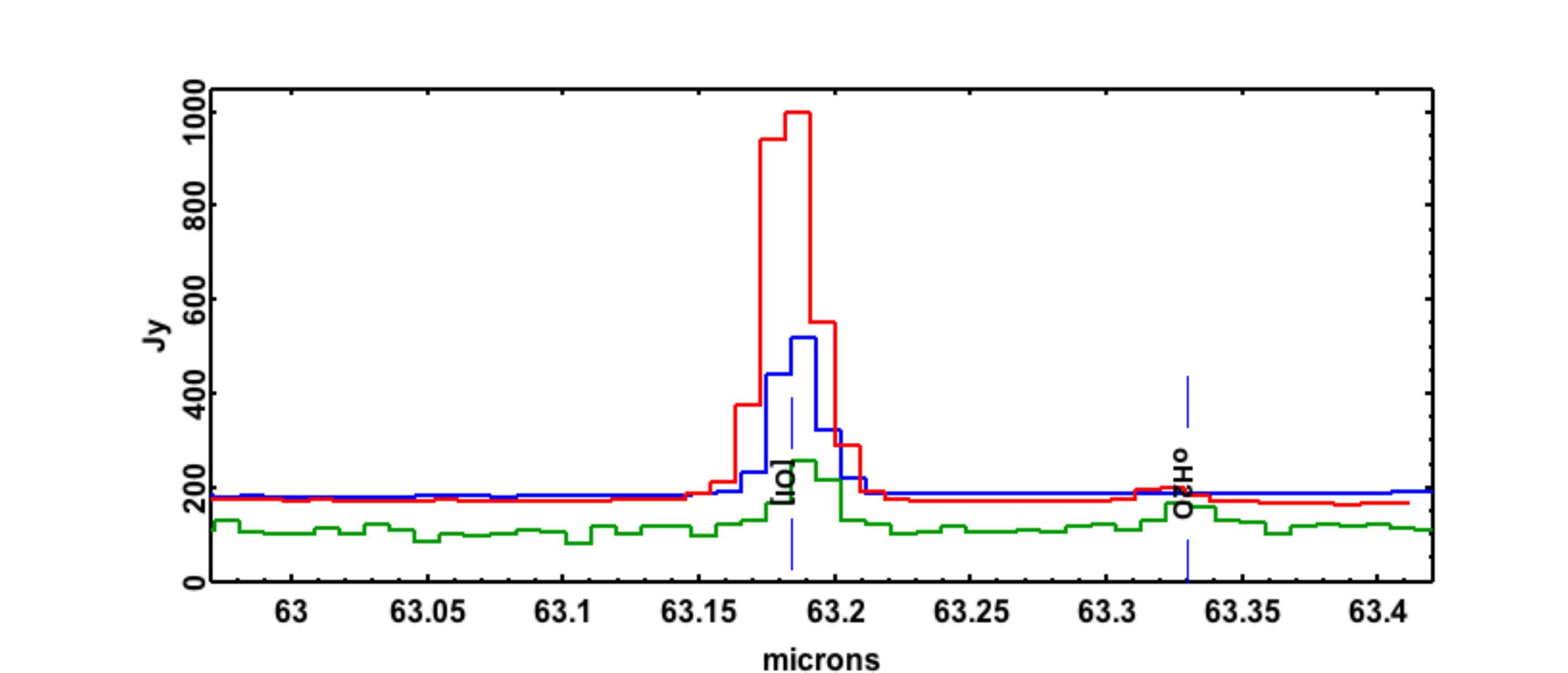}
\plotone {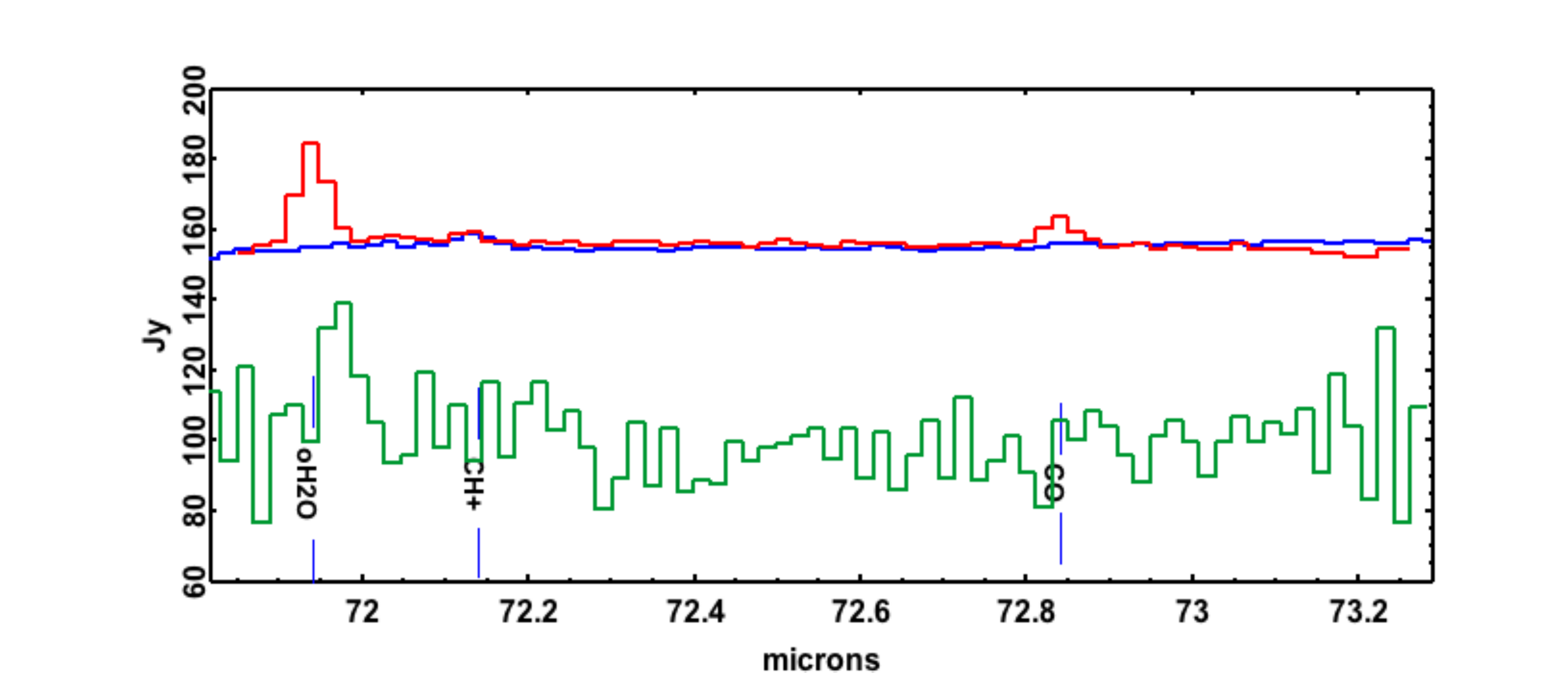}
\caption{Spectra of HD~100546 (blue), T~Tau (red) and AA~Tau (green), taken from the central PACS spaxel in the two shortest wavelength observations. Fluxes of T~Tau and AA~Tau fluxes are scaled by 2 and 150 respectively to facilitate comparison of the spectra. Lines found in any of the datasets are identified (although not all the lines are seen in all objects) - see Table~4 for full details of the transitions. Note that small wavelength errors are sometimes apparent in these early reductions of the AA~Tau spectra.}
\end{figure}

\begin{figure}[htbp]
\plotone{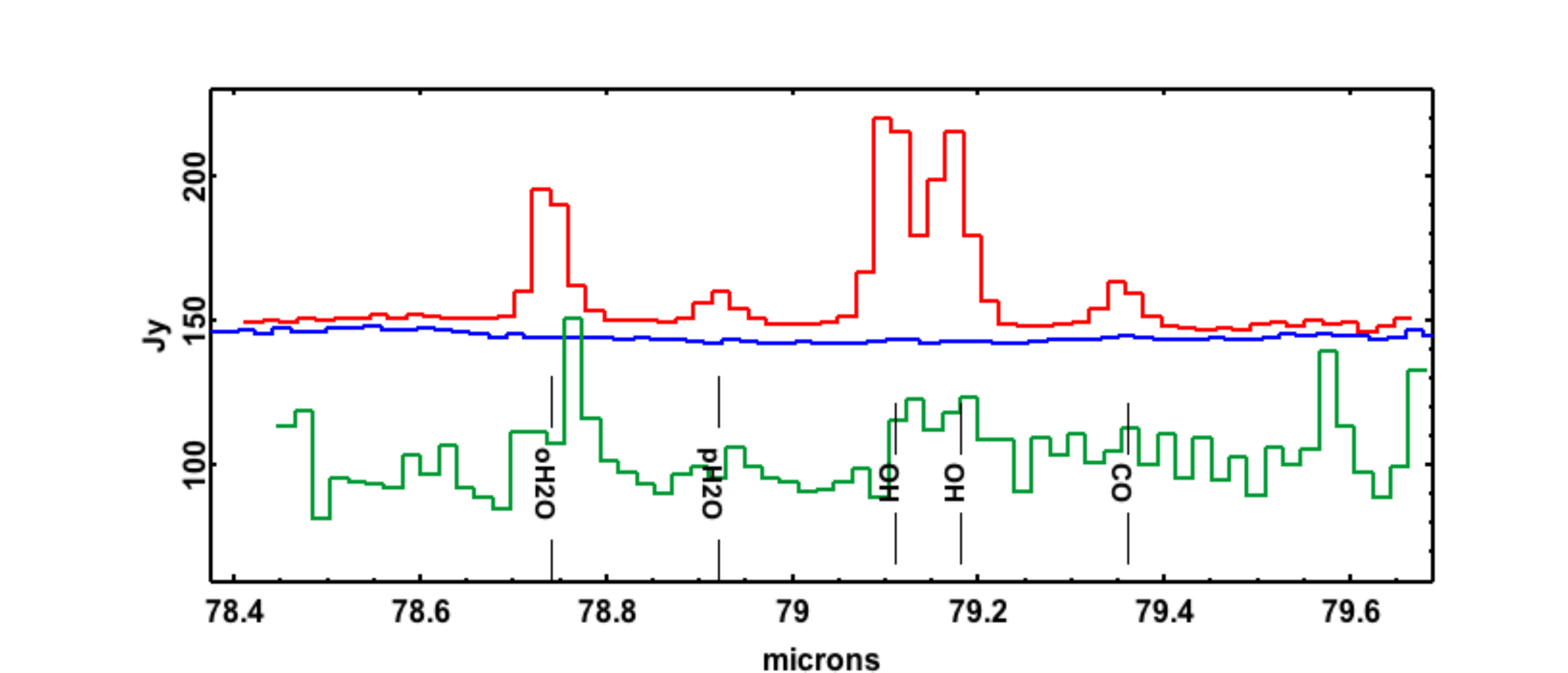}
\plotone{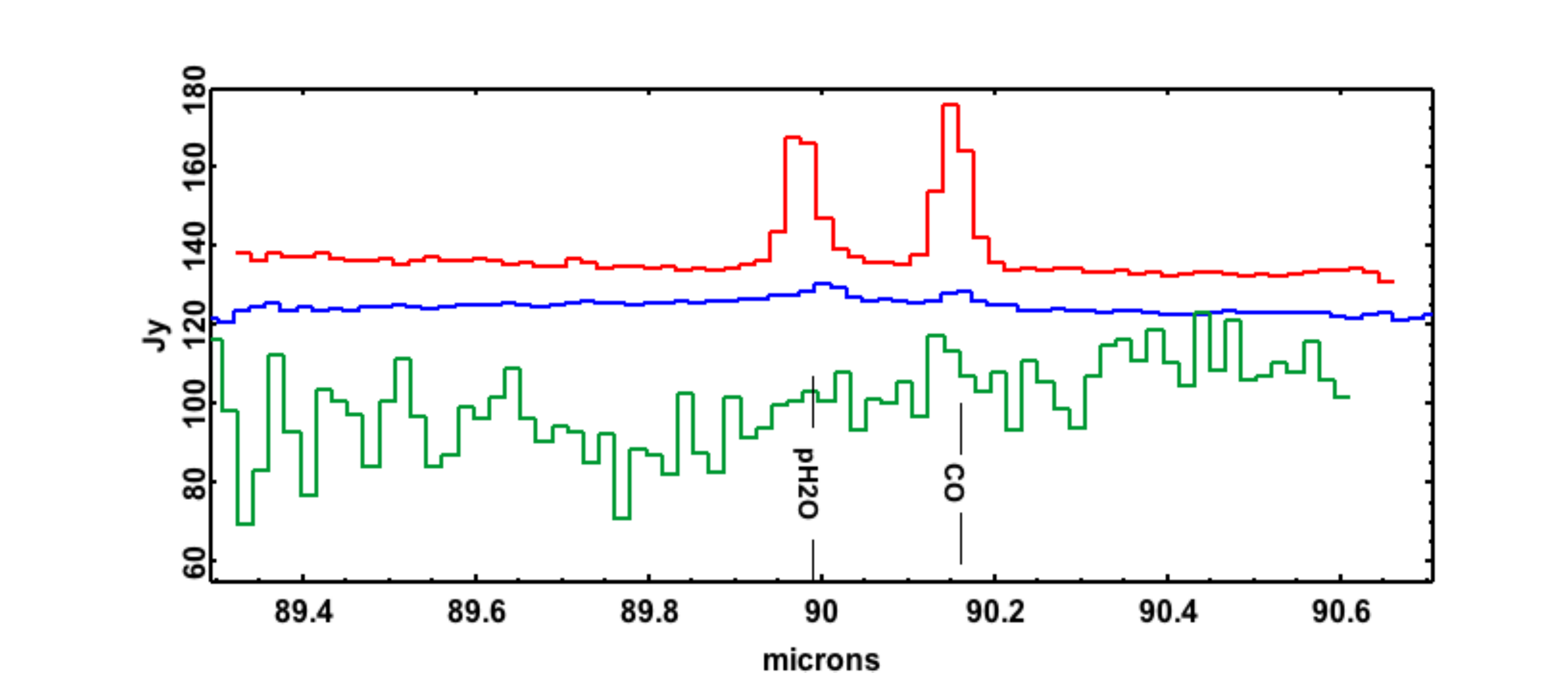}

\caption{Same as Figure~2, for regions around 79 and 90$\micron$. The spectra of T~Tau and AA~Tau have been multiplied by 2 and 150. The emission close to 90$\micron$ is a blend of H$_2$O and CH$^+$, and in HD~100546 is thought to be mostly from CH$^+$.}
\end{figure}

\begin{figure}[htbp]
\plotone{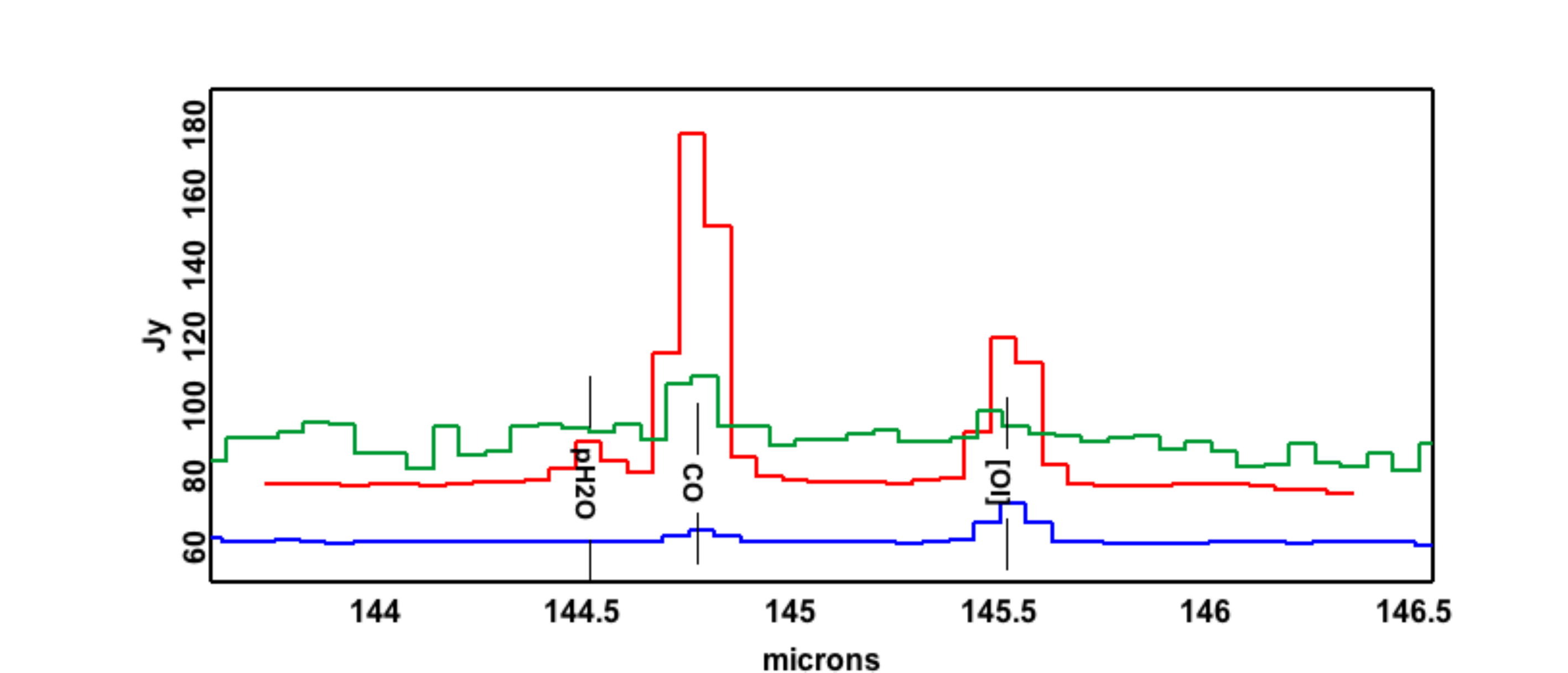}
\plotone{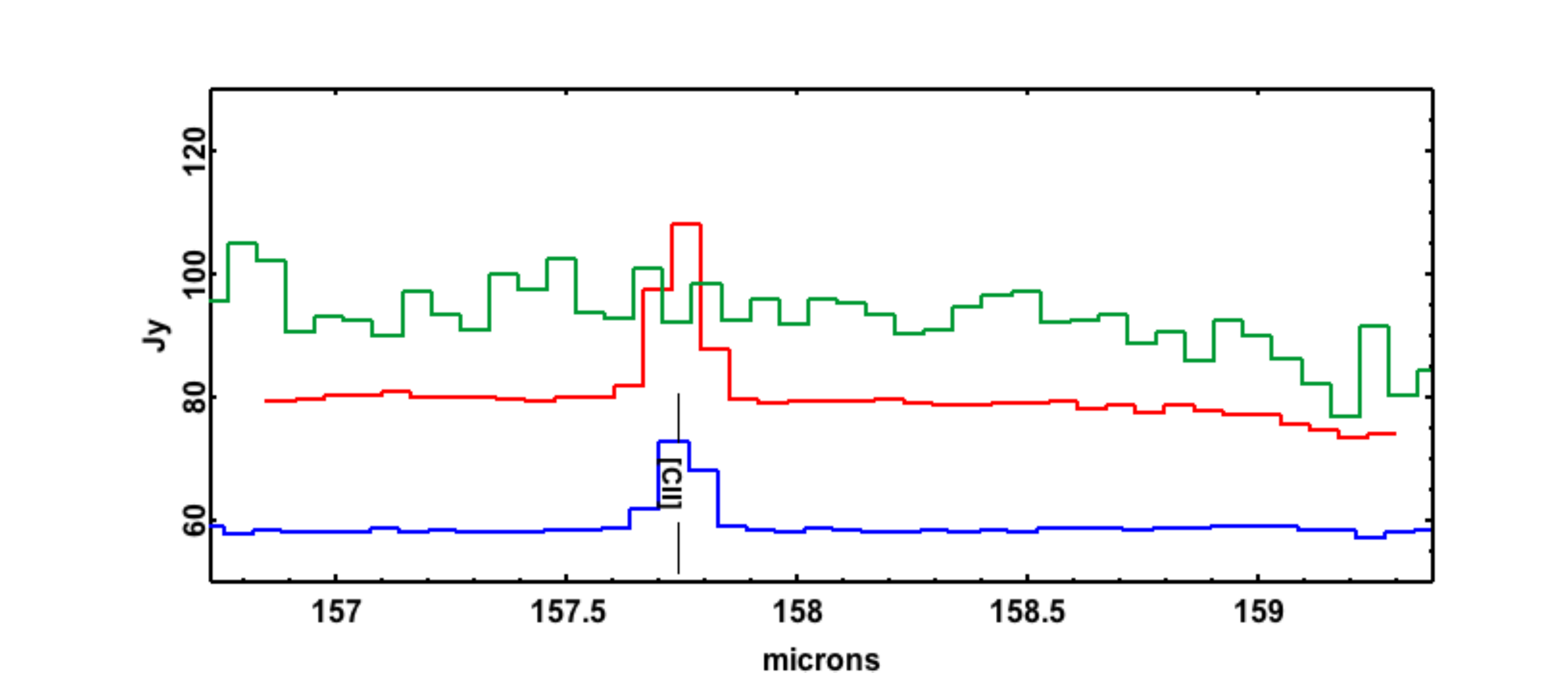}
\caption{Same as Figure~2 for regions around 145 and 158$\micron$. The spectra of T~Tau and AA~Tau have been multiplied by 2 and 150.}
\end{figure}

\begin{figure}[htbp]
\plotone{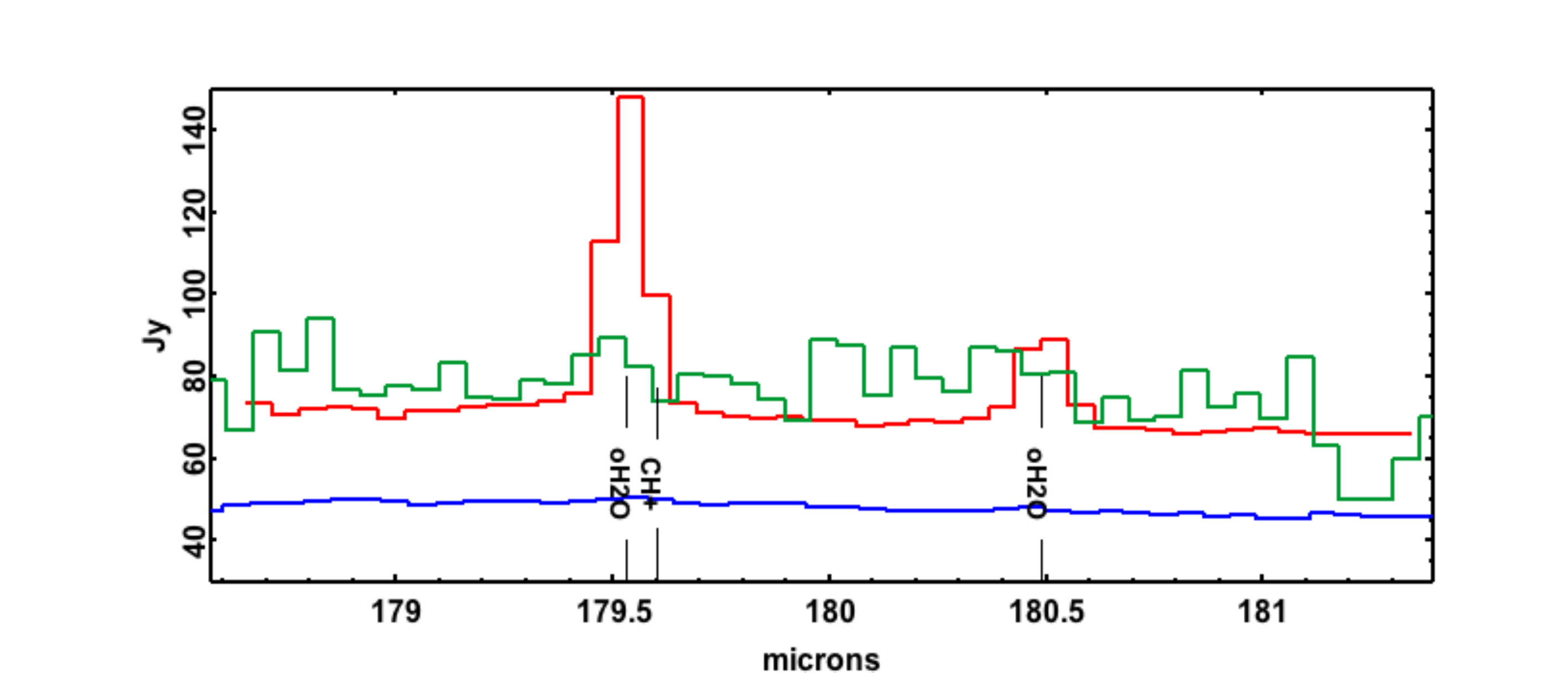}
\caption{Same as Figure~2 for region around 180$\micron$. The spectra of T~Tau and AA~Tau have been multiplied by 2 and 150.}
\end{figure}

\subsection{Primary line detection statistics}

Column 12 of Table~\ref{tab:slist} shows whether each of the four primary species ([OI]63$\micron$, [CII]157$\micron$, CO J=18-17 and H$_2$O 63.3$\micron$) were detected in the GASPS targets. Based on this, the overall detection statistics are given in Table~5. As noted above, observations of the range-scan observations were normally only performed if a target was already found or expected to be detected in [OI]63$\micron$, so the detection rates of [OI]145$\micron$, [CII] and CO in this table are biased towards those with known [OI]63$\micron$ emission. Of targets observed in multiple lines, only one remained undetected in [OI]63$\micron$ yet shows emission in one of the other lines. Based on this result and our modeling, it is thought unlikely that a significant number of the [OI]-unobserved objects would show emission in these other lines. The H$_2$O rates are the fraction of targets seen at 63.3$\micron$, which was observed as part of the [OI]63$\micron$ line-scan observations. The CO rates are the fraction of targets detected in the brightest line covered by GASPS (CO J=18-17). 

The main similarities and differences between line emission from the two types of objects are:

\begin{enumerate}
\item{Of the sample of 164 objects observed in spectroscopy at 63$\micron$, approximately 49\% were detected in [OI].}
\item{A biased subset of the brighter objects from (1) were observed in [OI]145$\micron$, [CII]157$\micron$ and CO J=18-17, and the detection rates in this subset were 25-40\% in each of these lines. Assuming that [OI]63$\micron$ is always the easiest to detect (see above), then an unbiased sample of all 164 targets from (1) would have had a detection rate of $\sim$14\% in these other lines.}
\item{All HAeBe stars were detected in [OI]63$\micron$ - a significantly higher detection rate than T Tauri systems. (Note that the statistics of HAeBe stars in Table~5 include 5 known A-star debris disks).}
\item{The [OI]145 detection rate is a factor of $\sim$2 higher in the T Tauri stars observed compared with HAeBe systems. This may reflect a higher [OI] 63/145$\micron$ line ratio in HAeBe disks.}
\item{The [CII] detection rate is similar (26\%) in both T Tauri and HAeBe stars. If this is envelope material (see \S\ref{sec:cii}), it indicates that compact envelopes of atomic gas can be maintained around both high and low-luminosity stars. Note, however, that in some cases the [CII] emission may be confused by ambient gas.}
\item{ One (possibly two) HAeBe stars were detected in H$_2$O. Although in the small number regime, the H$_2$O detection rate is formally similar to that of T Tauri systems. However, considering the HAeBe's are relatively bright in continuum compared with the T Tauri sample, this suggests that, on average, HAeBe systems are weaker in H$_2$O compared with T Tauri systems.}
\item{The fraction of objects with detectable warm CO (based on the J=18-17 transition) is similar (40\%) in disks around both types of stars.}

\end{enumerate}

\begin{table*}[htbp]
\begin{center}
\caption{\sf  Detection statistics of primary atomic and molecular species. Each entry gives the number of targets detected and number observed. For [OI]145, [CII] and CO, observations were mostly carried out only if the lines were detected (or likely to be detected) in [OI]63$\micron$. }
\begin{tabular}[t]{llllll}
\hline\
 & [OI]63 & [OI]145 & [CII]157 & H$_2$O(63) & CO 18-17  \\
\hline
\hline\\[-3mm]
Total & 80/164 & 24/61 & 19/72 & 12/164 & 24/58  \\
HAeBe stars$^1$ & 20/25 & 5/23 & 6/25 & 2/25 & 10/24 \\
T Tauri stars$^2$ & 60/139 & 19/38 & 13/47 & 10/139 & 14/34 \\
\hline
\end{tabular}
\end{center}

\footnotesize{(1) Includes 5 young A stars classed as debris disks.\\
(2) This includes all stars observed which were not part of the HAeBe group.}
\end{table*}

\subsection{[OI] line emission}
As is clear from the example spectra, [OI]63$\micron$ is normally several times brighter than any of the other FIR lines observed by GASPS, with an overall detection rate in the survey of $\sim$49\%. In most cases, it is the best tracer (in the far-IR) of whether gas is present. This is true for almost all GASPS sources. To help understand the origin of the emission we can look at the data in more detail.

\begin{figure}[htbp]
\plotone{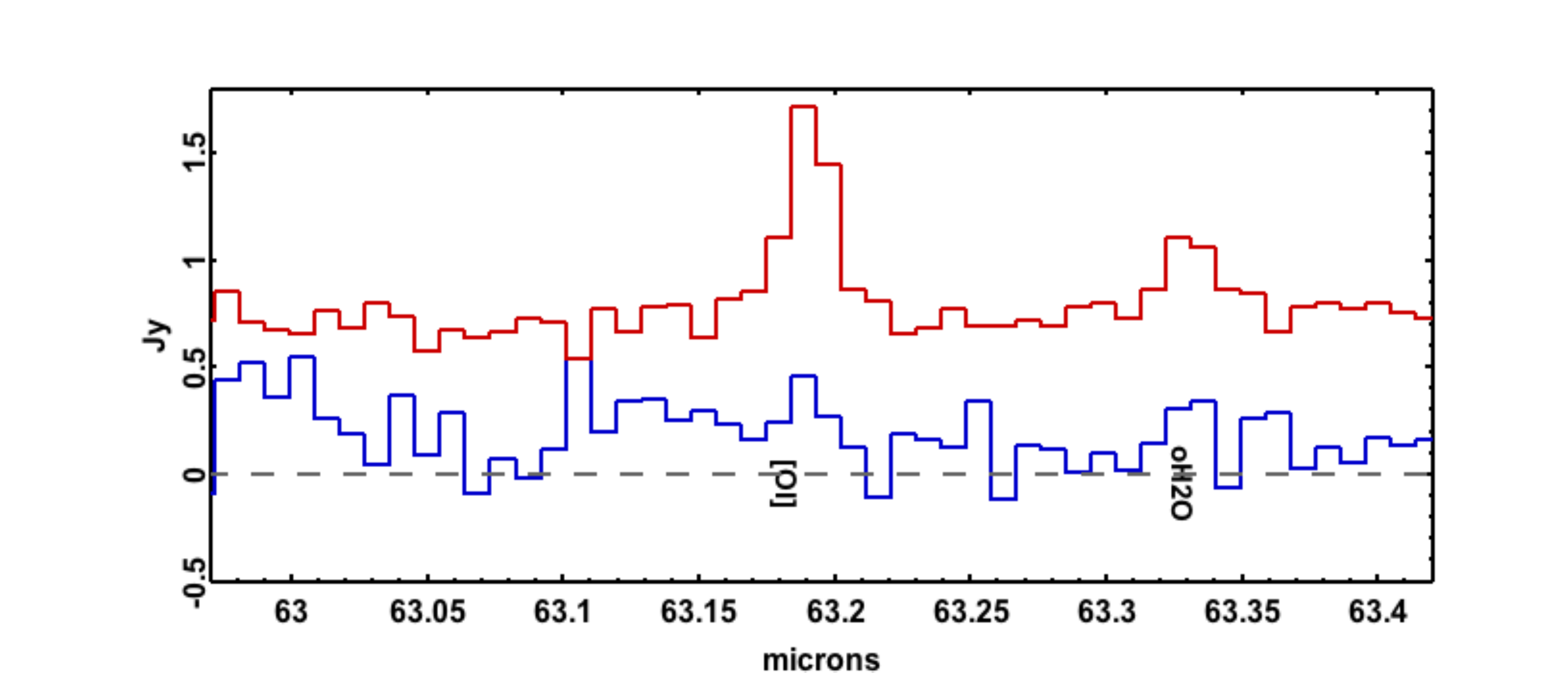}
\caption{Spectra covering the [OI]63$\micron$ line from the compact source AA Tau, in the central spaxel (red histogram), and an average of the 8 adjacent spaxels (in blue). The adjacent pixel spectrum has been scaled up by a factor of 5 for clarity. Both the [OI]63$\micron$ and nearby H$_2$O line are detected only in the central spaxel.}
\label{fig:AAT-OI}
\end{figure}

Most objects were unresolved in both line and continuum emission. An example is AA~Tau (Figure~\ref{fig:AAT-OI}), where the ratio of flux in the centre to average of adjacent spaxels is $\sim$20. This is consistent with an unresolved source, where we would expect the adjacent pixel average to be a few \% of the centre, given an inter-spaxel spacing of 9.4 arcsec, a PSF Gaussian equivalent width of $\sim$5.4 arcsec at 63$\micron$, the asymmetric sidelobes from PACS of a few percent, and taking into account possible pointing uncertainties of a few arcsec in some datasets (PACS User Manual, 2011).
This lack of extended emission indicates a line emitting region of radius~$\leq 500AU$.

For a number of individual unresolved objects with low accretion rates and no evidence of outflow we have assumed a disk origin, and combined the [OI] fluxes with data at other wavelengths to estimate disk properties. Initial ProDiMo models of the relatively large disk in TW Hya (several 100AU radius) indicate a gas mass of a few $10^{-3}M_{\odot}$ with gas:dust ratio a factor of $\sim$10 lower than the ISM value \citep{Thi10}, although some models suggest the gas mass an order of magnitude larger, with a more ISM-like gas:dust ratio \citep{Gor11}. ET Cha, by contrast, has a compact disk of modeled radius of only 10AU, a low dust mass of a few $10^{-8}M_{\odot}$ and gas mass of a few $10^{-4}M_{\odot}$ \citep{Woit11}, suggesting either the gas/dust ratio is enhanced or there may be another contribution to the line flux. The HAeBe stars HD~169142 and HD~163296 both show emission consistent with disks and ISM-like values of the gas/dust ratio \citep{Meus10, Till12}.

\subsubsection{Spatially and spectrally resolved [OI]63$\micron$ emission: outflow jets.}
\label{sec:spatial}

\begin{figure}[htbp]
\plotone{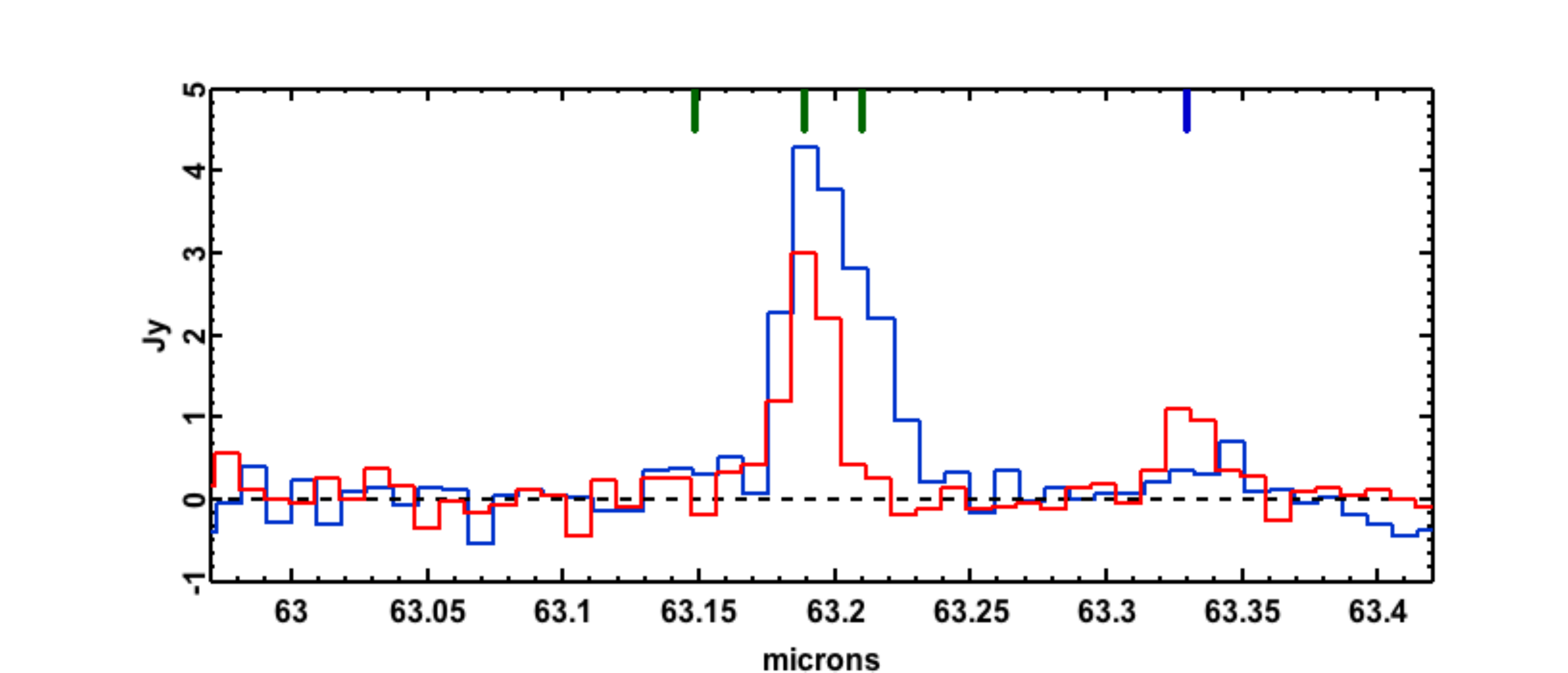}
\caption{Spectrum of [OI]63$\micron$ and H$_2$O from the jet source RW Aur (blue histogram), compared with AA Tau (red histogram, scaled up by a factor of 3 for easier comparison). These spectra are continuum-subtracted, and only the central spaxel is shown. AA Tau is unresolved whereas RW Aur has a prominent red-shifted wing in [OI]. The velocities of the three components which dominate the [OI]6300\AA~line are shown by the green tick marks, at the stellar velocity (heliocentric velocity +23 km s$^{-1}$) and at -190 and +100 km s$^{-1}$ \citep{Hart95}. Also shown is the wavelength of the H$_2$O line, at the stellar velocity.}
\label{fig:OI-RWAA}
\end{figure}

Although most objects in GASPS remain unresolved by PACS, five targets (identified in Table~\ref{tab:slist}) in Taurus were found to have clearly extended [OI]63$\micron$ emission along known optical jets \citep{Pod12}. Two of these also have broad line profiles in the centre. Figure~\ref{fig:OI-RWAA} compares the spectrum of one example (RW Aur) with the unresolved line from AA Tau (in red), revealing a prominent red-shifted wing in RW Aur extending as much as +200 km s$^{-1}$ from the stellar velocity. By contrast, AA Tau has emission centred at the stellar velocity, with a fitted linewidth of 93 km s$^{-1}$ (FWHM) - similar to the measured PACS resolution of 88 km s$^{-1}$ at this wavelength (see PACS User Manual). The optical [OI] 6300\AA~line from RW Aur is known to originate from highly-excited gas in a jet of length a few arcsec \citep{Mel09}, and the line profile is dominated by three components (marked in Figure~\ref{fig:OI-RWAA}), two at high velocities (+100 and -190 km s$^{-1}$) and one at the stellar velocity \citep{Hart95}. The brighter red-shifted optical component corresponds with the [OI]63$\micron$ wing, suggesting this is also from the shocked outflow gas (see \S\ref{sec:outflows}). However, the FIR line profile is dominated by emission centred approximately on the star, whereas this velocity component in the optical line is relatively weak \citep{Hart95}. This low-velocity gas may be from the disk or disk wind (see above).

In the GASPS data we have also identified five other objects with evidence of either broadened lines or spatially-extended [OI]63$\micron$ emission: HL Tau and XZ Tau (in the same PACS field), DO Tau, UZ Tau and DK Cha (for the latter source, see \citet{vanK10}). All targets resolved in [OI] are identified in Table~\ref{tab:slist} by the note `ext.OI', and it is likely that [OI] emission is dominated by outflow gas in these cases. 

\subsubsection{Objects with uncertain origin of [OI]63$\micron$}
\label{sec:jet}
In addition to the 10 resolved objects above, a further $\sim$17 objects (noted as `jet' sources in Table~\ref{tab:slist}) were identified as having published evidence of a high-velocity jet or outflow by \citet{Ken08}, \citet{Pod12} and Howard et al. (submitted).  These are sources with a jet imaged in optical lines, a high velocity molecular outflow, or a broad ($>$50 km s$^{-1}$), typically blue-shifted, emission line profile in [O I] 6300\AA~(see e.g. \citet{Hart95}). Three of these were HAeBe stars (HD~163296, MWC480 and HD~100546), leaving 14 T Tauri 'jet' sources. As noted in \S4.2, the survey sensitivity should allow us to detect [OI]63$\micron$ emission from outflows shocks with mass loss rates
$\dot{M} > 3\times10^{-9}$M$_{\odot}$/yr. Estimates from \cite{Hart95} suggest that 4 of the jet sources in the Taurus sample have mass loss rates exceeding this limit (indicated in Table~\ref{tab:slist} by the note 'high $\dot M$'). However, their [OI]63$\micron$ emission is neither spatially nor spectrally extended in the PACS data. This suggests that the outflow shock contribution may be small compared with the low-velocity gas, and the origin of the unresolved [OI]63$\micron$ emission in these remaining 'jet' sources is not clear from the PACS data alone.

\subsection{[CII] emission}
\label{sec:cii}
The detection rate of [CII]157$\micron$ in the survey was relatively low. For example, neither of the disks around AA Tau and HD~135344 were seen, yet both of these are among the most massive disks in the survey (total masses of $\sim 10^{-2}M_{\odot}$), with relatively rich spectra at other wavelengths.
\citet{Woit10} predicted that the {[CII] disk detection rate for Herschel/GASPS, assuming a wide range of grid parameters, should be 10 -- 55\%, and would be highly dependent on the UV excess (\S\ref{sec:grids}). Table~5 indicates a detection rate at the low end of this range: the brightest 44\% of [OI]-detected objects were targetted for [CII] and of those, only $\sim$26\% were detected in [CII]. This might indicate that the low-UV models are more applicable to the sample. However, this is not supported by the [OI] detection rates, which are more consistent with moderate UV excesses (\S\ref{sec:dmass}). Further investigation of this discrepancy is warranted.

A few objects showed extended [CII], or evidence of emission from the chop reference position, but in general problems from such confusion were limited (\S2.1.4).
There were, however, clear cases of both high and low-mass objects with [CII] emission centred on the star, examples being UY Aur and HD~100546. Figure~\ref{fig:CII-UYAur} compares the spectra of UY Aur from the central spaxel with the middle ring of 8 and the outer 16 spaxels in the PACS IFU. Both line and continuum are centrally peaked, with average fluxes consistent with the instrumental PSF (11$\arcsec$ FWHM) at 157$\micron$.
However, published coronograph images shows that these objects also have scattered light extending over 5-10arcsec \citep{Hiok07, Ard07} with a complex scattering morphology. This is larger than typical disk sizes and suggests emission may be from a compact envelope. The origin of the [CII] line in these objects and whether it arises from the disk, compact envelope or unresolved outflow is underway. But it suggests that the [CII] detection rate from the disks themselves might be even lower than indicated in Table~5.

\begin{figure}[htbp]
\plotone{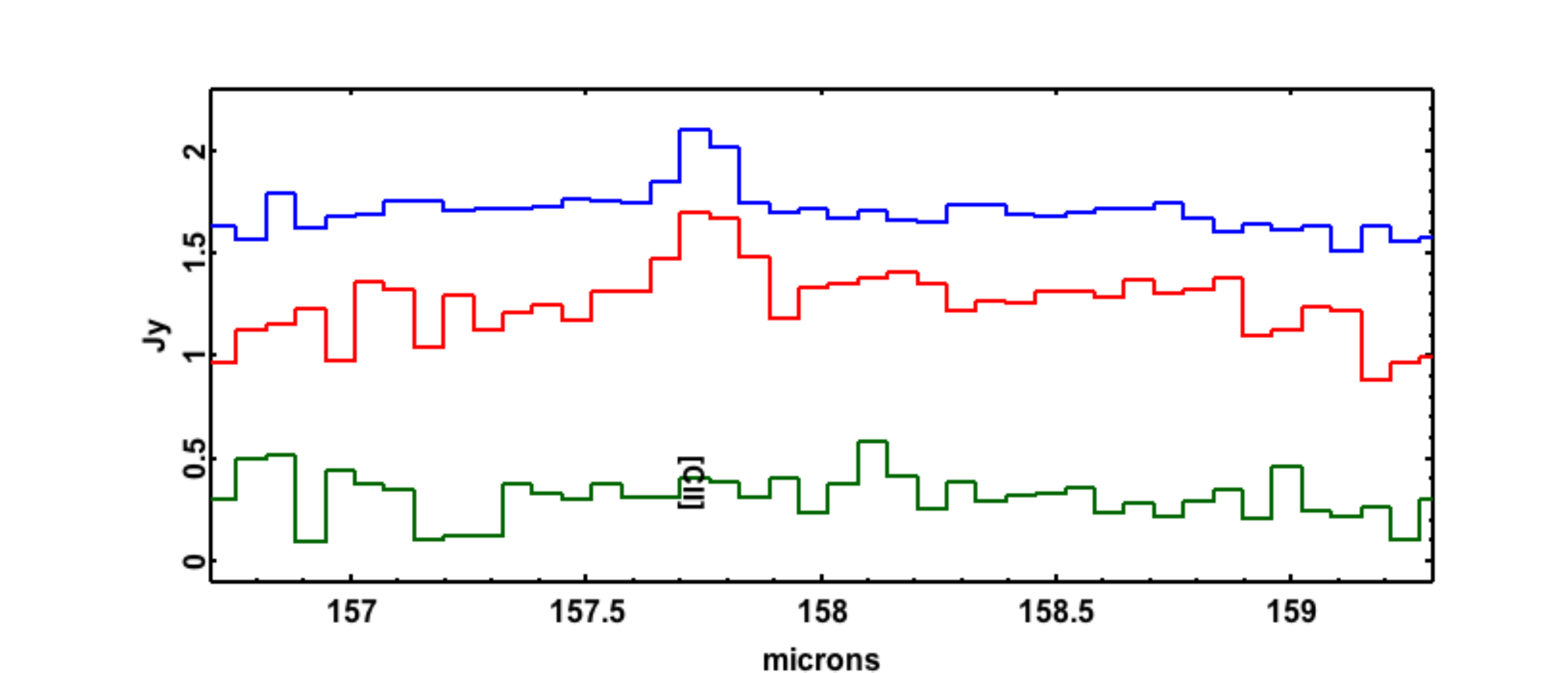}
\caption{Spectra of [CII]157 from UY Aur in the central spaxel (blue histogram), and average of the 8 adjacent and 16 outer spaxels (red and green histograms, both scaled up by a factor of 6 for clarity). The emission in both line and continuum is centrally-peaked and consistent with the PSF response and the line/continuum ratio is similar in the central and first ring of spaxels, indicating that neither the line nor continuum are spatially extended compared with the beam.}
\label{fig:CII-UYAur}
\end{figure}

\subsection{Effect of system parameters on line detectability}

The GASPS target list (Table~\ref{tab:slist}) comprises a rich sample of Class II-III objects in the $\sim$0.3-30~Myr age range, and the survey detects [OI]63$\micron$ from half of the targets observed.
In the following, we investigate preliminary trends in line detectability vs. other directly-observed parameters. Results from GASPS papers on the individual associations (both published and in preparation) are combined to look at overall detection statistics. A minimum detection limit is 3$\sigma$ and, although these were not all reduced with the same version of HIPE, the criteria for detection/non-detection is considered robust in this study. A more detailed investigation of correlations of line fluxes using systematically-calibrated data obtained from the same software version is left for a later paper.

\subsubsection{Disk dust mass and [OI] detections}
\label{sec:dmass}

The probability of [OI] detection in the GASPS survey is a strong function of the disk dust mass, $M_d$. This is illustrated in Figure \ref{fig:diskmass}, as a histogram of the detection rates as a function of distance-normalised dust mass, $M'_d = M_d.(140/D)^2$, where $D$ is the distance in pc. In this figure, we normalised the mass to an equivalent object giving the same flux at the distance of Taurus. We almost always detect the [OI]63$\micron$ line when $M'_d$ reaches a threshold of $\geq10^{-5} M_{\odot}$: 84$\pm$10\% of targets were detected above this mass (where the uncertainty is the statistical error). This is comparable with the mass detection limit of sub-mm continuum surveys (e.g. \citet{AW05}), assuming a standard mass opacity, $\kappa_{\nu}$. Assuming also an ISM-like gas/dust ratio can be used for all disks, this implies a {\em total} mass detection threshold for [OI]63$\micron$ of $\sim 10^{-3} M_{\odot}$. If the gas:dust ratio is more typically 10$\times$ lower (as has been suggested for TW Hya), then the [OI] observations are detecting disks with total masses $\geq 10^{-4} M_{\odot}$.
As noted in \S\ref{sec:spatial}, some [OI]63$\micron$ emission can be from outflows; on the plot we indicate in yellow the targets with spatially or spectrally-resolved [OI]. Additional targets (shown in green shading) are
those with published evidence of a high-velocity jet, although the contribution of this to the [OI]63$\micron$ emission flux is unclear (see \S\ref{sec:jet}).

\begin{figure}[htbp]
\plotone{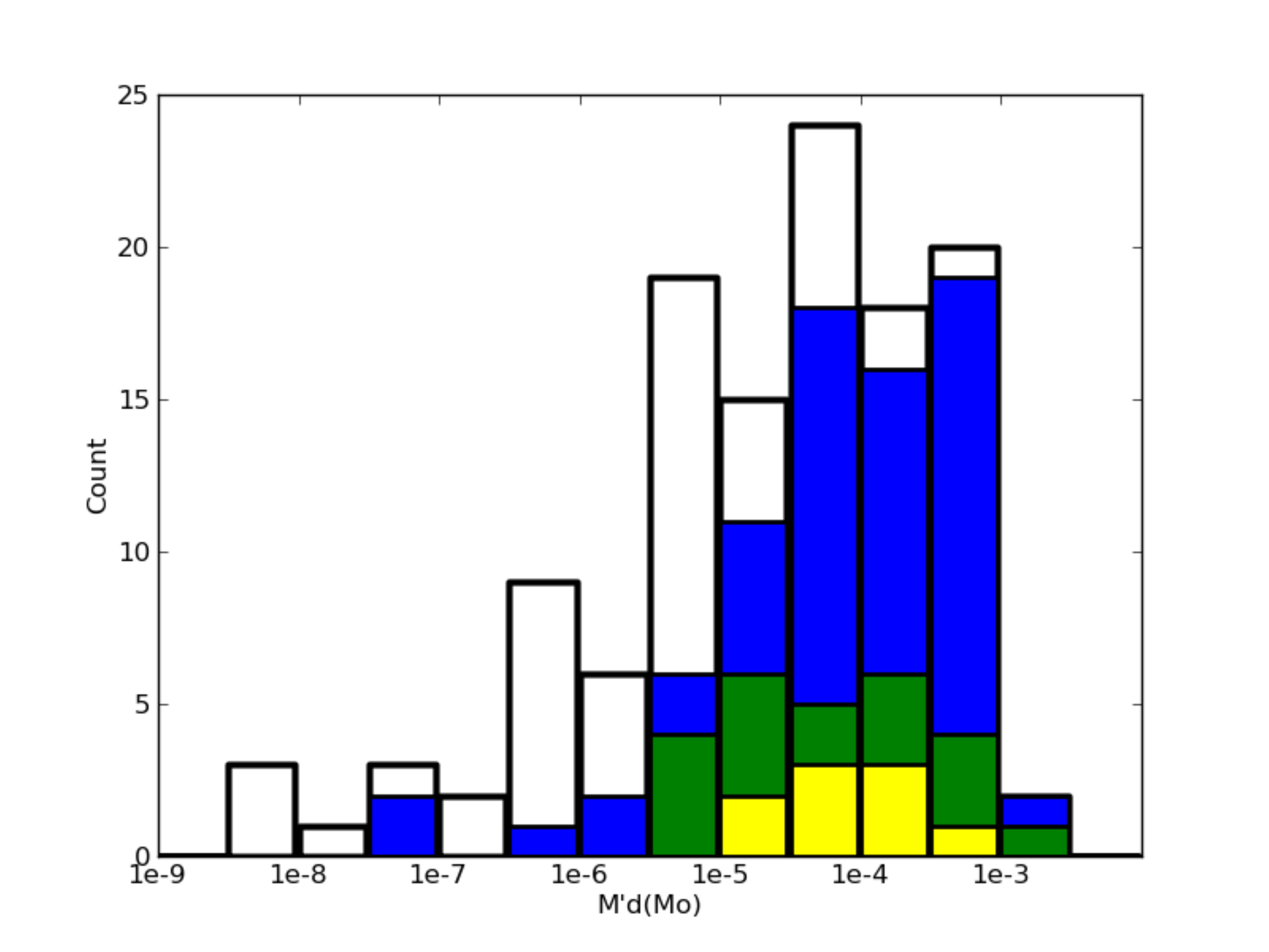}
\caption{Distribution of normalised dust mass in the GASPS sample, illustrating the mass threshold for [OI]63$\micron$ detection of a disk at the fiducial distance of 140pc. Targets with [OI] detections are shown in shaded colours. The dust masses (in Solar units) are mostly based on published mm continuum observations, normalised to the distance of Taurus (140pc), and assume a standard dust mass opacity (see text). Yellow shading indicates objects with extended energetic outflows, where the [OI] line is spatially or spectrally extended and includes some contribution from the jet (see \S5.3.1 and 5.3.2). Green shading indicates objects with evidence of an optical jet, but without spatially or spectrally-resolved [OI]63$\micron$ emission; the contribution to the line from the jet in these objects is unclear. }
\label{fig:diskmass}
\end{figure}

The 84\% detection rate for systems of $M'_d\geq10^{-5}M_{\odot}$ drops to 32$\pm$12\% for $10^{-6} \leq M'_d \leq10^{-5}M_{\odot}$. \citet{Woit10} constructed a large grid of disk models covering a parameter space similar to the that of the GASPS sample, and predicted overall [OI]63$\micron$ detection rates of 51-70\% for disks with dust masses of $10^{-7} - 10^{-3}M_{\odot}$ in systems with a high UV excess, and 17-30\% for this mass range in the case of low UV. Restricting the model grid of citet{Woit10} to systems with 
$M'_d\geq10^{-5}M_{\odot}$, we find that the observed 84\% detection rate is achieved for moderate UV excesses ($0.01 \leq f_{uv} \leq 0.1$) and disk flaring ($1.0 < \beta < 1.2$). This suggests these ranges are typical of most systems in the survey.

There are some notable exceptions to the mass detection threshold, where we detected [OI] in systems with $M'_d<3\times10^{-6}M_{\odot}$:
\begin{itemize}
\item{HD~172555, an unusual warm debris system with no evidence of molecular gas in mm lines, but with some indication that [OI] may be secondary gas released in collisions \citep{Riv12b}.}
\item{ ET Cha, an apparently compact disk in the relatively old $\eta$~Cha association \citep{Woit11}.} 
\item{J130521.6-773810, although the classification of this target in ChaII is uncertain.}
\item{51 Oph, a warm compact disk with notable hot and compact molecular gas component (Thi et al., submitted).}
\item{HD141569, a diffuse disk with spiral structure around a HAeBe star \citep{Clam03}.}
\end{itemize}
The number of disks with published dust masses as low as $10^{-8}M_{\odot}$ is relatively small (only $\sim$10 in GASPS have measured values), and further mm-wavelength measurements of such disks would be interesting to improve the statistics. 
At the opposite extreme, three relatively massive disks ($M'_d \geq10^{-4}M_{\odot}$) have no evidence of [OI]63$\micron$: GO Tau, V836 Tau and TWA03. \citet{Woit10} indicate that disks of this mass which have low flaring ($\beta \leq 1.0$) can have [OI]63$\micron$ fluxes too low to be detected by GASPS.

\subsubsection{Dependence on spectral type}
\label{sec:sptype}

It is already clear from Table~5 that [OI]63$\micron$ is significantly easier to detect around HAeBe stars than T~Tauri stars. Is this simply because HAeBe disks in the sample are more massive and the detection threshold is more commonly reached? Figure \ref{fig:mdTeff} shows the distribution of normalised disk dust masses ($M'_d$) in GASPS as a function of stellar T$_{eff}$. Systems detected in [OI] and [CII] are indicated by the filled black and red symbols respectively.  This shows that both early and late-type stars have a similar range of disk dust masses in this sample. As noted above, the [OI] detection rate is high for disks with $M'_d>10^{-5}M_{\odot}$, and Figure~10 shows that this is independent of T$_{eff}$ for T$_{eff} >4000$K. However, approximately half of the low-luminosity stars (T$_{eff}<4000K$, or M type) with $M'_d$ in the range $10^{-5}$ to $10^{-4}M_{\odot}$ were not detected. Clearly the spectral type has some effect on the [OI] line emission threshold for the lowest-luminosity stars.

\begin{figure}[htbp]
\plotone{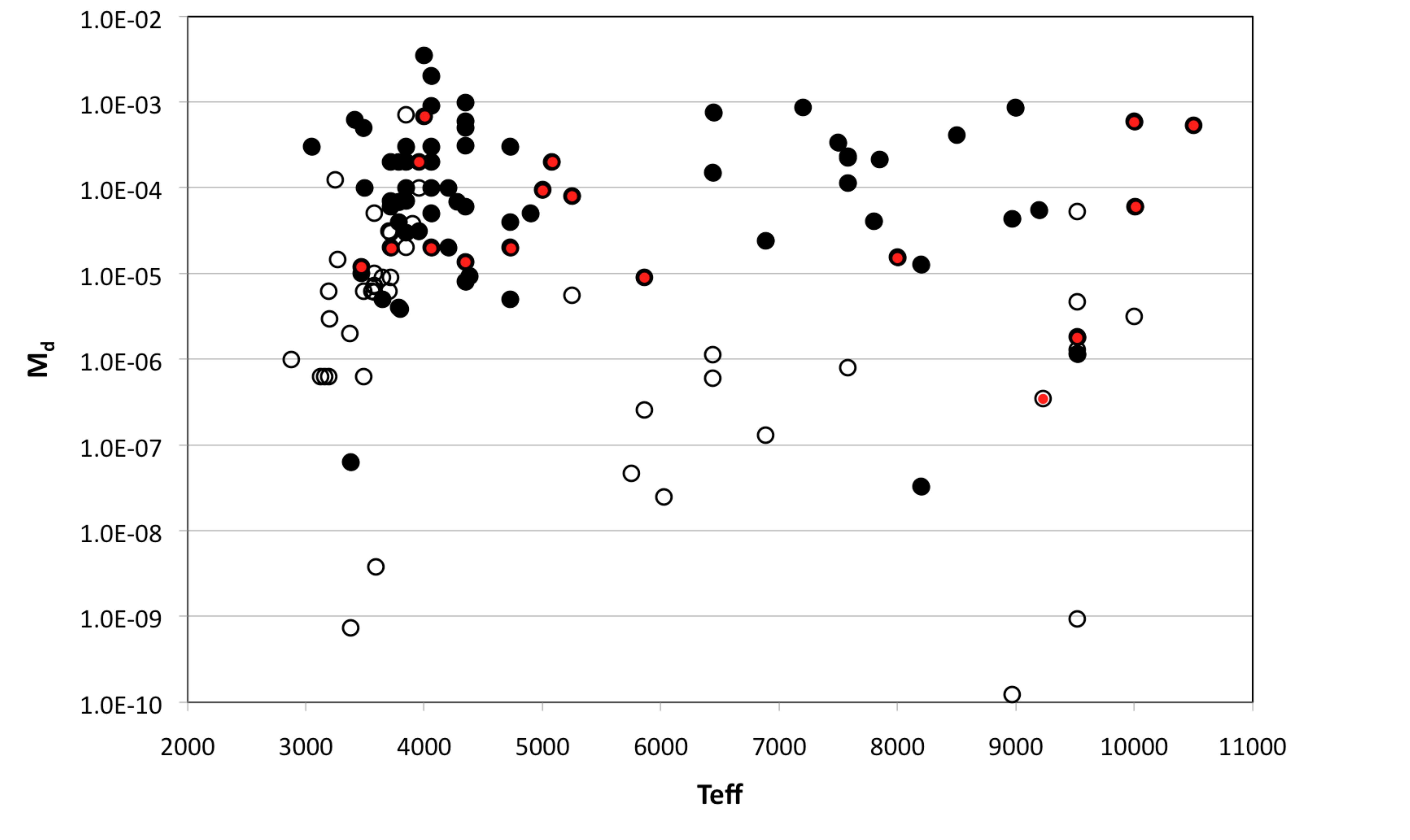}
\caption{Disk dust mass of the GASPS sample, normalised to a distance of 140pc (units of M$_{\odot}$), plotted as a function of the stellar effective temperature (in K). This illustrates the line detectability of a disk of a given dust mass at the distance of the Taurus star forming region.  [OI]63$\micron$ detections are shown as filled circles, and open circles depict [OI] upper limits. An additional red dot indicates systems which were  detected in [CII]. }
\label{fig:mdTeff}
\end{figure}

In the case of the [CII]157 line, the detectability in Figure \ref{fig:mdTeff} seems to be independent of the spectral type and disk mass, with [CII] detections (shown as filled red symbols) broadly distributed over the $M'_d$ -- T$_{eff}$ parameter space. Unlike [OI], there is no clear threshold with disk mass, or an increase in detection rate among HAeBe stars. If most [CII] arises from a compact envelope rather than the disk, this suggests that such gas may be retained around these stars independent of the mass of the inner disk or stellar type.

\subsubsection{Other observational parameters: binarity, H$\alpha$ and X-ray luminosity}

The histograms in Figure \ref{fig:hist_Lx} and \ref{fig:hist_sep} show the detection statistics for targets searched in [OI]63$\micron$ with published X-ray luminosity, H$\alpha$ equivalent width (EW) and binary separation. The numbers of targets with [OI] detections are shaded. Those which have additional extended [OI]63$\micron$ emission from a jet are shaded yellow. Figure \ref{fig:hist_Lx}a shows that line emission is detected in systems covering the full range of X-ray luminosity in the survey, with no clear trend of increased detectability for higher X-ray fluxes. The H$\alpha$ EW used in Figure \ref{fig:hist_Lx}b is linked with the accretion rate, although later-type K stars may have significant chromospheric contribution and the accretion luminosity may be lower than Figure \ref{fig:hist_Lx}b might suggest. But there is a trend of increasing [OI] detection probability for higher accretion rates: including all stars observed, the detection fraction is 70\% for EW$>30$\AA~(or 67\% excluding the stars with extended jet emission) compared with only 29\% for those with lower EW.  A systematic derivation of accretion luminosity and line flux over the whole survey would be interesting to study further correlations between the [OI]63$\micron$ flux and accretion rates.

\begin{figure}[htbp]
\plotone{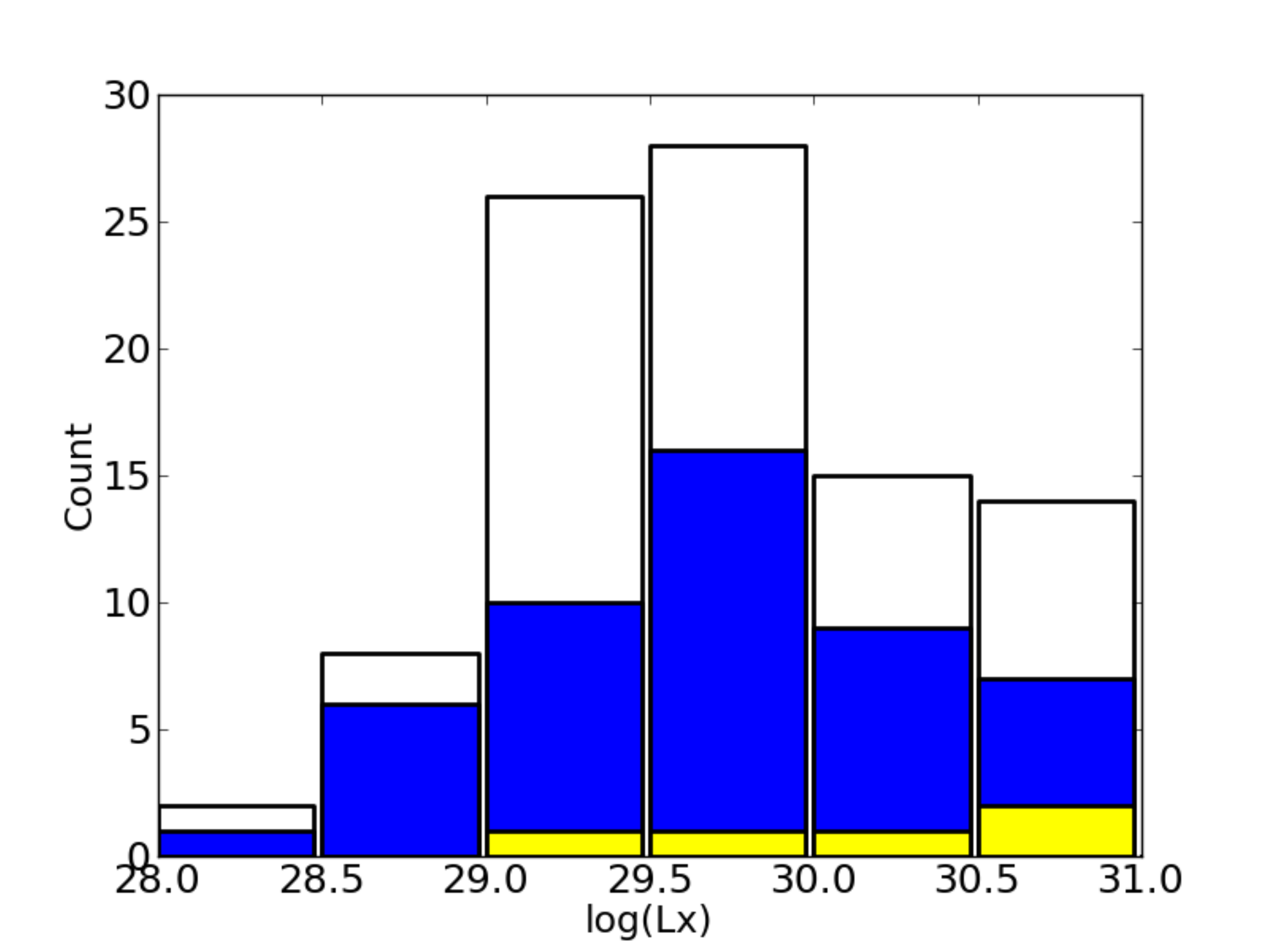}
\plotone{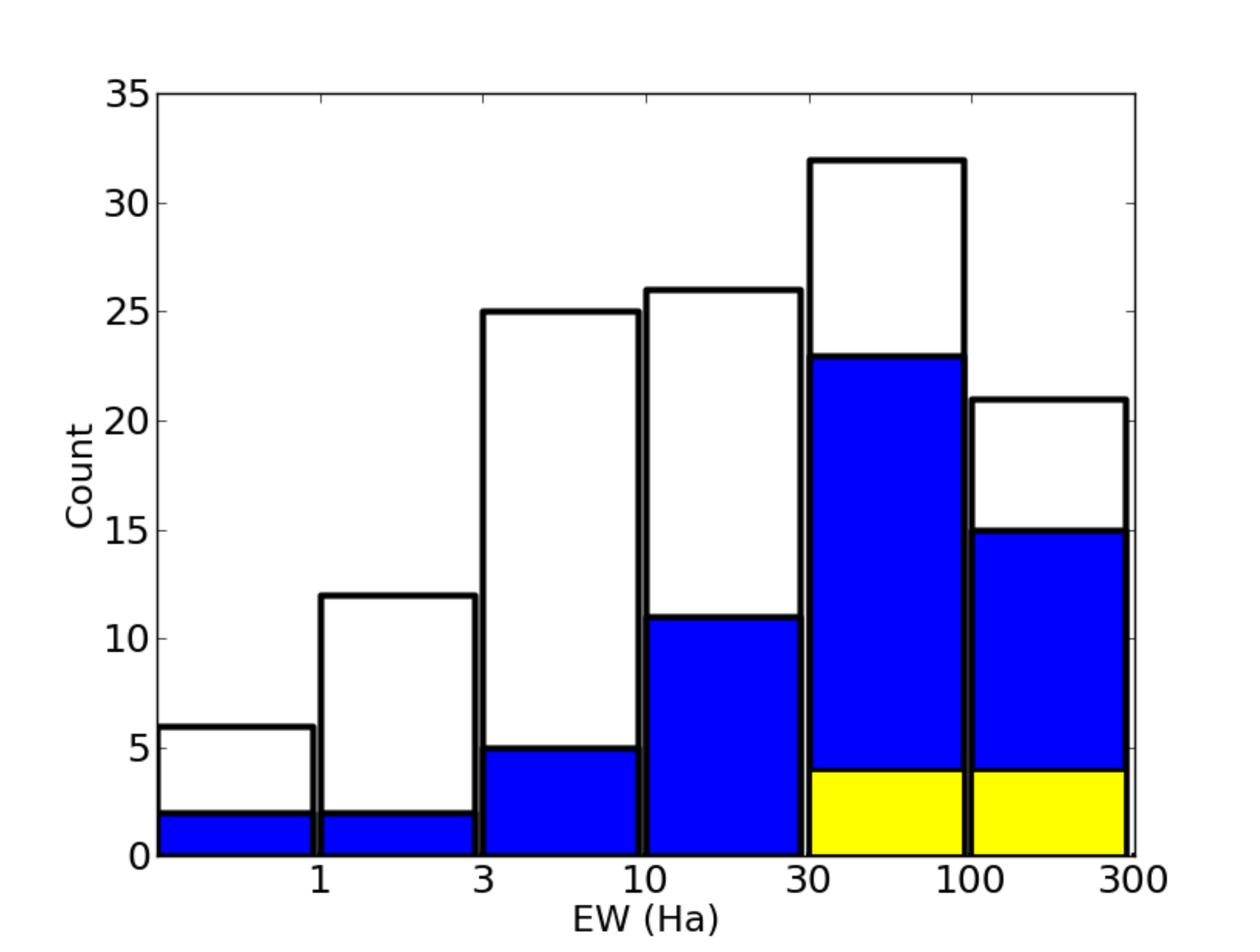}
\caption{Histogram of the distribution of X-ray luminosity ($erg~s^{-1}$) (upper plot) and H$\alpha$ EW (\AA)  (lower) of the GASPS sample observed in [OI]63$\micron$. The [OI] detections are shaded in blue, with the yellow shading indicating those with an extended [OI] component. There is no clear dependence of [OI] detectability on X-ray luminosity, but detection rates are higher for larger H$\alpha$ EW.}
\label{fig:hist_Lx}
\end{figure}

Figure \ref{fig:hist_sep} illustrates the [OI] detection rates distributed over binary separation. For hierarchical multiples we have used the separation of the widest component within the PACS beam. There is marginal evidence for a drop in detection rates in multiples of separation $<$300AU, from 64\% for the wider binaries to 40\% for the closer systems (with statistical errors of $\sim$10\%). By comparison, the [OI] detection fraction of single stars in the sample was 47\%. This would suggest that most [OI] emission arises from radii of $<$300AU - similar to the [OI]-emitting region suggested by the models in Figure \ref{fig:models}. By comparison, samples of T Tauri stars observed in mm dust indicate that M$_d$ typically drops by a factor of 5 for binary separations of $<$300AU \citep{Har12}.

\begin{figure}[htbp]
\plotone{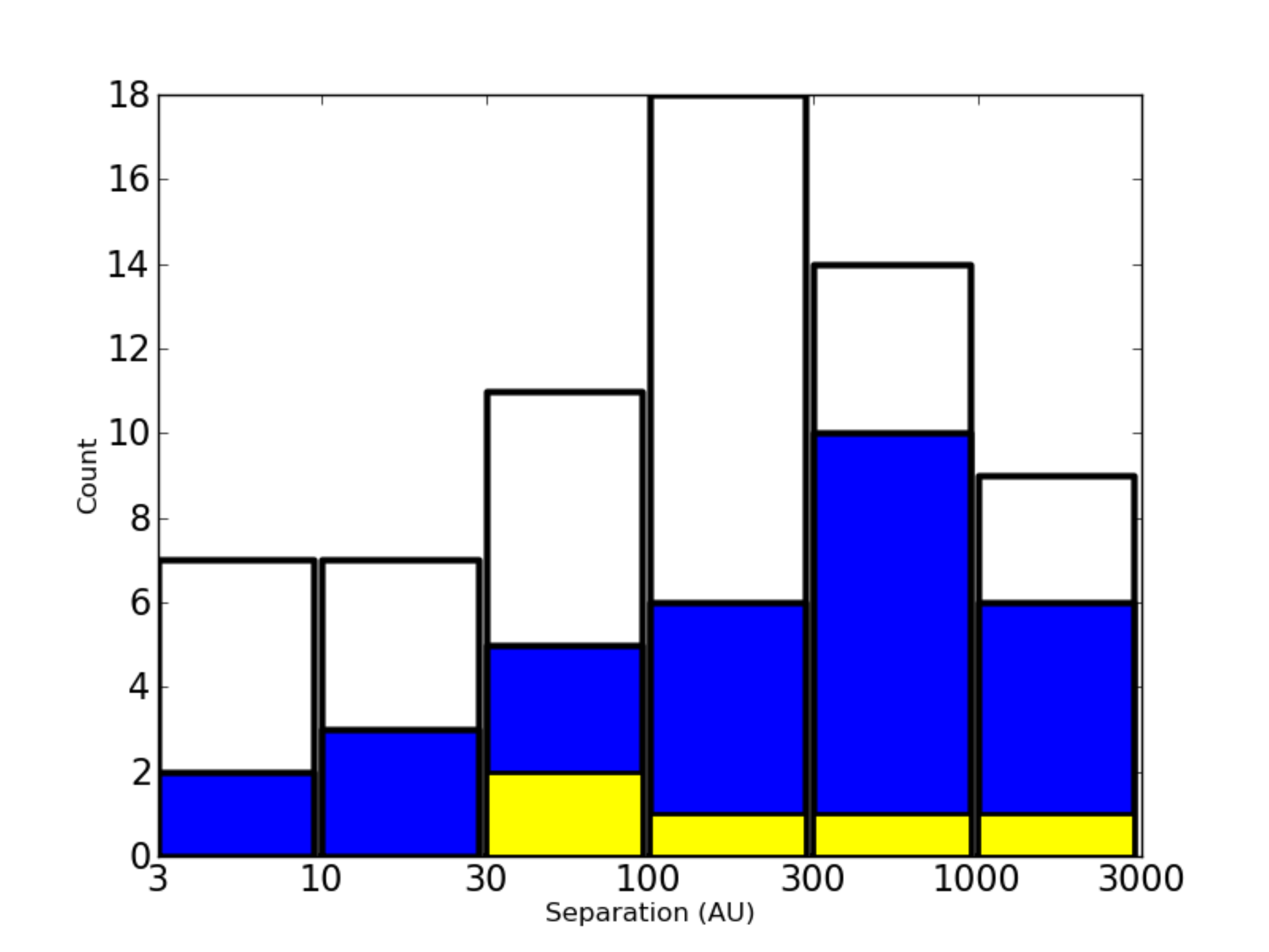}
\caption{Histogram of the distribution of binary separation (in AU) in the sample observed in [OI]63$\micron$, with line detections shaded in blue. Binary stars with known extended jet contribution to the [OI]63$\micron$ flux are shaded further in yellow.}
\label{fig:hist_sep}
\end{figure}

\subsubsection{Detection rates in different associations and dependence on age}

In Table~6 we give the [OI]63$\micron$ detection rates for the different associations observed in GASPS, for targets with and without published dust masses (with masses based on continuum photometry, although only two disks were detected which had no published estimates). As noted above, the required dust mass ($M'_d$) for [OI]63$\micron$ to be detected at a distance of 140pc is $\sim10^{-5}M_{\odot}$, or a total (gas+dust) mass of $1 M_{Jupiter}$ assuming an ISM gas/dust ratio. The detection rates are somewhat dependent on spectral type (M stars have a higher disk mass threshold for detection - see Figure \ref{fig:mdTeff}), binarity (lower for separations $<$300AU), and accretion rates (more for H$_{\alpha}>30$\AA). However, if we use [OI]63$\micron$ detections as a proxy for 1$M_{Jupiter}$ disks at the fiducial 140pc distance, it is possible to compare the detection rates in different associations, modulo the distances and assuming we are sampling most of the brighter detectable disks.   Each of the three intermediate-age (5-10Myr) associations ($\eta$ Cha, TWA and upper Sco) has $\sim$2-3 such disks. TWA is significantly closer, and only one of these would be detected by GASPS if moved to the fiducial distance of 140pc. The {\em total} number of stars in the upper Sco group is $\sim$200 \citep{Carp09}, with 20-50 in the two other associations in this age range \citep{Tor08}, giving a gas-detected disk fraction of 1-7\% at 5-10Myr. For the older systems, there are 2 unusual [OI]-emitting disks in the 10-20Myr $\beta$ Pic moving group but neither would have been detected at 140pc, and no disks were detected in [OI] in the 30-Myr old Tuc Hor association. Each of these contains $\sim$40 stars \citep{Tor08}. For the two younger groups, the detection rate in GASPS was $\sim$50\%, however, the total number of stars is estimated to be $\sim$250 in Taurus and $\sim$48 in Cha II \citep{Reb10, Alc08}, giving massive disk fractions of $\sim$18 and 17\% respectively. For the HAeBe stars in the sample (excluding debris disks), the derived ages are mostly in the range 2-10Myr \citep{Mo09, Meu12}, and the [OI] fraction is $\sim$100\%, much higher than the equivalent-aged FGKM star. While not a statistically-complete sample of AeBe stars in this age range, it suggests either that these more luminous stars are more commonly able to retain disks, or that their ages are overestimated. Overall, for T Tauri stars the fraction with Jupiter-mass, gas-rich disks is $\sim$18\% at ages of 0.3-4Myr, 1-7\% at 5-10Myr, and none are detected beyond 10Myr age. It is unclear why some particular stars can retain these gas-rich disks for up to 10Myr, and whether planets formed in such long-lived disks would be different from those around other stars.

\begin{table*}[htbp]
\begin{center}
\caption{\sf  Detection statistics of [OI]63$\micron$ for the associations in GASPS, with and without known disk masses. }
\begin{tabular}[t]{lcccccc}
\hline
& \multicolumn{2}{c} {M$_d$} & \multicolumn{2}{c}{No M$_d$$^1$} & \multicolumn{2}{c}{Total}\\
Association & Detected  & Observed &Detected &Observed & Detected & Observed\\
\hline\\[-3mm]
Taurus & 44 & 56 & 1 & 17 &45 & 73\\
Cha II & 7 & 17 & 1 & 2  &8 & 19\\
$\eta$ Cha & 2 & 7 & 0 & 6 &2 & 13\\
TW Hya & 3 & 5 & 0 & 3 &3& 8 \\
Upper Sco & 2 & 7 & 0 & 12  &2&19\\
$\beta$ Pic & 1 & 4 & 0 & 1 & 2$^2$ &6 \\
Tuc Hor & 0 & 2 & 0 & 0  &0 &2 \\
HAeBe stars & 20 & 25$^3$ & 0 & 0&20&20\\
\hline
\end{tabular}
\end{center}
\label{tab:det-assoc}
\footnotesize{(1) No disk masses were available from the literature.\\
(2) Includes detection of $\beta$ Pic itself \citep{Bran11}.\\
(3) 5 non-detected systems are those classed as possible debris disks.}
\end{table*}

\section{Conclusions}

In this paper we describe the GASPS far-infrared survey of gas and dust in young stellar systems. This Herschel Key Project observed selected lines and continuum with the PACS instrument, targetting 250 young SED-Class II-III systems, Herbig AeBe stars and young debris disks. The far-IR line emission can arise from the hot surface of gas disks around 30-300AU, high-velocity jet shocks, disk winds and/or compact remnant envelopes. We outline the models used to predict disk line strengths and their dependence on disk parameters, and look at the possible emission from these other mechanisms, in particularly from extended outflow jets in a few objects. The master list of targets with basic system parameters is presented, and we also indicate in this list which objects have detections in the primary GASPS lines. 

The FIR lines are identified and we show preliminary line detection statistics, referring to the published papers which analyse individual sources and associations in more detail. Rich spectra were seen in a number of sources, including fine-structure atomic lines of [OI] at 63 and 145$\micron$, and [CII] at 157$\micron$, as well as molecules including high-J transitions of CO, H$_2$O, OH and CH$^+$. In most systems [OI]63$\micron$ was the brightest line, by a factor of $\sim$10, and is associated in most cases with disk emission. Line and continuum emission was, in all but 10 systems, both spatially and spectrally unresolved and centred on the star. For the extended sources, emission is thought to arise from jet shocks, along with a disk contribution. The [CII]157 line was significantly weaker than [OI], resulting in a relatively low detection rate. However, in a small number of objects unresolved emission was found centred on the star, and may arise from a compact remnant envelope.

49\% of all targets observed were detected in [OI]63$\micron$, with an 84\% detection rate for those having disk dust masses of $\geq10^{-5}M_{\odot}$.  Comparison with statistics from a grid of model implies that most systems have moderate UV excesses and disk flaring. Assuming an ISM gas:dust ratio of 100 and typical mm-wavelength mass opacity, this represents an [OI] detection threshold for the {\em total} disk mass of $\sim1M_{Jupiter}$. Going against this trend, we find five unusually low disk mass systems with [OI]63$\micron$ emission, and a few high-mass systems which remained undetected. The [OI]63$\micron$ detection rates were independent of X-ray luminosity, but there was evidence of a decreased rate in binaries of separation $<$300AU, in stars with H$\alpha$ EW $<$30\AA, and for M-type stars. Based on [OI]63$\micron$ line detections, the results show that $\sim$18\% of stars in each association can retain gas-rich disks of minimum mass $10^{-3}M_{\odot}$ for up to 4Myr, a few \% of stars keep these disks for $\sim$5-10Myr, but none are detected beyond 10-20Myr.

\section{Acknowledgments}
{We wish to acknowledge the Herschel Helpdesk for their timely and useful responses. The Grenoble group thanks ANR (contracts BLAN-0221, 0504-01 and 0505-01),   European Commission's 7$^\mathrm{th}$ Framework Program 
(contract PERG06-GA-2009-256513), CNES, and PNPS of CNRS/INSU, France for support. PW, IK, and WFT acknowledge funding from the EU FP7-2011 under Grant Agreement nr. 284405. WFT acknowledges a Scottish University Physics Alliance fellowship (2006-2009).
PACS has been developed by a consortium of institutes led by MPE (Germany) and including UVIE (Austria); KU Leuven, CSL, IMEC (Belgium); CEA, LAM (France); MPIA (Germany); INAF-IFSI/OAA/OAP/OAT, LENS, SISSA (Italy); IAC (Spain). This development has been supported by the funding agencies BMVIT (Austria), ESA-PRODEX (Belgium), CEA/CNES (France), DLR (Germany), ASI/INAF (Italy), and CICYT/MCYT (Spain).}

}

{\it Facilities:} \facility{Herschel}.



\appendix

\section{Target list}
\begin{deluxetable}{llccclclrrccl}
\tabletypesize{\scriptsize}
\tablenum{A.1}
\rotate
\tablecaption{Initial target list\label{tab:slist}}
\tablewidth{0pt}
\label{tab:slist}
\tablehead{
\colhead{ID} & \colhead{Name}  & \colhead{RA} & \colhead{DEC} & \colhead{Dist.} &
\colhead{Sp.T.\tablenotemark{a}}  &
\colhead{Sep.\tablenotemark{b}} & \colhead{log(L$_x$) \tablenotemark{c}}&
\colhead{W$_{H\alpha}$\tablenotemark{d}} &\colhead{SED\tablenotemark{e}} & \colhead{M$_{dust}$\tablenotemark{f}} & \colhead{OI/CII/CO/H$_2$O\tablenotemark{g}} & \colhead{Notes/refs.\tablenotemark{h}}
}
\startdata
 T-1  & HBC 347   & 03 29 38.37  & +24 30 38.0  &140 & K1  &...&...&0.2 &III&$<$4.0e-6 & 0/:/:/0&  \\ 
 T-2  & HBC 351  & 03 52 02.24  & +24 39 47.9  &140 & K5   &0.61 &...&2.8  &III&$<$5.0e-6 & :/:/:/:& \tablenotemark{i} \\ 
 T-3  & HBC 352/353  & 03 54 29.51  & +32 03 01.4  &140 & G0 & 8.6 &30.4 &$<$&III& $<$5.0e-6 & :/:/:/:& \tablenotemark{i} \\ 
 T-4  & HBC 354/355   & 03 54 35.56  & +25 37 11.1  &140 & K3    & 6.3  &...&$<$&III&$<$4.0e-6 & :/:/:/:& \tablenotemark{i} \\ 
 T-5  & SAO 76411   & 04 02 53.57  & +22 08 11.8  &140 & G1   &...&...&1 &III& $<$5.0e-6 & :/:/:/:&  \\ 
 T-6  & HBC 356/357   & 04 03 13.96  & +25 52 59.8  &140 & K2   & 2.0  &...&1 &III& $<$4.0e-6 & 0/:/:/0& \tablenotemark{i} \\ 
 T-7  & HBC 358/359  & 04 03 50.84  & +26 10 53.2  &140 & M2   &  1.6 &29.8 &7 &III& $<$5.0e-6 & 0/:/:/0&  \\ 
 T-8  & SAO 76428 & 04 04 28.49  & +21 56 04.6  &140 & F8  &...&...&1.3 &III& $<$6.0e-6 & :/:/:/:&  \\ 
 T-9  & HBC 360/361   & 04 04 39.37  & +21 58 18.6  &140 & M3   &7.2 &...&6.6 &III& $<$7.0e-6 & :/:/:/:&  \\ 
 T-10  & HBC 362   & 04 05 30.88  & +21 51 10.7  &140 & M2   &...&...&6.6 &III& $<$4.0e-6 & :/:/:/:&  \\ 
 T-11  & LkCa 1   & 04 13 14.14  & +28 19 10.8  &140 & M4    &...&29.3 &3.5 & III & $<$4.0e-6 & 0/:/:/0&  \\ 
 T-12  & Anon 1  & 04 13 27.23  & +28 16 24.8  &140 & M0   & 0.015 &30.5 &2.5 &III& $<$4.0e-6 & 0/:/:/0&  \\ 
 T-13  & V773 Tau A-D  & 04 14 12.92  & +28 12 12.4  &140 & K2   &0.2 &30.9 &3.0 &II&5.0e-6 & 1/:/:/0& Jet. Hierachical\\ 
 T-14  & FM Tau   & 04 14 13.58  & +28 12 49.2  &140 & M0   &...&29.7 &76 &II&2.0e-5 & 0/0/0/0 & 37" from V773Tau \\ 
 T-15  & CW Tau   & 04 14 17.0  & +28 10 57.8  &140 & K3    &...&30.4 &140 &II&2.0e-5 & 1/1/:/0 & Jet, high$\dot M$\\ 
 T-16  & CX Tau  & 04 14 47.86  & +26 48 11.0  &140 & M3   &...&...&18 &II&1.0e-5 & 1/:/:/0&  \\ 
 T-17  & LkCa 3 AB   & 04 14 47.97  & +27 52 34.6  &140 & M1   & 0.47  &29.8 &2.7 &III& $<$4.0e-6 & 0/:/:/0&  \\ 
 T-18  & FO Tau AB   & 04 14 49.29  & +28 12 30.6  &140 & M2   &0.15 &28.7 &126 &II&6.0e-6 & 0/0/0/0 &  \\ 
 T-19  & CIDA-2   & 04 15 05.16  & +28 08 46.2  &140 & M5.5  &...&29.1 &6 &III& $<$7.0e-6 & 0/:/:/0&  \\ 
 T-20  & LkCa 4  &  04 16 28.11  & +28 07 35.8  &140 & K7  &...&30.0 &3.2 &III& $<$2.0e-6 & 0/:/:/0&  \\ 
 T-21  & CY Tau   & 04 17 33.73  & +28 20 46.8  &140 & M1.5 &...&29.3 &63 &II&6.0e-5 & 1/:/:/0&  \\ 
 T-22  & LkCa 5   & 04 17 38.94  & +28 33 00.5  &140 & M2   &  0.048 &29.7 &3.8 &III& $<$2.0e-6 & 0/:/:/0&  \\ 
 T-23  & HBC 372   & 04 18 21.47  & +16 58 47.0  &140 & K5   &...&...&$<$ &III& $<$4.0e-6 & :/:/:/:&  \\ 
 T-24  & HBC 376   & 04 18 51.70  & +17 23 16.6  &140 & K7   &...&...&1.9 &III& $<$3.0e-6 & :/:/:/:&  \\ 
 T-25  & 04158+2805   & 04 18 58.14  & +28 12 23.5  &140 & M3  &...&...&175 &I-II&3.0e-4 & 1/0/1/0& Jet \\ 
 T-26  & FQ Tau AB   & 04 19 12.81  & +28 29 33.1  &140 & M3  &0.76 &28.8 &97 &II&1.0e-5 & 0/0/0/0 &  \\ 
 T-27  & BP Tau   & 04 19 15.84  & +29 06 26.9  &140 & K7  &...&30.2 &66 &II&2.0e-4 & 1/:/:/0&  \\ 
 T-28  & V819 Tau AB  & 04 19 26.26  & +28 26 14.3  &140 & K7   & 10.5  &30.3 &2.5 &III& $<$4.0e-6 & 0/0/:/0&  \\ 
 T-29  & LkCa 7 AB  & 04 19 41.27  & +27 49 48.5  &140 & K7   &1.04 &29.9 &3.9 &III& $<$4.0e-6 & 0/:/:/0&  \\ 
 T-30  & DE Tau   & 04 21 55.64  & +27 55 06.1  &140 & M2   &...& - &59 &III&5.0e-5 & 0/0/:/0&  \\ 
 T-31  & RY Tau   & 04 21 57.40  & +28 26 35.5  &140 & K1   &...&30.7 &13 &II&2.0e-4 & 1/1/0/1& Jet \\ 
 T-32  & HD 283572   & 04 21 58.85  & +28 18 06.6  &140 & G5  &...&31.1 &$<$ &III& $<$4.0e-6 & 0/:/:/0& - \\ 
 T-33  & T Tau NS  & 04 21 59.43  & +19 32 06.4  &140 & K0  & 0.7 &30.9 &40 &I-II&8.0e-5 & 1/1/1/1 & Ext.OI. Po12 \\ 
 T-34  & FS Tau AB  & 04 22 02.18  & +26 57 30.5  &140 & M0   &0.25 &30.9 &69 &II&2.0e-5 & 1/1/1/1 & Ext.OI. Po12 \\ 
 T-35  & LkCa 21  & 04 22 03.14  & +28 25 39.0  &140 & M3   & 0.044   &30.9 &6.1 &III& $<$5.0e-6 & :/:/:/:&  \\ 
 T-36  & FT Tau  & 04 23 39.19  & +24 56 14.1  &140 & M3  &...&...& 254 &II&1.0e-4 & 1/:/:/0&  \\ 
 T-37  & IP Tau   & 04 24 57.08  & +27 11 56.5  &140 & M0   &...&...&10.5 &II&3.0e-5 & 1/:/:/0&  \\ 
 T-38  & J1-4872 AB  & 04 25 17.68  & +26 17 50.4  &140 & K7   &3.3 &29.7 &2.9 &III& $<$4.0e-6 & 0/:/:/0&  \\ 
 T-39  & DG Tau B  & 04 27 02.56  & +26 05 30.7  &140 & $<$K6   &...& 31.0 & 270 &I-II&6.8e-4 & 1/1/:/0& Ext.OI. Po12\\ 
 T-40  & DF Tau AB  & 04 27 02.80  & +25 42 22.3  &140 & M0.5  &0.07 &29.1 &54.5 &II&4.0e-6 & 1/0/0/0 & Jet, high$\dot M$ \\ 
 T-41  & DG Tau A  & 04 27 04.70  & +26 06 16.3  &140 & K6    &...& 29.4 &90 &I-II&2.0e-4 & 1/1/1/0 & Ext.OI, high$\dot M$. Po12 \\ 
 T-42  & HBC 388   & 04 27 10.57  & +17 50 42.6  &140 & K1  &...&...&$<$&III& $<$3.0e-6 & :/:/:/:&  \\ 
 T-43  & J1-507   & 04 29 20.71  & +26 33 40.7  &140 & M4  &  0.08  &29.6 &5.1 &III& $<$3.0e-6 & :/:/:/:&  \\ 
 T-44  & FW Tau ABC & 04 29 29.71  & +26 16 53.2  &140 & M4   & 0.2  &...&17 &III&2.0e-6 & 0/:/:/0&  \\ 
 T-45  & DH/DI Tau  & 04 29 42.02  & +26 32 53.2  &140 & M2/M2  & 2.3 &30.9 &35/2 &II/III&3.0e-5 & 0/:/:/0& Jet. Mult.\\ 
 T-46  & IQ Tau   & 04 29 51.56  & +26 06 44.9  &140 & M0.5  &...&29.5 &12 &II&2.0e-4 & 1/0/0/0 &  \\ 
 T-47  & UX Tau B/AC  & 04 30 04.00  & +18 13 49.4  &140 & K1   & 2.7 &29.9 &4 &II/III&5.0e-5 & 1/:/:/0& Mult. \\ 
 T-48  & FX Tau AB & 04 30 29.61  & +24 26 45.0  &140 & M1    &0.9 &29.6 &12 &II&9.0e-6 & 0/:/:/0&  \\ 
 T-49  & DK Tau AB  & 04 30 44.25  & +26 01 24.5  &140 & K7   &2.3 &30.0 &40& II&5.0e-5 & 1/0/:/0 &  \\ 
 T-50  & ZZ Tau  & 04 30 51.38  & +24 42 22.3  &140 & M3    &0.04 &...&14 &III&$<$4.0e-6 & 0/:/:/0& \\ 
 T-51  & JH 56   & 04 31 14.44  & +27 10 18.0  &140 & M0.5    &...&...&2.2 &III& $<$4.0e-6 & :/:/:/: &  \\ 
 T-52  & V927 Tau AB  & 04 31 23.82  & +24 10 52.9  &140 & M5.5   &0.29 &29.2 &10 &III& $<$5.0e-6 & 0/:/:/0&  \\ 
 T-53  & HBC 392   & 04 31 27.17  & +17 06 24.9  &140 & K5   &...&...&1.1 &III& $<$3.0e-6 & :/:/:/:&  \\ 
 T-54  & XZ Tau AB  & 04 31 40.07  & +18 13 57.2  &140 & M1.5   &0.3 &29.9 &274 & II &1.2e-5 & 1/1/1/1 & Jet, ext.OI. +HL Tau\\ 
 T-55  & HK Tau AB & 04 31 50.57  & +24 24 18.1  &140 & M0.5    &2.4 &28.9 &42&I-II &4.0e-5 & 1/0/1/0&  \\ 
 T-56  & V710 Tau AB  & 04 31 57.79  & +18 21 38.1  &140 & M0.5   &3.1 &30.1 &61 &II&7.0e-5 & 1/:/:/0 &  \\ 
 T-57  & J1-665   & 04 31 58.44  & +25 43 29.9  &140 & M5  &...&28.8 &5.2 & III& $<$4.0e-6 & :/:/:/: &  \\ 
 T-58  & L1551-51   & 04 32 09.27  & +17 57 22.8  &140 & K7   &...&30.2 &1.5 &III& $<$6.0e-6 & :/:/:/: &  \\ 
 T-59  & V827 Tau   & 04 32 14.57  & +18 20 14.7  &140 & K7   &  0.09 &30.6 &3 &III& $<$3.0e-6 & 1/:/:/0&  \\ 
 T-60  & Haro 6-13   & 04 32 15.41  & +24 28 59.7  &140 & M0  &...&29.2 &61 &I-II&1.0e-4 & 1/0/1/0 &  Jet \\ 
 T-61  & V826 Tau AB & 04 32 15.84  & +18 01 38.7  &140 & K7   & 0.014 &30.6 &3 &III& $<$4.0e-6 & :/:/:/:&  \\ 
 T-62  & V928 Tau AB   & 04 32 18.86  & +24 22 27.2  &140 & M0.5   &0.2 &30.0 &1.5 &III& $<$4.0e-6 & :/:/:/:&  \\ 
 T-63  & GG Tau AB & 04 32 30.35  & +17 31 40.6  &140 & K7  & 0.25 &28.6 &50 &II&2.0e-3 & 1/0/0/0 & Mult. \\ 
 T-64  & UZ Tau EW  & 04 32 43.04  & +25 52 31.1  &140 & M1  & 3.5 &29.9 &73 &II&2.0e-4 & 1/0/0/0 & Jet, ext.OI, high$\dot M$, mult.\\ 
 T-65  & L1551-55   & 04 32 43.73  & +18 02 56.3  &140 & K7  &...&29.8 &1.0 &III& $<$3.0e-6 & :/:/:/:& \\ 
 T-66  & GH/V807 Tau   & 04 33 06.43  & +24 09 44.5  &140 & M2/K7 & 0.3 &29.1/30.0 &20/10 &II/III&7.0e-6 & 0/0/:/0 &Mult.\\ 
 T-67  & V830 Tau   & 04 33 10.03  & +24 33 43.4  &140 & K7  &...&30.7 &2.5 &III& $<$3.0e-6 & :/:/:/:& - \\ 
 T-68  & GI/GK Tau   & 04 33 34.31  & +24 21 11.4  &140 & K6/K7  &  13  &29.9/30.1 &18/25 &II/II&2.0e-5 & 1/:/:/1 &  Jet. GK is binary \\ 
 T-69  & DL Tau   & 04 33 39.06  & +25 20 38.2  &140 & K7  &...&...&101 &II&9.0e-4 & 1/0/:/1 &  \\ 
 T-70  & HN Tau AB  & 04 33 39.35  & +17 51 52.4  &140 & K5   &3.1 & 29.2 &145 &II&8.0e-6 & 1/0/0/0 & Jet, high$\dot M$ \\ 
 T-71  & DM Tau   & 04 33 48.72  & +18 10 10.0  &140 & M1   &...& 29.3 &114 &II&2.0e-4 & 1/0/:/0&  \\ 
 T-72  & CI Tau   & 04 33 52.00  & +22 50 30.2  &140 & K7  &...&29.5 &90 &II&3.0e-4 & 1/0/0/0 &  \\ 
 T-73  & J2-2041  & 04 33 55.47  & +18 38 39.1  &140 & M3.5  &  0.42 &...&4.7 &...& ... & :/:/:/:&  \\ 
 T-74  & JH 108   & 04 34 10.99  & +22 51 44.5  &140 & M1   &...&30.0 &3 &III& $<$4.0e-6 & :/:/:/:&  \\ 
 T-75  & HBC 407  & 04 34 18.04  & +18 30 06.6  &140 & G8   &  0.14  &...&$<$ &III& $<$4.0e-6 & :/:/:/:&  \\ 
 T-76  & Wa Tau/1   & 04 34 39.29  & +25 01 01.0  &140 & K0   &...&...&0.5 &III& $<$3.0e-6 & :/:/:/:&  \\ 
 T-77  & AA Tau   & 04 34 55.42  & +24 28 53.2  &140 & K7    &...&30.0 &42&II&1.0e-4 & 1/0/:/1& - \\ 
 T-78  & HO Tau AB   & 04 35 20.20  & +22 32 14.6  &140 & M0.5   &6.9 &29.5 &108 &II&2.0e-5 & 0/:/:/0 &  \\ 
 T-79  & FF Tau AB  & 04 35 20.90  & +22 54 24.2  &140 & K7  &0.03 &29.8 &2 &III& $<$2.0e-6 & 0/:/:/0&  \\ 
 T-80  & HBC 412 AB  & 04 35 24.51  & +17 51 43.0  &140 & M2  &0.7 &...&9 &III& $<$4.0e-6 & :/:/:/:&  \\ 
 T-81  & DN Tau   & 04 35 27.37  & +24 14 58.9  &140 & M0   &...&30.0 &45 &II&3.0e-4 & 1/0/:/0&  \\ 
 T-82  & LkCa 14  & 04 36 19.09  & +25 42 59.0  &140 & M0   &...&...&1.1 &III& $<$4.0e-6 & :/:/:/:&  \\ 
 T-83  & HD 283759   & 04 36 49.12  & +24 12 58.8  &140 & F2   &...&...&...&III& $<$4.0e-6 & :/:/:/:&  \\ 
 T-84  & DO Tau & 04 38 28.58  & +26 10 49.4  &140 & M0   &...&...&101 &II&7.0e-5 & 1/0/1/0&Ext.OI, jet, high$\dot M$ \\ 
 T-85  & HV TauABC & 04 38 35.28  & +26 10 38.6  &140 & M1/M4  & 4.0 &29.57 &10 &III/II&2.0e-5 & 1/:/:/0& Jet, mult. \\ 
 T-86  & VY Tau AB  & 04 39 17.41  & +22 47 53.4  &140 & M0  &0.66 &...&7.3 &II&$<$5.0e-6 & 0/:/:/0&  \\ 
 T-87  & LkCa 15   & 04 39 17.80  & +22 21 03.5  &140 & K5   &...&...&18.5 &II&5.0e-4 & 1/0/0/0 &  \\ 
 T-88  & JH 223& 04 40 49.51  & +25 51 19.2  &140 & M2    &  2.1 &28.8 &4 &II& $<$3.0e-6 & :/:/:/:&  \\ 
 T-89  & IW TauAB   & 04 41 04.71  & +24 51 06.2  &140 & K7   &0.28 &30.0 &4 &III& $<$4.0e-6 & :/:/:/:&  Jet \\ 
 T-90  & CoKuTau/4  & 04 41 16.81  & +28 40 00.1  &140 & M1.5  & 0.05 &...&3 &II&5.0e-6 & 1/:/:/0&  \\ 
 T-91  & 04385+2550(Haro6-33)  & 04 41 38.8  & +25 56 26.8  &140 & M0  & 19  &29.6 &17 &I-II &...& 1/:/:/0&  \\ 
 T-92  & DP Tau  & 04 42 37.70  & +25 15 37.5  &140 & M0.5   & 0.11 & 29.0 &87 & II&$<$5.0e-6 & 1/1/1/0 & Jet \\ 
 T-93  & GO Tau   & 04 43 03.10  & +25 20 18.7  &140 & M0    &...&29.4 &80 &II &7.0e-4 & 0/:/:/0& \\ 
 T-94  & DQ Tau AB  & 04 46 53.04  & +17 00 00.5  &140 & K5 & 0.0004  &...&102 &II &2.0e-4 & 1/0/:/0& \\ 
 T-95  & Haro 6-37 AB  & 04 46 58.98  & +17 02 38.2  &140 & K7   &2.6/0.3 &...&13 &II &1.0e-4 & 1/:/:/0& Mult.\\ 
 T-96  & DS Tau   & 04 47 48.11  & +29 25 14.4  &140 & K5    &7.1 &...&38 &II&6.0e-5 & 1/:/:/0&  \\ 
 T-97  & UY Aur AB  & 04 51 47.37  & +30 47 13.5  &140 & K7   &0.88 &...&63 &II&2.0e-5 & 1/1/1/1 & Jet, high$\dot M$ \\ 
 T-98  & St 34   & 04 54 23.68  & +17 09 53.5  &110 & M3  &  1.2 &...&90&II& $<$5.0e-6 & :/:/:/:&  \\ 
 T-99  & GM Aur  & 04 55 10.99  & +30 21 59.2  &140 & K3   &...&29.8 &79 &II&3.0e-4 & 1/0/0/0&  \\ 
 T-100  & LkCa 19   & 04 55 36.96  & +30 17 55.3  &140 & K0   &...&30.7 &1.2 &III& 5.0e-6 & :/:/:/:&  \\ 
 T-101  & AB Aur  & 04 55 45.83  & +30 33 04.4  &140 & A0    &...&29.5 &33 &II& 4.0e-5 & 1/1/1/0& \\ 
 T-102  & SU Aur   & 04 55 59.38  & +30 34 01.6  &140 & G2    &  ... &31.1 &4 &II&9.0e-6 & 1/1/1/0& Jet \\ 
 T-103  & HBC 427   & 04 56 02.02  & +30 21 03.7  &140 & K7    & 0.03  &30.5 &1.4 &III& $<$7.0e-6 & :/:/:/:&  \\ 
 T-104  & V836 Tau  & 05 03 06.60  & +25 23 19.7  &140 & K8.5   &...&30.0 &7.7 &II&1.0e-4 & 0/:/:/0&  \\ 
 T-105  & CIDA-10   & 05 06 16.75  & +24 46 10.2  &140 & M4    &  0.08  &29.0 &9 &III& $<$5.0e-6 & :/:/:/:&  \\ 
 T-106  & RW Aur AB  & 05 07 49.54  & +30 24 05.1  &140 & K1   &1.4 &...&75 &II&4.0e-5 & 1/0/1/0& Ext.OI. Po12\\
 S-1 &  HIP 76310    &     15 35 16.10 &     -25 44 03.1 & 150 &  A0V    &...&...& $<$ &  D  &  3.6e-6 & 0/:/:/0 &  \\ 
   S-2 &  J153557.8-232405              &     15 35 57.80 &     -23 24 04.6 & 145 &           K3     &     0.05 &   30.04 &  $<$ &III& $<$4.2e-6 & :/:/:/:  &    \\ %
   S-3 &  J154413.4-252258           &     15 44 13.34 &     -25 22 59.1 & 145 &           M1    &...&   30.0 &   3.2 &III &$<$4.2e-6 & :/:/:/:&      \\ %
   S-4 &  HIP 77815                            &     15 53 21.93 &     -21 58 16.5 & 171 &          A5V     &...&...&...& - & $<$3.0e-6 & :/:/:/:&      \\ %
   S-5 &  HIP 77911                           &     15 54 41.60 &     -22 45 58.5 & 147 &      B9V     &     7.96 &...&$<$& D &$<$3.5e-6 & 0/:/:/0&   \\ %
   S-6 &  J155624.8-222555      &     15 56 24.77 &     -22 25 55.3 & 145 &           M4     &...&...&   5.4 &II& $<$3.7e-6 & :/:/:/:&   \\ %
   S-7 &  HIP 78099                          &     15 56 47.85 &     -23 11 02.6 & 140 &      A0V     &...&...&$<$& ... & $<$4.2e-6 & :/:/:/: &   \\ %
   S-8 &  J155706.4-220606      &     15 57 06.42 &     -22 06 06.1 & 145 &           M4     &...&...&   3.6 &II& $<$4.3e-6 & :/:/:/: &    \\ %
   S-9 &  J155729.9-225843      &     15 57 29.86 &     -22 58 43.8 & 145 &           M4     &...&...&   7.0 &II& $<$3.7e-6 & 0/:/:/0 &    \\ %
  S-10 &  J155829.8-231007        &     15 58 29.81 &     -23 10 07.7 & 145 &           M3     &...&...& 250 &II& $<$3.4e-6 & 0/:/:/0 &   \\ %
  S-11 &  RXJ1600.7-2343  &       16 00 44.60 &     -23 43 12.0 & 145 &           M2  &     1.46 &   30.4 &...&III & $<$3.8e-6 & 0/:/:/0&   \\ %
  S-12 &  J160108.0-211318            &     16 01 08.01 &     -21 13 18.5 & 145 &           M0     &...&   30.3 &  2.4 &III& $<$4.0e-6 & :/:/:/: &    \\ %
  S-13 &   J160210.9-200749     &     16 02 10.96 &     -20 07 49.6 & 145 &           M5     &...&...&   3.5&III& $<$3.7e-6 & :/:/:/: &    \\ %
  S-14 &   J160245.4-193037        &     16 02 45.45 &     -19 30 37.8 & 145 &           M5     &    28.2 &...&   1.1 &III& $<$3.6e-6 & :/:/:/: &    \\ %
  S-15 &  J160357.6-203105              &     16 03 57.68 &     -20 31 05.5 & 145 &           K5    &...&...&  12 &II& $<$3.7e-6 & 0/0/0/0 &    \\ %
  S-16 &   J160357.9-194210       &     16 03 57.94 &     -19 42 10.8 & 145 &           M2     &...&...&   3.0 &II& $<$3.7e-6 & 0/:/:/0 & \\ %
  S-17 &  J160421.7-213028              &     16 04 21.66 &     -21 30 28.4 & 145 &           K2     &    16.22 &   30.3 &   0.6 &II-III&  1.1e-4 & 1/1/0/0  &    \\ %
  S-18 &   J160525.5-203539        &     16 05 25.56 &     -20 35 39.7 & 145 &           M5    &...&...&   6.1 &III& $<$5.4e-6 & :/:/:/: &   \\ %
  S-19 &   J160532.1-193315        &     16 05 32.15 &     -19 33 16.0 & 145 &           M5     &...&...&  26 &III&$<$3.9e-6 & 0/:/:/0 &  \\ %
  S-20 &   J160545.4-202308      &     16 05 45.40 &     -20 23 08.8 & 145 &           M2     &...&...&  35 & II&  7.7e-6 &  0/0/0/0 &     \\ %
  S-21 &   J160600.6-195711        &     16 06 00.62 &     -19 57 11.5 & 145 &           M5    &...&...&   7.5 &II& $<$4.9e-6 & 0/:/:/0 &   \\ %
  S-22 &  ScoPMS 31                            &     16 06 21.96 &     -19 28 44.6 & 145 &        M0.5V     &     0.58 &   30.1 &  21 &II&      4.1e-6 & 0/0/:/0 &    \\ %
  S-23 &   J160622.8-201124       &     16 06 22.78 &     -20 11 24.4 & 145 &           M5  &...&...&   6.0 &II& $<$4.3e-6 & :/:/:/: &    \\ %
  S-24 &   J160643.8-190805     &     16 06 43.86 &     -19 08 05.6 & 145 &           K6     &...&...&   2.4 &...&...& :/:/:/: &  \\ %
  S-25 &  J160654.4-241610            &     16 06 54.36 &     -24 16 10.8 & 145 &           M3     &     1.50 &   29.9 &   3.6 &...&...& :/:/:/: &  \\ %
  S-26 &   J160702.1-201938        &     16 07 02.12 &     -20 19 38.8 & 145 &           M5     &     1.63 &...&  30 &II& $<$3.7e-6 & :/:/:/: &      \\ %
  S-27 &  HIP 78996                          &     16 07 29.93 &     -23 57 02.3 & 108 &      A9V     &...&...&$<$&D& $<$4.2e-6 & :/:/:/: &    \\ %
  S-28 &   J160801.4-202741       &     16 08 01.42 &     -20 27 41.7 & 145 &           K8    &...&   29.9 &   2.3 &...&...& :/:/:/: &    \\ %
  S-29 &   J160823.2-193001        &     16 08 23.25 &     -19 30 00.9 & 145 &           K9     &...&...&   6.0 &II&  4.4e-5 & 0/0/0/0 &   \\ %
  S-30 &   J160827.5-194904       &     16 08 27.52 &     -19 49 04.7 & 145 &           M5     &...&...&  12 &III& $<$5.3e-6 & :/:/:/: &    \\ %
  S-31 &  J160856.7-203346              &     16 08 56.73 &     -20 33 46.0 & 145 &           K5     &...&   30.1 &   0.5 &...&...& :/:/:/: &     \\ %
  S-32 &   J160900.7-190852     &  16 09 00.39 &     -19 08 44.8  & 145 &           K9 &...&   30.0&  13 & II &   2.5e-5 & :/:/:/: &  +J160900.0-190836  \\ 
  S-33 &  HIP 79156                             &     16 09 20.89 &     -19 27 25.9 & 170 &      A0V    &     0.89 &...&$<$&D& $<$3.3e-6 & :/:/:/: &   \\ %
  S-34 &   J160953.6-175446        &     16 09 53.62 &     -17 54 47.4 & 145 &           M3     &...&...&  22 &II& $<$4.5e-6 & :/:/:/: &    \\ %
  S-35 &   J160959.4-180009        &     16 09 59.33 &     -18 00 09.1 & 145 &           M4     &...&...&   4.0 &II& $<$5.1e-6 & 0/0/0/0 &    \\ %
  S-36 &   J161115.3-175721    &     16 11 15.34 &     -17 57 21.4 & 145 &           M1     &...&   30.2 &   2.4 &II& $<$6.3e-6 & :/:/:/: &    \\ %
  S-37 &  HIP 79410                              &     16 12 21.83 &     -19 34 44.6 & 140 &      B9V   &...&...&$<$&D& $<$4.6e-6 & :/:/:/: &  \\ %
  S-38 &  HIP 79439                              &     16 12 44.11 &     -19 30 10.2 & 131 &      B9V     &...&...&$<$&D& $<$3.9e-6 & 0/:/:/0 &   \\ %
  S-39 &  J161402.1-230101               &     16 14 02.12 &     -23 01 02.2 & 145 &           G4     &...&   30.2 & $<$ &...&...& :/:/:/: &    \\ %
  S-40 &  J161411.0-230536                &     16 14 11.08 &     -23 05 36.2 & 145 &           K0    &     0.22 &   30.8 &  0.8 &II&    6.0e-6 & 0/:/:/0 &   \\ %
  S-41 &  J161420.3-190648 &       16 14 20.30 &     -19 06 48.1 & 145 &           K5  &...&   29.3 &  52 &II&  1.7e-5 & 1/1/1/0 &  \\ %
  S-42 &  HIP 79878                            &     16 18 16.17 &     -28 02 30.1 & 129 &      A0V     &...&...&$<$&D& $<$4.2e-6 & 0/:/:/0 &   \\ %
  S-43 &  HIP 80088                             &     16 20 50.23 &     -22 35 38.7 & 139 &      A9V     &...&...&$<$&D& $<$3.7e-6 & 0/:/:/0 &    \\ %
  S-44 &  HIP 80130                             &     16 21 21.15 &     -22 06 32.3 & 144 &      A9V     &...&...&$<$&...& $<$4.4e-6 & :/:/:/: &    \\ %
E-1 	& RECX18 & 08 36 10.7 & -79 08 18.4 &97 & M5.3/M5.3   &$<$0.04 &30.6 &...&III&...& :/:/:/: & \\ 
E-2 & RECX1 (EG Cha) & 08 36 56.24 & -78 56 45.5 &97 & K7/M0   &0.2 &30.6 &1.4 &III& ... & 0/:/:/0 & \\
E-3 & RECX17 & 08 38 51.50 & -79 16 13.7 &97 & M5.0/M5.0   &$<$0.04 &...&...&III&...& :/:/:/: & \\ 
E-4 & RECX14 (ES Cha) & 08 41 30.3 & -78 53 06.5 &97 & M4.7   &...&...& 12 &TO&3e-7& 0/:/:/0 &  \\ 
E-5 & RECX3 (EH Cha) & 08 41 37.04 & -79 03 30.4 &97 & M3.0   &...&29.1 &2.2 & TO &3.5e-10& 0/:/:/0 & \\ 
E-6 & RECX13 (HD75505) & 08 41 44.72 & -79 02 53.2 &97 & A5   &...&...&...&III&...& :/:/:/: & \\
E-7 & RECX4 (EI Cha) & 08 42 23.73 & -79 04 03.0 &97 & M1.3  &...&30.1 &2.3 & TO &2e-9& 0/:/:/0 &  \\ 
E-8 & RECX5 (EK Cha)& 08 42 27.11 & -78 57 47.9 &97 & M3.8   &...&29.0 &35 & TO &7e-6& 0/:/:/0&  \\ 
E-9 & RECX6 (EL Cha) & 08 42 38.80 & -78 54 42.8 &97 & M3.0   &...&29.5 &3.6 &III&...& 0/:/:/0 & \\ 
E-10 & RECX7 (EM Cha) & 08 43 07.24 & -79 04 52.5 &97 & K6.9/M1   & 0.001 &30.3 &0.4 &III&...& :/:/:/: & \\ 
E-11 & RECX8 (RS Cha AB)& 08 43 12.23 & -79 04 12.3 &97 & A7/A8  & e &29.8 & $<$ &III&...& 0/:/:/0 &  \\ 
E-12 & RECX15 (ET Cha) & 08 43 18.58 & -79 05 18.2 &97 & M3.4   & ... & 28.8 & 90 &II& 2.5e-8 & 1/0/0/0 & Woi11\\ 
E-13 & RECX16 (J0844.2-7833) & 08 44 09.15 & -78 33 45.7 &97 & M5.5 &...&...&...& II &...& 0/:/:/0 &  \\ 
E-14 & RECX9 (EN Cha) & 08 44 16.38 & -78 59 08.1 &97 & M4.4/M4.7  &0.2 &28.5 &10 & TO &1.4e-6& 0/:/:/0 &  \\ 
E-15 & RECX10 (EO Cha) & 08 44 31.88 & -78 46 31.2 &97 & M0.3 &...&30.0 &1.0 &III&...& 0/:/:/0 &  \\ 
E-16 & RECX11 (EP Cha) & 08 47 01.66 & -78 59 34.5 &97 & K6.5   &...&30.1 &3.2 & II &3.3e-5& 1/:/:/0 &  \\ 
E-17 & RECX12 (EQ Cha) & 08 47 56.77 & -78 54 53.2 &97 & M3.2/M3.2   &0.04 &30.1 &4.2 &III&...& 0/:/:/0 &\\ 
 W-1  & TWA21    & 10 13 14.76  & -52 30 54.1  &69 & K3   &...&30.2& 3 & III& $<$1.0e-6 & :/:/:/:&  \\ 
 W-2  & TWA07    & 10 42 30.11  & -33 40 16.2  &38 & M1    &...&29.6 &5 & TO &6.0e-7 & 0/:/:/0& \\ 
 W-3  & TWA01 (TW Hya)  & 11 01 51.92  & -34 42 17.0  &58 & M2.5   &...& $<$30.4 &220 &II&6e-4 & 1/0/0/1&Thi10 \\ 
 W-4  & TWA02AB   & 11 09 13.8 & -30 01 39.8  &52 & M2 &2 &29.4 &2 &...& $<$1.0e-6 & 0/:/:/0& \\ 
 W-5  & TWA03A (Hen3-600A)  & 11 10 27.88  & -37 31 52.0  &42 & M3e  &10 &29.2 &22 & TO &1.1e-5 & 0/:/:/0&  \\ 
 W-6  & TWA12    & 11 21 05.50  & -38 45 16.3  &32 & M2    &...& 29.1 &51 &...& $<$1.0e-6 & :/:/:/:&  \\ 
 W-7  & TWA13AB    & 11 21 17.24  & -34 46 45.5  &38 & M2e    &5.1 & 29.4 &4 & TO & ... & 0/:/:/0& confused region \\ 
 W-8  & TWA04AB  (HD98800AB) & 11 22 05.30  & -24 46 39.3  &46 & M5    & 0.8 & 29.9 &$<$&D &1e-6 & 1/0/:/0& Mult. \\ 
 W-9  & TWA05Aab    & 11 31 55.26  & -34 36 27.2  &50 & M2    &2 &29.8 &13.4 &...& $<$1.0e-6 & :/:/:/:&  \\ 
 W-10  & TWA23    & 12 07 27.38  & -32 47 00.3  &37 & M1   &...&29.2&$<$ &...& $<$1.0e-6 & 0/:/:/0&  \\ 
 W-11  & TWA25    & 12 15 30.72  & -39 48 42.6  &44 & M0    &...&29.8&2 &...& $<$1.0e-6 & :/:/:/:&  \\ 
 W-12  & TWA16    & 12 34 56.30  & -45 38 07.6  &66 & M1.5    & 0.7 & 29.6 &4 &...& $<$1.0e-6 & :/:/:/:&  \\ 
 W-13  & TWA10    & 12 35 04.25  & -41 36 38.6  &57 & M2.5    &...&29.6 &11 &...& $<$1.0e-6 & 0/:/:/0 &  \\ 
 B-1  & HD 203   & 00 06 50.09  & -23 06 27.1  &39 & F2IV   &...& 28.9 &...& D &9e-9 &:/:/:/: & \\ 
 B-2  & HD 14082B    & 02 17 25.02  & +28 44 36.3  &39 & F5V    &  (10) & 30.0 &...&D& 1.5e-8 &:/:/:/: &SN03  \\ 
 B-3  & AG Tri    & 02 27 29.25  & +30 58 24.6  &42 & K8    &(22)&...&...&D& $>$1.0e-10 &:/:/:/: &SN03 \\ 
 B-4  & HIP 12545    & 02 41 25.89  & +05 59 18.4  &41 & M0    &sb&...& 0.6 &...&...&:/:/:/: & \\ 
 B-5   & 51 Eri (HD 29391)  & 04 37 36.13  & -02 28 24.8  &30 & F0V    &66 &...&...&...&...&:/:/:/: &\\ 
 B-6  & GJ 3305   & 04 37 37.47  & -02 29 28.4  &30 & M0.5    &...& 30.2 & 2.2 &...&...&:/:/:/: &  \\ 
 B-7  & AF Lep (HD 35850)  & 05 27 04.76  & -11 54 03.5  &27 & F7V  &sb& 30.3 & $<$ &...&...&:/:/:/: & \\ 
 B-8  & AO Men    & 06 18 28.21  & -72 02 41.5  &39 & K7  &...& 30.2 & 0.6 &...&...&:/:/:/: &\\ 
 B-9  & HD 139084AB   & 15 38 57.23  & -57 42 22.7  &40 & K0V    &10.7 &...&...&...&...&:/:/:/: &  \\ 
 B-10  & HD 146624 (HR6070)   & 16 18 17.90  & -28 36 50.5  &43 & A0V    &...&...&...&...&...&:/:/:/: &  \\ 
 B-11  & HD 164249    & 18 03 03.41  & -51 38 56.4  &47 & F5V    &16& 30.6 & $<$ &D& $>$4e-10 &0/:/:/0 &Nil09 \\ 
 B-12  & HD 172555    & 18 45 26.90  & -64 52 16.5  &29 & A7V    &68.5 & 28.8 &...&D& 2e-9 &1/:/:/0 & Nil09,Riv12\\ 
 B-13  & CD-64 1208   & 18 45 37.03  & -64 51 46.1  &29 & K7   &  0.2 & 29.9 &...&...&...&:/:/:/: & \\ 
 B-14  & PZ Tel (HD 174429)   & 18 53 05.87  & -50 10 49.9  &50 & K0Vp   &  0.3 & 30.6 &$<$ & D &(1e-8) &:/:/:/: & \\ 
 B-15  & $\eta$ Tel AB (HD 181296)    & 19 22 51.21  & -54 25 26.1  &48 & A0V    &4.2 &$<$28.9&...&D&3e-8 &0/:/:/0 &  \\ 
 B-16  & AT Mic AB (GJ 799A)  & 20 41 51.16  & -32 26 06.8  &10.2 & M4.5e    &3.3 & 29.4 & 10.9 &...&...&0/:/:/0 & \\ 
 B-17  & HD 199143AB    & 20 55 47.67  & -17 06 51.0  &48 & F8V    & 1.1   & 30.6 &$<$&...&...&:/:/:/: &  \\ 
 B-18  & HD 181327    & 19 22 58.94  & -54 32 17.0  &51 & F5.5V      &...& $<$29.4  &...&D & 1.5e-7 &0/0/0/0 & Nil09,Leb12\\
 H-1  & HD 105   & 00 05 52.55  & -41 45 11.0  &40 & G0V             &...&...&...&D & $>$2e-9 & 0/:/:/0& Nil10,M06\\ 
 H-2  & HD 1466   & 00 18 26.12  & -63 28 39.0  &41 & F9V           &...&29.6    &...& D &$>$3e-10& :/:/:/:&  \\ 
 H-3  & HD 2884/5    & 00 31 33.07  & -62 57 41.3  &43 & B9V/F2V      & 2.4/0.4 & $<$28.8 &...&D&...& :/:/:/:& Unassociated pair\\ 
 H-4  & HD 3003   & 00 32 43.91  & -63 01 53.4  &46 & A0V            & 0.1  & $<$28.7  &...& D &$>$1e-10& 0/:/:/0& \\ 
 H-5  & HD 3221   & 00 34 51.20  & -61 54 58.1  &46 & K5V          &...&30.0     & 0.7 &...&...& :/:/:/:&  \\ 
 H-6  & HIP 3556  & 00 45 28.15  & -51 37 33.9  &39 & M1.5                 &...&...& 0.8 &...&...& :/:/:/:&  \\ 
 H-7  & HD 12039 (DK Cet)  & 01 57 48.98  & -21 54 05.3  &42 & G3/5V             &   0.2 &29.6&...&D &$>$5e-10& :/:/:/:&  \\
 H-8  & GSC8056-482   & 02 36 51.54  & -52 03 04.4  &25 & M3Ve                &...&    29.7  & 5.3 &...&...& :/:/:/:&  \\ 
 H-9  & HD 16978 ($\epsilon$ Hyi)  & 02 39 35.36  & -68 16 01.0  &47 & B9V      &...&    ...     &...&...&...& :/:/:/:&  \\ 
 H-10  & HD 30051    & 04 43 17.20  & -23 37 42.0  &58 & F2/F3IV/V   &...&   29.8  &...& D &...& :/:/:/:& \\ 
 H-11  & HD 44627 (AB Pic) & 06 19 12.91  & -58 03 15.5  &46 & K2V           &5.5 &   30.0      &...& D &...& :/:/:/:&  \\ 
 H-12  & HD 53842    & 06 46 13.54  & -83 59 29.5  &57 & F5V     &...&...&...& D   &...& :/:/:/:& \\ 
 H-13  & HD 55279    & 07 00 30.49  & -79 41 46.0  &64 & K3V      &...&29.9&...&...&...& :/:/:/:&  \\ 
 H-14  & HD 202917   & 21 20 49.96  & -53 02 03.1  &46 & G5V    &...&30.1     &...& D &$>$5e-9& :/:/:/:& \\ 
 H-15  & HIP 107345   & 21 44 30.12  & -60 58 38.9  &42 & M1                  &...&29.4    & 1.4 &...&...& :/:/:/:&  \\ 
 H-16  & HD 224392 ($\eta$ Tuc)  & 23 57 35.08  & -64 17 53.6  &49 & A1V       &...& $<$29.3 &...&...&...& :/:/:/:& \\ 
 A-1  & HD 9672 (49 Cet)   & 01 34 37.78  & -15 40 34.9  &59 & A4V   &...&...&...&D?&3e-7 &0/1/0/0& Hu08,Zu12\\ 
 A-2  & HD 31648 (MWC480) & 04 58 46.27  & +29 50 37.0  &137 & A5V  &...& 29.4 &...&gr.II&3.6e-4 &1/0/1/1&Jet \\ 
 A-3  & HD 32297   & 05 02 27.44  & +07 27 39.7  &112 & A0     &...&...&...&D & 3e-6 &0/0/0/0& Ma08 \\ 
 A-4  & HD 35187   & 05 24 01.17  & +24 57 37.6  &114 & A2V/A7V  & 1.4&...&...&gr.II &5e-5 &1/0/0/0&\\ 
 A-5  & HD 36112 (MWC758) & 05 30 27.53  & +25 19 57.1  & 279 & A5IV   &...&...&...&gr.I & 3e-5 &1/0/1/0& Cha08\\ 
 A-6  & HD 36910 (CQTau)  & 05 35 58.47  & +24 44 54.1  &113 & F2Ve   &...&...&...&gr.II&1e-5 &1/0/1/0& Cha08 \\ 
 A-7  & HR 1998 ($\zeta$ Lep)  & 05 46 57.34  & -14 49 19.0  &22 & A2V   &...&...&...&D &$>$3e-12 &0/0/0/0& Mo07\\ 
 A-8  & HD 97048 (CU Cha) & 11 08 03.34  & -77 39 17.5  &158 & A0   &...& 29.5 &...&gr.I &9.2e-4 &1/1/1/0& Ski04\\ 
 A-9  & HD 100453   & 11 33 05.58  & -54 19 28.5  &121 & A9V   & 1.1 & 28.8 &...&gr.I &2.1e-4 &1/0/0/0& Co09\\ 
 A-10  & HD 100546   & 11 33 25.44  & -70 11 41.2  &97 & B9V     &...& 28.9 &...&gr.I&2.9e-4 &1/1/1/0& Fei03,Gra05\\ 
 A-11  & HD 104237 (DX Cha) & 12 00 05.08  & -78 11 34.6  &115 & A8    & sb  & 30.2 &...&gr.II &7.8e-5 &1/0/0/0& Jet. Fei03\\ 
 A-12  & HR 4796A (TWA11)  & 12 36 01.0  & -39 52 10.2  &73 & A0  & 7.8 &29.4&...&D &1e-5 &0/0/0/0& Aug99\\ 
 A-13   & HD 135344B (SAO 206462)   & 15 15 48.4  & -37 09 16.0  &142 & F4V   &...& 29.7 &...&gr.I & 1.6e-4  &1/0/0/0& Pon08\\ 
 A-14  & HD 139614   & 15 40 46.38  & -42 29 53.5  &140 & A7V  &...&...&...&gr.I &2.7e-4 &1/0/0/0& Ack04\\ 
 A-15  & HD 141569   & 15 49 57.75  & -03 55 16.4  & 116 & B9.5V  & (6?) &$<$28.1&...&TO? & 1e-6 &1/1/0/0& Ste06\\ 
 A-16  & HD 142666 (V1026Sco) & 15 56 40.02  & -22 01 40.0  & 145 & A8V    &...&...&...&gr.II &1.6e-4 &1/0/0/0&  \\ 
 A-17  & HD 142527  & 15 56 41.89  & -42 19 23.3  &233 & F6III   &...&...&...&gr.I &1.5e-3 &1/0/0/0& Ack04\\ 
 A-18  & HD 144668 (HR 5999)  & 16 08 34.29  & -39 06 18.3  &163 & A7IVe   & 1.2 &28.3&...&gr.II &9e-5 &1/0/1/0& Ste10 \\ 
 A-19  & HD 150193 (MWC863)  & 16 40 17.92  & -23 53 45.2  & 216 & A2IVe    & 1.1 & 29.6 &...&gr.II & 2e-5 &1/0/0/0& Ste06\\ 
 A-20  & KK OphAB & 17 10 08.06  & -27 15 18.2  & 260 & A6/G5V   & 1.6 &...&...&gr.II&2e-5 &1/1/1/0& \\ 
 A-21  & HD 158352 (HR 6507) & 17 28 49.65  & +00 19 50.2 &60 & A7Vp    &...&...&...&D & 2e-7 &0/0/:/0& Rh07 \\ 
 A-22  & HD 158643 (51 Oph)  & 17 31 24.95  & -23 57 45.5  &124 & B9.5V   &...&$<$29.0&...&gr.II &1e-6 &1/0/0/0&vdA01\\ 
 A-23  & HD 163296 (MWC275) & 17 56 21.29  & -21 57 21.9  &119 & A1V    &...& 29.6 &...&gr.II& 6.5e-4 &1/0/1/1& Jet. GS09,Till12 \\ 
 A-24  & HD 169142 (MWC925) & 18 24 29.78  & -29 46 49.4  & 145 & A7V   &9.3 & 29.1 &...&gr.I&2.4e-4 &1/0/1/0& Gra07,Meu10 \\ 
C-1  &  DK Cha  &   12 53 17.23  &  -77 07 10.7   &  178  &   F0    &...& $<$29.0  &   88    & II  &   4.0e-3& 1/:/:/0& Ext.OI, jet. vK10\\
C-2  &  IRAS12500-7658          &    12 53 42.86   & -77 15 11.5  &   178    & M6.5   &...&   $<$29.3   &  20    &  I   &...& 1/:/:/0 & \\
C-3  &  Sz46N                   &     12 56 33.66  &  -76 45 45.3 &    178  &   M1    &...&  29.3   &  16  &    II  &   5.0e-5& 0/:/:/0& \\
C-4 &   IRAS12535-7623  &  12 57 11.77  &  -76 40 11.3  &   178  &   M0       &...&  29.3 &    15  &    II &   1.6e-4 & 0/:/:/0& \\
C-5 &  ISO-ChaII 13           &      12 58 06.78  & -77 09 09.4  &   178  &   M7       &...&...&  101   &   II   &  1.6e-6  & 0/:/:/0& \\
C-6 &   Sz50 &  13 00 55.36 &  -77 10 22.1   &  178  &   M3     &  ... &   29.5 &    29   &   II  &  1.0e-3  & 1/:/:/0& \\
C-7 &   Sz51               &          13 01 58.94  &  -77 51 21.7  &   178  &   K8.5     &...&   $<$29.5   & 102   &   II   &  5.0e-5 & 1/:/:/0 & \\ 
C-8 &   C50                 &         13 02 22.85  &  -77 34 49.3   &  178  &   M5     &...&...&   36  &    II &   1.0e-6& 0/:/:/0& \\
C-9  &   Sz52               &          13 04 24.92   & -77 52 30.1  &   178  &   M2.5       &...&   $<$30.9 &    48 &    II  &   8.0e-4& 1/:/:/0& \\
C-10 &  Hn25            &             13 05 08.53 &   -77 33 42.4 &    178  &   M2.5     &...&   $<$29.2   &  24  &   II  &   1.0e-5& 0/:/:/0& \\
C-11  & Sz53          &               13 05 12.69  &  -77 30 52.3  &   178  &   M1      &...&  $<$29.0   &  46  &    II  &  1.0e-5& 0/:/:/0& \\
C-12 &   Sz54            &             13 05 20.68 &   -77 39 01.4  &   178  &   K5       &...&    28.6  &   23    &  II &  5.0e-4& 1/:/:/0& \\
C-13 &  J130521.6-773810       &                     13 05 21.66  &  -77 38 10.0   &  178  &   ...  &...&   ...    & 29   &   I-II   &  1.0e-6& 1/:/:/0& embedded? \\
C-14 &  J130529.0-774140           &                13 05 29.04  &  -77 41 40.1  &   178  &...&...&...&...&    II &...& 0/:/:/0& \\
C-15 &  C62     &                     13 07 18.05&    -77 40 52.9  &   178   &  M4.5    &...&...&     34  &    II   &  1.0e-5& 0/:/:/0& \\
C-16  & Hn26 &                        13 07 48.51&    -77 41 21.4  &   178 &    M2     &...& $<$29.1   &  10   &   II  &   1.0e-5& 0/:/:/0& \\
C-17  & Sz61  &                       13 08 06.28  &  -77 55 05.2 &    178  &   K5     &...& $<$31.0  &   84   &   II &   1.6e-3 & 1/:/:/0&\\
C-18 &  C66  &                        13 08 27.17  &  -77 43 23.2  &   178  &   M4.5    &...&...&    30  &    II   & 1.0e-6 & 0/:/:/0& \\
C-19 &  Sz62  &                       13 09 50.38  &  -77 57 23.9  &   178 &    M2.5     & 1.1 &  $<$32.4   & 150    &  II &    1.0e-6 & 0/:/:/0& \\
       
\enddata
\tablecomments{Table \ref{tab:slist} is ordered by association or group and then RA. Coordinates are the pointing position of the observations, and may be centred between multiple systems. Key to the naming convention is: T=Taurus, S=Upper Sco, E= $\eta$ Cha, W=TW Hya, B=$\beta$ Pic Moving group, H=Tuc Hor, A=Herbig AeBe stars and A stars with debris disks (noted in column 10 as D) and C=ChaII.}

\tablerefs{Ack04 - \citet{Ack04}, Aug99 - \citet{Aug99}, Cha08 - \citet{Cha08}, Co09 - \citet{Co09}, Fei03 - \citet{Fei03}, Gra05 - \citet{Grad05}, Gra07 - \citet{Grad07}, GS09 - \citet{GS09}, Hu08 - \citet{Hu08}, Leb12 - \cite{Leb12}, Ma08 - \citet{Ma08}, Meu10 - \citet{Meus10}, Mo07 - \citet{Mo07}, M04 - \citet{Mam04}, M06 - \citet{M06}, Nil09 - \citet{Nil09}, Nil10 - \citet{Nil10}, Po12 - \citet{Pod12}, Pon08 - \citet{Pont08}, Rh07 - \citet{Rh07}, Riv12 - \citet{Riv12b}, Ski04 - \citet{Ski04}, Smi06 - \citet{Smi06}, SN03 - \citet{SN03}, Ste06 - \citet{Ste06}, Till12 - \citet{Till12}, Thi10 - \citet{Thi10}, vdA01 - \cite{vdA01}, vK10 - \cite{vanK10}, Woi11 - \citet{Woit11}, Zu12 - \citet{Zu12}.
Additional references are given in the text.}

\tablenotetext{a}{ Spectral type. For multiple systems, we give either only the primary, or the brightest components when these are separable in PACS.}
\tablenotetext{b}{ For multiple systems, separation, in arcsec, between the two main components. 'sb' indicates a spectroscopic binary; 'e' means eclipsing binary.}
\tablenotetext{c}{X-ray luminosity, between 0.3-10keV, in erg s$^{-1}$}
\tablenotetext{d}{ H$\alpha$ EW is given as positive for an emission component, `$<$' for photospheric absorption and blank for no measurement. An average from published values for the most luminous component in indicated, or both components when they are separable with PACS. Note that the H$\alpha$EW is highly variable in many objects.}
\tablenotetext{e}{ SED Class from the literature, if available: either I-II, II (defined as $-0.3>\alpha_{IR}>-1.6$), III ($\alpha_{IR}<-1.6$, which includes stars without a measured IR excess), TO (Transition Object, defined as a Class III object with an additional excess at wavelengths longer than 10$\micron$), and D (classed as a debris disk). Note that for the Herbig AeBe stars, the SED classification is instead based on the Meeus SED groups I and II - see $\S$5.0.9 for details. Blank means no excess is known in either the IR or longer wavelengths.}
\tablenotetext{f}{ Disk {\it dust} mass, in Solar units, derived in most cases from published mm/sub-mm observations (see section $\S$5.0 for references). Upper limits are 3-$\sigma$, and a dashed line indicates no published values were available. Lower limits are normally based on fits to FIR photometry, where no sub-mm datapoint is available. Note that disk mass for protoplanetary disks is normally assumed to be 100.$M_{dust}$.}
\tablenotetext{g}{ Summary of line detections from PACS of [OI]63$\micron$, [CII]157$\micron$, CO J-18-17 and H$_2$O (any transition detected, most commonly the line at 63.3$\micron$ - see text). '1' indicates detection, '0' means not detected, ':' means not observed.}
\tablenotetext{h}{ 'Jet' indicates published evidence of an optical jet, 'Ext.OI' indicates the [OI]63$\micron$ emission appears extended and 'high$\dot M$' indicates high published
mass loss rate (see \S5.2.1 and 5.2.2 for details). 'Mult.' indicates a multiple hierarchical system.}
\tablenotetext{i}{ Membership is now in doubt - see \citet{Luh09}.}
\label{tab:targets}
\end{deluxetable}


\clearpage



\clearpage

\clearpage


\clearpage






\end{document}